%

\font\sevenrm=cmr7  

\pageno=1
\hyphenation{Hipparcos pa-ra-llax-es con-si-de-ra-ble de-ter-mi-na-tion
tri-go-no-met-ric pho-to-me-try dis-tri-bu-tion con-si-der-ed ob-ser-va-tion
re-pre-sen-ta-tive Astro-me-tric mag-ni-tu-des summa-ri-sed pro-vi-de}

\def\ddeg{\hbox{.\hskip-3pt $^\circ$}}

\def\pmb#1{\setbox0=\hbox{#1}%
  \kern-.025em\copy0\kern-\wd0
  \kern.05em\copy0\kern-\wd0
  \kern-.025em\raise.0433em\box0 }


\def\la{\mathrel{\mathchoice {\vcenter{\offinterlineskip\halign{\hfil
$\displaystyle##$\hfil\cr<\cr\sim\cr}}}
{\vcenter{\offinterlineskip\halign{\hfil$\textstyle##$\hfil\cr
<\cr\sim\cr}}}
{\vcenter{\offinterlineskip\halign{\hfil$\scriptstyle##$\hfil\cr
<\cr\sim\cr}}}
{\vcenter{\offinterlineskip\halign{\hfil$\scriptscriptstyle##$\hfil\cr
<\cr\sim\cr}}}}}

\def\ga{\mathrel{\mathchoice {\vcenter{\offinterlineskip\halign{\hfil
$\displaystyle##$\hfil\cr>\cr\sim\cr}}}
{\vcenter{\offinterlineskip\halign{\hfil$\textstyle##$\hfil\cr
>\cr\sim\cr}}}
{\vcenter{\offinterlineskip\halign{\hfil$\scriptstyle##$\hfil\cr
>\cr\sim\cr}}}
{\vcenter{\offinterlineskip\halign{\hfil$\scriptscriptstyle##$\hfil\cr
>\cr\sim\cr}}}}}%

\MAINTITLE{The Hyades: distance, structure, dynamics, and age
\FOOTNOTE{Based on observations made with the ESA Hipparcos astrometry 
satellite. Table~2 is also available in electronic form at the CDS 
via anonymous ftp to cdsarc.u-strasbg.fr (130.79.128.5) or via 
http://cdsweb.u-strasbg.fr/Abstract.html}}

\AUTHOR=
{M.A.C.$\,$Perryman@{1,2}, 
A.G.A.$\,$Brown@{1},
Y.$\,$Lebreton@{3},
A.$\,$G\'omez@{3},
C.$\,$Turon@{3}, 
G.$\,$Cayrel de Strobel@{3},
J.C.$\,$Mermilliod@{4}, 
N.$\,$Robichon@{3},
J.$\,$Kovalevsky@{5}, 
F.$\,$Crifo@{3}}


\INSTITUTE={
    @{\phantom{1}1}\ Sterrewacht Leiden, Postbus 9513, 2300RA Leiden, The Netherlands

    @{\phantom{1}2}\ Astrophysics Division, European Space Agency,
         ESTEC, Noordwijk 2200AG, The Netherlands

    @{\phantom{1}3}\ DASGAL/URA CNRS 335, Section d'Astrophysique, 
	Observatoire de Paris, F--92195 Meudon Cedex, France

    @{\phantom{1}4}\ Universit\'e de Lausanne, Chavannes-des-Bois, CH-1290,
	Switzerland

    @{\phantom{1}5}\ Observatoire de la C\^ote d'Azur, CERGA, Avenue Copernic,
         F--06130 Grasse, France
}

\ABSTRACT={We use absolute trigonometric parallaxes from the Hipparcos
Catalogue to determine individual distances to members of the Hyades
cluster, from which the 3-dimensional structure of the cluster can be
derived. Inertially-referenced proper motions are used to rediscuss
distance determinations based on convergent-point analyses. A
combination of parallaxes and proper motions from Hipparcos, and
radial velocities from ground-based observations, are used to
determine the position and velocity components of candidate members
with respect to the cluster centre, providing new information on
cluster membership: 13 new candidate members within 20~pc of the
cluster centre have been identified. Farther from the cluster centre
there is a gradual merging between certain cluster members and field
stars, both spatially and kinematically. Within the cluster, the
kinematical structure is fully consistent with parallel space motion
of the component stars with an internal velocity dispersion of about
0.3~km~s$^{-1}$. The spatial structure and mass segregation are
consistent with $N$-body simulation results, without the need to
invoke expansion, contraction, rotation, or other significant
perturbations of the cluster.  The quality of the individual distance
determinations permits the cluster zero-age main sequence to be
accurately modelled.  The helium abundance for the cluster is
determined to be Y~=~$0.26\pm0.02$ which, combined with isochrone
modelling including convective overshooting, yields a cluster age of
$625\pm50$~Myr.  The distance to the observed centre of mass (a
concept meaningful only in the restricted context of the cluster
members contained in the Hipparcos Catalogue) is $46.34\pm0.27$~pc,
corresponding to a distance modulus $m-M=3.33\pm0.01$~mag for the
objects within 10~pc of the cluster centre (roughly corresponding to
the tidal radius).  This distance modulus is close to, but
significantly better determined than, that derived from recent
high-precision radial velocity studies, somewhat larger than that
indicated by recent ground-based trigonometric parallax
determinations, and smaller than those found from recent studies of
the cluster convergent point. These discrepancies are investigated and
explained.  }

\KEYWORDS={astrometry -- parallaxes -- HR Diagram -- Hyades -- distance scale}

\THESAURUS={ 05.01.1; 08.04.1; 08.08.1; 10.15.2; 12.4.3}

\DATE={Received: }          

\maketitle

\AUTHORRUNNINGHEAD{M.A.C.~Perryman et al.}

\MAINTITLERUNNINGHEAD{The Hyades: distance, structure, dynamics, and age}

\setbox1=\vbox
{\sevenrm\baselineskip 8.6pt
\hbox{ \kern 0.0truecm
\vbox {
\halign {
\tabskip 7pt
#\hfil& 
\tabskip 7pt
\hfil#\hfil& 
\tabskip 2pt
#\hfil& 
\tabskip 0pt
#\hfil \cr
\noalign{\hrule}
\noalign {\smallskip}
D.M.& Year& Author& Method \cr
\noalign {\smallskip}
\noalign{\hrule}
\noalign {\smallskip}
2.75&          1939& Smart		&Convergent point \cr
2.91&          1945& Seares 		&Convergent point \cr
3.03$\pm$0.06& 1952& van Bueren		&Convergent point (GC, N30, NWZC) \cr
2.85&          1955& Pearce 		&Convergent point \cr
3.08&          1956& Heckmann...	&Convergent point \cr
3.04&          1965& Wayman...		&Convergent point \cr
3.23$\pm$0.12& 1967& Wallerstein...	&Dynamical parallaxes \cr
3.14$\pm$0.19& 1967& Eggen		&Trigonometric parallaxes (Yale) \cr
3.37&          1967& Iben		&Stellar interiors \cr
3.08$\pm$0.07& 1969& Sears...		&Stebbins photometric parallaxes \cr
3.10$\pm$0.06& 1969& Eggen		&R$-$I photometric parallaxes \cr
3.25$\pm$0.20& 1969& Helfer		&Wilson-Bappu \cr
3.09$\pm$0.06& 1970& Upton		&pm gradient (FK4, N30, Yale, SAO) \cr
3.23$\pm$0.25& 1970& Lutz		&Wilson-Bappu \cr
3.19$\pm$0.06& 1971& Upton		&UBV photometric parallaxes \cr
3.25$\pm$0.20& 1972& Golay		&Geneva photometric parallaxes \cr
3.30&          1972& Iben...		&Stellar interiors \cr
3.23&          1973& Koester...		&Stellar interiors \cr
3.21&          1974& van Altena		&Mean of secondary indicators \cr
3.29$\pm$0.20& 1974& Upgren		&Trigonometric parallaxes (van Vleck) \cr
3.29$\pm$0.08& 1975& Hanson		&Compilation of methods to date \cr
3.19$\pm$0.15& 1975& Klemola...		&Trigonometric parallaxes (Lick) \cr
3.19$\pm$0.04& 1975& Corbin... 		&Proper motions from meridian circles \cr
3.42$\pm$0.20& 1975& Hanson		&Absolute pm's wrt extragalactic \cr
3.18$\pm$0.16& 1977& McAllister		&Absolute pm's corrected \cr
3.10$\pm$0.17& 1977& Buchholz		&GCTSP + systematic corrections \cr
3.32$\pm$0.06& 1977& Hanson		&Proper motion gradients \cr
3.30$\pm$0.04& 1980& Hanson		&Weighted mean of geometric methods \cr
3.25$\pm$0.08& 1980& Hanson		&Trigonometric parallaxes \cr
3.40$\pm$0.29& 1981& Hauck		&Gliese/field + Lutz-Kelker correction \cr
3.30&          1981& Hardorp		&Masses of visual binaries \cr
3.47$\pm$0.05& 1982& McClure		&Masses of visual binaries \cr
3.20&          1982& Eggen		&Photoelectric photometry of 72 stars \cr
3.30&          1983& Morris...		&Convergent point \cr
3.45$\pm$0.05& 1984& VandenBerg...	&Stellar evolution theory \cr
3.23&          1984& Detweiler...	&Revised radial velocity \cr
3.26$\pm$0.11& 1985& Cameron		&Main sequence versus Gliese stars \cr
3.33&          1985& Stefanik...	&Vrad (212 stars) + Hanson pm \cr
3.42$\pm$0.10& 1987& Loktin...		&Proper motion geometry in FK4 \cr
3.36$\pm$0.05& 1987& Peterson...	&McClure data plus new photometry \cr
3.28$\pm$0.10& 1988& Gunn...		&Vrad from Griffin + bulk Hanson pm \cr
3.35$\pm$0.07& 1988& Heintz		&5 binaries \cr
3.42$\pm$0.10& 1989& Loktin...		&Proper motion gradient \cr
3.37$\pm$0.07& 1990& Schwan		&Proper motions from 44 FK5 \cr
3.30$\pm$0.10& 1990& Upgren...		&Parallaxes (van Vleck, 23 stars) \cr
3.18$\pm$0.09& 1991& Patterson...	&Parallaxes (McCormick, 10 stars) \cr
3.40$\pm$0.04& 1991& Schwan		&Proper motions from 145 FK5/PPM \cr
3.45$\pm$0.06& 1992& Morris		&Convergent point \cr
3.16$\pm$0.10& 1992& Gatewood...	&Parallax of 51 Tauri \cr
3.20$\pm$0.06& 1992& Gatewood...	&Mean parallaxes to date \cr
3.2\phantom{0}$\pm$0.1& 1994& Turner... &Combined methods \cr
3.40$\pm$0.07& 1997a& Torres...		&Orbital parallax 51 Tau (propagated)\cr
3.38$\pm$0.11& 1997b& Torres...		&Orbital parallax 70 Tau (propagated)\cr
3.39$\pm$0.08& 1997c& Torres...		&Orbital parallax 78 Tau (propagated)\cr
3.42$\pm$0.09& 1997a& van Altena...	&HST FGS observations of 7 objects \cr
3.32$\pm$0.06& 1997b& van Altena...	&Mean ground parallaxes to date \cr
\noalign {\smallskip}
\noalign{\hrule}
}}}}

\def\tabtwocap{
{\bf Table 2.}
{Data on the membership of the Hyades for the 282 stars
in our sample, listed by various authors. Membership or non-membership 
inferred by the relevant authors are indicated by `1' or `0' in the 
corresponding column respectively (see text). Entries with `--' in columns
(b--m) inclusive are new candidates proposed in this paper. Columns have the 
following meaning:
{\bf(a)}~Hipparcos Catalogue (HIP) number;$\;$
{\bf(b)}~van Bueren number (1952, BAN, 11, 385);$\;$
{\bf(c)}~Membership according to van Bueren;$\;$
{\bf(d)}~van Altena number (1969, AJ, 74, 2);$\;$
{\bf(e)}~Membership according to van Altena;$\;$
{\bf(f)}~Hanson number (1975, AJ, 80, 379);$\;$
{\bf(g)}~Membership according to Hanson;$\;$
{\bf(h)}~Pels et al.\ (Leiden) number (1975, A\&A, 43, 423);
	van Bueren stars have the vB number + 1000;$\;$
{\bf(i)}~Membership according to Pels et al.;$\;$
{\bf(j)}~Sequential number in Table~4 of Griffin et al.\ 
	(1988, AJ, 96, 172);$\;$
{\bf(k)}~Membership according to Griffin et al.;$\;$
{\bf(l)}~Schwan number (1991, A\&A, 243, 386);$\;$
{\bf(m)}~Membership according to Schwan;$\;$
{\bf(n)}~Hipparcos parallax (mas);$\;$
{\bf(o)}~Hipparcos parallax standard error (mas);$\;$
{\bf(p)}~Radial velocity (km~s$^{-1}$);$\;$
{\bf(q)}~Error in radial velocity (km~s$^{-1}$; \# preceding the error 
 indicates SB/RV (column~s) with undetermined $\gamma$~velocity);$\;$
{\bf(r)}~Source of radial velocity;$\;$
{\bf(s)}~SB = spectroscopic binary, RV = radial velocity (possibly) variable;$\;$
{\bf(t)}~H, I, M = star was previously known, or classified by 
Hipparcos, to have resolved components (from Field~H56); this may overlap 
with the column~u flag, but may also indicate visual or wide binary
(see text for details);$\;$
{\bf(u)}~C, G, O, V, or X = relevant part of the Hipparcos 
	Double and Multiple Systems Annex, from Field~H59, supplemented
	by S = suspected binary in Hipparcos Catalogue, from Field~H61
	(see text for details);$\;$
{\bf(v)}~distance, $d$ (pc), from the cluster centre defined by the 
	134 stars within $r<10$~pc (see Table~3);$\;$
{\bf(w)}~kinematic statistic $c={\bf z'}\pmb{$\Sigma$}^{-1}{\bf z}$ 
	($c=14.16$ corresponding to $3\sigma$);$\;$
{\bf(x)}~Final membership assigned in this paper (0, 1); `?' 
	indicates possible new members unclassifiable due to unknown 
	radial velocities.\par

Sources of radial velocities:
{\bf(0)}~Radial velocity unknown;$\;$
{\bf(1)}~Griffin et al.\ AJ, 96, 172 (1988);
                  AJ, 90, 609 (1985);
                  AJ, 86, 588 (1981);
                  AJ, 83, 1114 (1978);
                  AJ, 82, 176 (1977);
                  A\&A, 106, 221 (1982);$\;$
{\bf(2)}~Hipparcos Input Catalogue (mainly from R.E.\ Wilson, 1953);$\;$
{\bf(3)}~Weighted mean of ref.~2 ($39.6\pm1.2$) and 
	Kraft, ApJ, 142, 681 (1965, $38.4\pm1.5$);$\;$
{\bf(4)}~Kraft, ApJ, 142, 681 (1965, $37.4\pm0.4$ and 
	$36.5\pm0.5$);
	Cheriguene, A\&A, 13, 447 (1971, $37.3\pm0.7$);$\;$
{\bf(5)}~McClure, ApJ, 254, 606 (1982);$\;$
{\bf(6)}~Torres et al., ApJ, 474, 256 (1997);$\;$
{\bf(7)}~Mayor \& Mazeh, A\&A, 171, 157 (1987);$\;$
{\bf(8)}~Kraft, ApJ, 142, 681 (1965);$\;$
{\bf(9)}~Margoni et al., A\&AS, 93, 545 (1992);$\;$
{\bf(10)}~Lucy \& Sweeney, AJ, 76, 544 (1971);$\;$
{\bf(11)}~Abt \& Levy, ApJS, 59, 229;$\;$
{\bf(12)}~Griffin, MNRAS, 155, 1 (1971);$\;$
{\bf(13)}~Andersen \& Nordstrom, A\&A, 122, 23 (1983);$\;$
{\bf(14)}~Morse et al., AJ, 101, 1495 (1991);$\;$
{\bf(15)}~Detweiler et al., AJ, 89, 1038 (1984);$\;$
{\bf(16)}~Weighted mean of data from Palmer et al., 
	Roy. Obs. Bull., 135 (1968) and Stillwell, PDAO, 7, 337 (1949);$\;$
{\bf(17)}~Tomkin et al., AJ, 109, 780 (1995);$\;$
{\bf(18)}~Heintz, ApJS, 46, 247 (1981);$\;$
{\bf(19)}~Abt, ApJS, 11, 429 (1965);$\;$
{\bf(20)}~Fekel, PASP, 92, 785 (1980);$\;$
{\bf(21)}~Perraud, Journal des Observateurs, 45, 361 (1962);$\;$
{\bf(22)}~Fouts \& Sandage, AJ, 91, 1189 (1986; star G83--18);$\;$
{\bf(23)}~Strassmeier et al., A\&AS, 72, 291 (1988);$\;$
{\bf(24)}~New Coravel observations provided by J.C. Mermilliod;$\;$
{\bf(25)}~Woolley et al., Royal Obs.\ Annals, 14, 1;$\;$
{\bf(26)}~Hanson \& Vasilevskis, AJ, 88, 844;$\;$
{\bf(27)}~Evans, Bull. Inf. CDS, 15, 121 (1978);$\;$
{\bf(28)}~Orbit recomputed by Mermilliod with period $=490\pm1$~days 
(from Batten).\par
}
}

\def\tablabels{
HIP& 
\hfill vB \hfill \span\omit& \hfill vA \hfill \span\omit& 
\hfill Hanson \hfill \span\omit&  \hfill Pels \hfill \span\omit& 
\hfill Griffin \hfill \span\omit&  \hfill Schwan \hfill \span\omit& 
Parallax\ \ \span\omit& 
Radial Velocity\ \ \span\omit\span\omit& 
Multiplicity\span\omit\span\omit& 
& \omit\hidewidth \hfill Membership \hfill \hidewidth&  \cr
& \#& & \#& & \#& & \#& & \#& & \#& & 
$\scriptstyle \pi$& $\scriptstyle \sigma_\pi$&
$\scriptstyle V_{\rm r}$\ \ & $\scriptstyle \sigma_V$& s& 
$\scriptstyle V_{\rm r}$\ H56\ H59\span\omit\span\omit& 
\omit \hfill $\scriptstyle d$~(pc) \hfill& 
\omit \hfill $\scriptstyle c$ \hfill& 
\omit \hfill S \hfill \cr
\noalign{\vskip 4pt}
(a)& (b)& (c)& \quad (d)& (e)& \quad (f)& (g)& \quad (h)& (i)&
\quad (j)& (k)& \quad (l)& (m)& (n)& \ \ \ \ (o)& \quad (p)\ \ & (q)& (r)& (s)& 
(t)& (u)& \omit \hfill (v) \hfill& \omit \hfill (w) \hfill& (x) \cr
}

\setbox3=\vbox
{
\sevenrm\baselineskip 7pt
\hbox{ 
\def\p{\phantom{0}}
\vbox { 
\halign {
\tabskip 10pt \hfil#\hfil& 
\tabskip 2pt \hfil#& \tabskip 8pt \hfil#& 
\tabskip 2pt \hfil#& \tabskip 8pt \hfil#& 
\tabskip 2pt \hfil#& \tabskip 8pt \hfil#& 
\tabskip 2pt \hfil#& \tabskip 8pt \hfil#& 
\tabskip 2pt \hfil#& \tabskip 8pt \hfil#& 
\tabskip 2pt \hfil#& \tabskip 14pt \hfil#& 
\tabskip 2pt \hfil#& \tabskip 12pt \hfil#& 
\tabskip 2pt \hfil#& \tabskip 4pt \hfil#& \tabskip 15pt \hfil#& 
\tabskip 2pt \hfil#\hfil& \hfil#\hfil& \tabskip 10pt \hfil#\hfil& 
\tabskip 2pt \hfil#& \hfil#& \hfil#\hfil \cr		
\noalign{\hrule}
\noalign {\smallskip}
\tablabels
\noalign {\smallskip}
\noalign{\hrule}
\noalign {\smallskip}
10540& 157&  0&   *& --&   *& --&    *& --&   *& --&   *& --&   24.93&   0.88& +26.0\p&     1.2\p&     2&  *& *&   *&   24.6\p&    43.09& 0\cr
10672&  --& --&  --& --&  --& --&   --& --&  --& --&  --& --&   15.37&   1.29& +26.40&   0.32&  24&  *& *&   *&   37.1\p&     9.81& 1\cr
12031&  --& --&  --& --&  --& --&   --& --&  --& --&  --& --&   13.44&   3.62& *&        *&      0&  *& *&   *&   41.8\p&     1.11& ?\cr
12709&   *& --&   *& --&   *& --&    *& --&   1&  0&   *& --&   53.89&   1.27& +32.15&   0.15&   1& SB& *&   O&   30.5\p&    13.10& 1\cr
13042&  --& --&  --& --&  --& --&   --& --&  --& --&  --& --&   11.18&  17.11& *&        *&      0&  *& I&   C&   51.5\p&     5.40& ?\cr
13117&  --& --&  --& --&  --& --&   --& --&  --& --&  --& --&   29.67&   9.34& +26.6\p&     0.49&  24&  *& H&   C&   23.8\p&     4.26& 1\cr
13600&  --& --&  --& --&  --& --&   --& --&  --& --&  --& --&   18.89&   1.29& +30.41&   0.23&  24&  *& *&   *&   20.1\p&     7.75& 1\cr
13684&   *& --&   *& --&   *& --&    *& --&   3&  0&   *& --&    5.84&   0.92& +30.67&   0.13&   1&  *& *&   *&  130.5\p&    37.89& 0\cr
13806& 153&  0&   *& --&   *& --&    *& --&   4&  1&   *& --&   25.77&   1.39& +26.62&   0.21&   1&  *& *&   *&   19.4\p&     0.30& 1\cr
13834& 154&  0&   *& --&   *& --&    *& --&   *& --&   5&  1&   31.41&   0.84& +28.1\p&     1.2\p&     2&  *& *&   *&   20.5\p&     0.26& 1\cr
\noalign{\vskip 5pt}
13976&  --& --&  --& --&  --& --&   --& --&  --& --&  --& --&   42.66&   1.22& +28.35&   0.18&  24&  *& *&   *&   26.5\p&     0.38& 1\cr
14792& 133&  1&   *& --&   *& --&    *& --&   5&  0& 177&  0&    5.13&   2.22& +25.99&   0.17&   1&  *& *&   *&  152.2\p&    11.74& 0\cr
14838&  --& --&  --& --&  --& --&   --& --&  --& --&  --& --&   19.44&   1.23& +24.70&   0.50&   2&  *& *&   S&   16.4\p&     8.05& 1\cr
14976&  --& --&  --& --&  --& --&   --& --&  --& --&  --& --&   23.73&   1.18& +27.27&   0.22&  24&  *& *&   *&   18.4\p&     0.68& 1\cr
15206& 158&  0&   *& --&   *& --&    *& --&   *& --&   *& --&   10.74&   1.12& +42.8\p&    \#0.9\p&    24& SB& *&   *&   51.1\p&    60.38& 0\cr
15300&   *& --&   *& --&   *& --&    *& --&   6&  1&   *& --&   29.49&   4.70& +29.84&   0.29&   1&  *& I&   C&   18.1\p&     0.79& 1\cr
15304&   1&  1&   *& --&   *& --& 1001&  1&   8&  1& 141&  1&   20.20&   1.18& +32.44&   0.21&   1&  *& I&   *&   16.5\p&     9.27& 1\cr
15310&   2&  1&   *& --&   *& --& 1002&  1&   9&  1& 149&  0&   21.64&   1.33& +33.00&   0.13&   1&  *& I&   *&   15.6\p&     7.12& 1\cr
15368&  --& --&  --& --&  --& --&   --& --&  --& --&  --& --&   13.76&   5.62& *&        *&      0&  *& H&   C&   32.6\p&     3.66& ?\cr
15374&  --& --&  --& --&  --& --&   --& --&  --& --&  --& --&   24.54&   3.95& *&        *&      0&  *& *&   *&   15.1\p&    11.05& ?\cr
\noalign{\vskip 5pt}
15532&   *& --&   *& --&   *& --&    2&  1&  13&  0&   *& --&    4.48&   2.24& +47.27&   0.22&   1&  *& *&   *&  179.4\p&   107.93& 0\cr
15563&   *& --&   *& --&   *& --&    *& --&   *& --& 158&  0&   34.18&   1.70& +30.45&   0.26&  24&  *& *&   *&   20.7\p&     0.51& 1\cr
15720&  --& --&  --& --&  --& --&   --& --&  --& --&  --& --&   29.75&   2.73& +28.9\p&     0.45&  24&  *& *&   *&   17.9\p&     0.47& 1\cr
16329&   3&  1&   *& --&   *& --&    3&  1&  14&  0& 164&  0&   21.61&   1.48& +26.67&   0.09&   1&  *& I&   G&   11.3\p&    36.71& 0\cr
16377&  --& --&  --& --&  --& --&   --& --&  --& --&  --& --&   10.48&   1.61& *&        *&      0&  *& *&   *&   55.2\p&    11.41& ?\cr
16529&   4&  1&   *& --&   *& --& 1004&  1&  16&  1&  46&  1&   22.78&   1.26& +32.72&   0.17&   1&  *& *&   *&   11.8\p&     0.27& 1\cr
16548&  --& --&  --& --&  --& --&   --& --&  --& --&  --& --&   17.20&   3.36& +26.6\p&     0.34&  24&  *& *&   *&   20.0\p&    10.22& 1\cr
16896& 159&  0&   *& --&   *& --&    *& --&  20&  0&   *& --&   11.73&   1.33& +47.83&   0.22&   1&  *& *&   *&   41.1\p&    70.94& 0\cr
16908&   5&  1&   *& --&   *& --& 1005&  1&  21&  1&  47&  1&   25.23&   1.58& +33.56&   0.21&   1&  *& *&   *&   11.7\p&     3.86& 1\cr
17128& 134&  1&   *& --&   *& --&    *& --&   *& --& 178&  0&    2.47&   1.59& +62.4\p&     0.4\p&    26&  *& *&   *&  360.7\p&   163.77& 0\cr
\noalign{\vskip 5pt}
17324&   *& --&   *& --&   *& --&    *& --&  24&  0&   *& --&    1.46&   1.13& +31.88&   0.26&   1&  *& *&   *&  640.1\p&     5.75& 0\cr
17605&   *& --&   *& --&   *& --&    *& --&  28&  0&   *& --&    6.63&   1.68& +92.10&   0.65&   1&  *& *&   *&  106.3\p&   710.47& 0\cr
17609&  --& --&  --& --&  --& --&   --& --&  --& --&  --& --&   68.62&   1.78& +32.20&   2.50&   2&  *& *&   *&   32.4\p&     8.39& 1\cr
17766&   *& --&   *& --&   *& --&    *& --&  30&  1&   *& --&   24.02&   2.27& +35.40&   0.25&   1&  *& *&   *&   11.4\p&     1.21& 1\cr
17779& 136&  1&   *& --&   *& --&    *& --&  31&  0& 180&  0&    7.65&   0.95& --1.48&    0.17&   1&  *& *&   *&   86.1\p&   299.33& 0\cr
17950&  --& --&  --& --&  --& --&   --& --&  --& --&  --& --&   22.22&   0.97& *&        *&      0&  *& I&   C&   16.3\p&     7.53& ?\cr
17962&  --& --&  --& --&  --& --&   --& --&  --& --&  --& --&   21.37&   1.62& +40.00&   5.00&   2&  *& *&   *&    7.3\p&     0.98& 1\cr
18018& 170&  0&   *& --&   *& --&    6&  1&  33&  1&   *& --&   24.72&   4.62& +35.30&   0.12&   1&  *& *&   X&   10.3\p&     0.08& 1\cr
18096&  --& --&  --& --&  --& --&   --& --&  --& --&  --& --&   11.19&   1.65& +40.02&   0.24&  24&  *& *&   *&   44.1\p&    11.14& 1\cr
18170&   6&  1&   *& --&   *& --& 1006&  1&   *& --&   6&  1&   24.14&   0.90& +35.0\p&     2.5\p&     2&  *& *&   *&    8.1\p&     0.27& 1\cr
\noalign {\smallskip}
\noalign{\hrule}
}}}}

\setbox4=\vbox
{
\tabcap{2}{Hyades membership compilation summary (2/4)}
\sevenrm\baselineskip 7pt
\hbox{ 
\def\p{\phantom{0}}
\vbox { 
\halign {
\tabskip 10pt \hfil#\hfil& 
\tabskip 2pt \hfil#& \tabskip 8pt \hfil#& 
\tabskip 2pt \hfil#& \tabskip 8pt \hfil#& 
\tabskip 2pt \hfil#& \tabskip 8pt \hfil#& 
\tabskip 2pt \hfil#& \tabskip 8pt \hfil#& 
\tabskip 2pt \hfil#& \tabskip 8pt \hfil#& 
\tabskip 2pt \hfil#& \tabskip 14pt \hfil#& 
\tabskip 2pt \hfil#& \tabskip 12pt \hfil#& 
\tabskip 2pt \hfil#& \tabskip 4pt \hfil#& \tabskip 15pt \hfil#& 
\tabskip 2pt \hfil#\hfil& \hfil#\hfil& \tabskip 10pt \hfil#\hfil& 
\tabskip 2pt \hfil#& \hfil#& \hfil#\hfil \cr		
\noalign{\hrule}
\noalign {\smallskip}
\tablabels
\noalign {\smallskip}
\noalign{\hrule}
\noalign {\smallskip}
18322&   *& --&   *& --&   *& --&    8&  1&  36&  1& 155&  0&   26.49&   1.98& +37.18&   0.22&   1&  *& *&   *&   10.8\p&     3.82& 1\cr
18327&   7&  1&   *& --&   *& --& 1007&  1&  37&  1&  65&  1&   24.16&   1.40& +36.79&   0.13&   1&  *& *&   *&    7.8\p&     0.21& 1\cr
18617&  --& --&  --& --&  --& --&   --& --&  --& --&  --& --&   10.38&   2.61& *&        *&      0&  *& M&   C&   54.6\p&     9.52& ?\cr
18658&   8&  1&   *& --&   *& --& 1008&  1&   *& --&  19&  1&   25.42&   1.05& +39.1\p&     1.1\p&     3&  *& *&   G&    9.9\p&     2.97& 1\cr
18692&   *& --&   *& --&   *& --&    *& --&  46&  0&   *& --&   10.93&   1.19& +37.94&   0.18&   1&  *& *&   *&   45.9\p&    13.67& 1\cr
18719&   9&  1&   *& --&   *& --& 1009&  1&   *& --&   *& --&   16.04&   1.33& +37.0\p&    \#2.5\p&     2& SB& *&   *&   17.5\p&    47.25& 0\cr
18735& 137&  1&   *& --&   *& --& 1137&  1&   *& --& 162&  0&   21.99&   0.81& +31.7\p&    \#1.1\p&     2& SB& I&   *&    5.4\p&     2.84& 1\cr
18946&   *& --&   *& --&   *& --&   11&  1&  55&  1& 146&  0&   23.07&   2.12& +36.93&   0.26&   1&  *& *&   *&    5.8\p&     0.70& 1\cr
18975& 160&  0&   *& --&   *& --&    *& --&   *& --&   *& --&   27.80&   0.95& +34.4\p&     1.5\p&    13&  *& *&   *&   12.7\p&    14.75& 0\cr
19082&   *& --&   *& --&   *& --&   12&  1&  57&  1&   *& --&   14.56&   3.17& +38.33&   0.22&   1&  *& *&   *&   23.1\p&     2.20& 1\cr
\noalign{\vskip 5pt}
19098&   *& --&   *& --&   *& --&   10&  1&  58&  1&   *& --&   19.81&   1.39& +37.61&   0.05&   1&  *& *&   *&    6.2\p&     1.10& 1\cr
19117&   *& --&   *& --&   *& --&    *& --&  60&  0&   *& --&   29.02&   2.12& +37.28&   0.13&   1&  *& *&   *&   12.9\p&    19.85& 0\cr
19148&  10&  1&   *& --&   *& --& 1010&  1&  62&  1&  66&  1&   21.41&   1.47& +38.04&   0.17&   1&  *& *&   *&    4.3\p&     0.36& 1\cr
19207&   *& --&   *& --&   *& --&   15&  1&  65&  1&   *& --&   23.57&   2.26& +38.95&   0.23&   1&  *& *&   *&    5.6\p&     1.01& 1\cr
19261&  11&  1&   *& --&   *& --& 1011&  1&  68&  1&  67&  1&   21.27&   1.03& +36.35&   0.26&   1&  *& I&   C&    4.2\p&     0.71& 1\cr
19263&   *& --&   *& --&   *& --&   16&  1&  70&  1&   *& --&   19.70&   1.68& +38.72&   0.05&   1&  *& *&   *&    6.0\p&     0.92& 1\cr
19316&   *& --&   *& --&   *& --&   14&  1&  75&  1&   *& --&   24.90&   2.59& +38.43&   0.28&   1&  *& *&   *&    7.9\p&     1.81& 1\cr
19365&   *& --&   *& --&   *& --&    *& --&  79&  0&   *& --&   10.68&   1.43& +37.92&   0.15&   1&  *& I&   *&   49.7\p&     4.75& 1\cr
19386&  --& --&  --& --&  --& --&   --& --&  --& --&  --& --&   15.37&   0.97& +33.6\p&     0.39&  24&  *& *&   *&   24.9\p&     6.62& 1\cr
19441&   *& --&   *& --&   *& --&    *& --&  84&  1&   *& --&   29.78&   1.90& +39.24&   0.16&   1&  *& *&   *&   14.1\p&     0.47& 1\cr
\noalign{\vskip 5pt}
19449&  --& --&  --& --&  --& --&   --& --&  --& --&  --& --&   12.14&   2.03& *&        *&      0&  *& *&   *&   38.7\p&     3.82& ?\cr
19472&   *& --&  14&  0&  18&  0&    *& --&   *& --&   *& --&   29.88&   2.67& *&        *&      0&  *& M&   C&   13.2\p&    55.80& 0\cr
19481&   *& --&  19&  0&  23&  0&    *& --&   *& --&   *& --&   23.85&   1.26& +38.0\p&     4.5\p&    25&  *& *&   G&    6.5\p&   175.60& 0\cr
19504&  13&  1&   *& --&   *& --& 1013&  1&   *& --&  68&  1&   23.22&   0.92& +37.1\p&     0.3\p&     4&  *& *&   *&    4.8\p&     0.14& 1\cr
19554&  14&  1&   *& --&   *& --& 1014&  1&   *& --&  11&  1&   25.89&   0.95& +36.6\p&    \#1.2\p&     2& SB& I&   *&   11.7\p&     0.69& 1\cr
19572& 138&  1&   *& --&   *& --& 1138&  1&  85&  0& 165&  0&   12.91&   1.19& +78.24&   0.27&   1&  *& *&   S&   34.2\p&   401.51& 0\cr
19591&   *& --&   *& --&   *& --&   20&  1&  86&  1&   *& --&   27.21&   2.11& +36.90&   0.26&   1& SB& M&   C&   11.2\p&     0.41& 1\cr
19641&   *& --&   *& --&   *& --&    *& --&  87&  0&   *& --&   11.42&   1.27& +26.97&   0.15&   1&  *& *&   *&   41.4\p&    22.37& 0\cr
19696&   *& --&  51&  1&  89&  1&    *& --&  90&  0&   *& --&   11.33&   1.61& --5.92&    0.33&   1&  *& *&   *&   42.1\p&   414.42& 0\cr
19767&   *& --&  59&  0& 100&  0&    *& --&   *& --&   *& --&   27.98&   1.18& +53.4\p&     7.2\p&    25&  *& *&   *&   11.3\p&    98.78& 0\cr
\noalign{\vskip 5pt}
19781&  17&  1&   *& --& 101&  1& 1017&  1&  93&  1&  69&  1&   21.91&   1.27& +39.24&   0.06&   1&  *& *&   *&    3.2\p&     1.15& 1\cr
19786&  18&  1&  60&  1& 105&  1& 1018&  1&  94&  1&  70&  1&   22.19&   1.45& +39.32&   0.14&   1&  *& *&   *&    4.5\p&     0.74& 1\cr
19789&  16&  1&   *& --&   *& --& 1016&  1&   *& --&  48&  1&   18.12&   0.92& +38.4\p&     1.2\p&     2&  *& *&   *&   10.6\p&     0.75& 1\cr
19793&  15&  1&   *& --&   *& --& 1015&  1&  92&  1&  49&  1&   21.69&   1.14& +38.21&   0.23&   1&  *& *&   *&    6.1\p&     0.35& 1\cr
19796&  19&  1&   *& --&   *& --& 1019&  1&  97&  1&  71&  1&   21.08&   0.97& +38.50&   0.15&   1&  *& *&   *&    5.7\p&     0.51& 1\cr
19808&   *& --&  68&  1& 111&  1&   23&  1&  98&  1&   *& --&   22.67&   2.30& +40.51&   0.15&   1&  *& *&   *&    4.4\p&     1.21& 1\cr
19834&   *& --&  72&  1& 115&  1&   24&  1&  99&  1&   *& --&   31.94&   3.74& +38.79&   0.36&   1&  *& *&   *&   15.2\p&     7.86& 1\cr
19862&   *& --&  75&  1& 119&  1&    *& --& 100&  1&   *& --&   31.11&   2.76& +38.96&   0.17&   1&  *& *&   *&   14.4\p&     7.74& 1\cr
19870& 162&  0&   *& --&   *& --& 1162&  1& 101&  1&  50&  1&   19.48&   0.99& +38.46&   0.12&   1& SB& *&   *&    6.6\p&     0.50& 1\cr
19877&  20&  1&  79&  1& 122&  1& 1020&  1&   *& --&  20&  1&   22.51&   0.82& +36.4\p&     1.2\p&     2&  *& I&   *&    3.2\p&     0.44& 1\cr
\noalign{\vskip 5pt}
19934&  21&  1&   *& --&   *& --& 1021&  1& 103&  1&  51&  1&   19.48&   1.17& +38.46&   0.19&   1&  *& *&   *&    7.1\p&     0.11& 1\cr
19981&   *& --&   *& --&   *& --&   28&  1& 106&  0&   *& --&   30.56&   1.52& +28.82&   0.20&   1&  *& I&   *&   14.3\p&    21.14& 0\cr
20019&  22&  1& 108&  1& 167&  0& 1022&  1& 111&  1&  72&  1&   21.40&   1.24& +38.18&   0.13&   1& SB& *&   *&    2.1\p&     0.83& 1\cr
20056&  23&  1& 123&  1& 178&  1& 1023&  1&   *& --&  73&  1&   21.84&   1.14& +37.7\p&     0.4\p&     5& SB& *&   *&    2.4\p&     0.00& 1\cr
20082&  25&  1& 133&  1& 185&  1& 1025&  1& 117&  1&  74&  1&   20.01&   1.91& +39.64&   0.08&   1&  *& *&   *&    4.1\p&     2.00& 1\cr
20086&   *& --& 135&  1& 187&  1&   30&  1& 118&  1&   *& --&   19.57&   1.86& +40.53&   0.04&   1&  *& *&   S&    5.2\p&     5.45& 1\cr
20087&  24&  1&   *& --&   *& --& 1024&  1&   *& --&  12&  1&   18.25&   0.82& +37.78&   0.12&   6& SB& I&   O&    9.7\p&     0.08& 1\cr
20130&  26&  1&   *& --&   *& --& 1026&  1& 120&  1&  75&  1&   23.53&   1.25& +39.58&   0.06&   1&  *& *&   *&    4.9\p&     1.10& 1\cr
20146&  27&  1& 156&  1& 198&  1& 1027&  1& 122&  1&  76&  1&   21.24&   1.32& +38.80&   0.08&   1&  *& *&   *&    2.0\p&     0.07& 1\cr
20187&   *& --& 171&  0& 210&  0&    *& --& 125&  0&   *& --&   20.13&   2.02& +37.99&   0.06&   1&  *& *&   *&    5.0\p&    10.97& 1\cr
\noalign{\vskip 5pt}
20197&   *& --& 174&  0&   *& --&    *& --&   *& --&   *& --&   12.93&   1.06& --19.10&   1.3\p&    12&  *& *&   *&   31.2\p&   657.25& 0\cr
20205&  28&  1& 175&  1&   *& --& 1028&  1& 127&  1&   1&  1&   21.17&   1.17& +39.28&   0.11&   1&  *& *&   *&    2.0\p&     0.22& 1\cr
20215&  29&  1& 179&  1& 212&  1& 1029&  1& 129&  1&  77&  1&   23.27&   1.14& +39.21&  \#0.27&   1& SB& I&   C&    3.7\p&     1.85& 1\cr
20219&  30&  1& 182&  1& 213&  1& 1030&  1&   *& --&  21&  1&   22.31&   0.92& +42.0\p&     2.5\p&     2&  *& I&   *&    3.0\p&     1.23& 1\cr
20226&   *& --&   *& --&   *& --&    *& --& 130&  0&   *& --&    4.91&   0.88& +8.78&    0.19&   1&  *& *&   *&  157.8\p&   182.99& 0\cr
20237&  31&  1&   *& --&   *& --& 1031&  1& 132&  1&  78&  1&   22.27&   0.93& +38.81&   0.18&   1&  *& *&   *&    2.9\p&     0.21& 1\cr
20255&  32&  1&   *& --&   *& --& 1032&  1&   *& --&  79&  1&   21.12&   0.77& +42.0\p&    \#1.2\p&     2& SB& *&   *&    2.5\p&     8.65& 1\cr
20261&  33&  1&   *& --&   *& --& 1033&  1&   *& --&  22&  1&   21.20&   0.99& +36.2\p&     1.2\p&     2&  *& *&   *&    2.1\p&     0.65& 1\cr
20284&  34&  1& 201&  1& 230&  1& 1034&  1&   *& --&  23&  1&   21.80&   0.85& +39.2\p&     0.3\p&     7& SB& *&   *&    2.7\p&     0.53& 1\cr
20319&  --& --&  --& --&  --& --&   --& --&  --& --&  --& --&   11.64&   3.73& *&        *&      0&  *& *&   *&   41.3\p&     9.00& ?\cr
\noalign{\vskip 5pt}
20349&  35&  1&   *& --&   *& --& 1035&  1&   *& --&  52&  1&   19.55&   0.89& +37.1\p&     1.2\p&     2&  *& *&   *&    6.2\p&     0.23& 1\cr
20350&  36&  1&   *& --&   *& --& 1036&  1&   *& --&  80&  1&   19.83&   0.89& +40.8\p&     2.4\p&    24&  *& *&   *&    4.5\p&     1.43& 1\cr
20357&  37&  1& 215&  1& 246&  1& 1037&  1& 137&  1&  81&  1&   19.46&   1.02& +39.20&   0.21&   1&  *& *&   *&    5.6\p&     0.54& 1\cr
20400&  38&  1& 229&  1& 257&  1& 1038&  1&   *& --&  24&  1&   21.87&   0.96& +37.8\p&     2.3\p&     9& SB& I&   *&    2.5\p&     0.24& 1\cr
20415& 139&  1&   *& --&   *& --&    *& --& 139&  0& 163&  0&   15.44&   1.28& +26.77&   0.20&   1&  *& *&   *&   22.6\p&    28.11& 0\cr
20419&   *& --&   *& --&   *& --&   33&  1& 142&  1&  82&  1&   19.17&   1.93& +40.77&  \#0.20&   1& SB& *&   *&    7.5\p&     1.63& 1\cr
20440&  40&  1& 249&  1& 271&  1& 1040&  1&   *& --&  83&  1&   21.45&   2.76& +37.4\p&     2.9\p&    10& SB& I&   C&    1.7\p&     0.21& 1\cr
20441&  39&  1& 248&  0& 270&  1& 1039&  1&   *& --&  25&  1&   26.96&   1.40& +34.8\p&    \#2.6\p&    24& SB& *&   G&    9.3\p&    12.16& 1\cr
20455&  41&  1& 256&  1&   *& --& 1041&  1& 148&  1&   2&  1&   21.29&   0.93& +39.65&   0.08&   1& SB& I&   *&    1.4\p&     0.17& 1\cr
20480&  42&  1&   *& --&   *& --& 1042&  1& 149&  1&  53&  1&   20.63&   1.34& +39.24&   0.24&   1&  *& *&   *&    4.5\p&     0.28& 1\cr
\noalign{\vskip 5pt}
20482&  43&  1&   *& --&   *& --& 1043&  1& 150&  1&  84&  1&   15.82&   1.44& +39.90&   0.09&   1& SB& *&   O&   17.1\p&     2.57& 1\cr
20484&  45&  1& 272&  1& 288&  1& 1045&  1&   *& --&  26&  1&   21.17&   0.80& +37.7\p&     0.3\p&    11& SB& *&   *&    1.3\p&     0.17& 1\cr
20485& 173&  0& 276&  1& 290&  1&   35&  1& 151&  1&  85&  1&   21.08&   2.69& +39.30&   0.21&   1&  *& *&   *&    1.6\p&     1.36& 1\cr
20491&  44&  1& 279&  1&   *& --& 1044&  1&   *& --&  54&  1&   20.04&   0.89& +35.9\p&     0.5\p&     8&  *& *&   *&    7.4\p&     0.81& 1\cr
20492&  46&  1&   *& --& 292&  1& 1046&  1& 152&  1&  86&  1&   21.23&   1.80& +40.29&   0.06&   1&  *& *&   *&    2.0\p&     0.34& 1\cr
20527&   *& --& 294&  1& 299&  1&   34&  1& 156&  1&   *& --&   22.57&   2.78& +40.64&   0.26&   1&  *& *&   *&    3.0\p&     0.83& 1\cr
20540&   *& --& 304&  0& 302&  0&    *& --&   *& --&   *& --&    6.58&   1.09& +59.3\p&     1.0\p&    12&  *& *&   *&  105.8\p&   113.29& 0\cr
20542&  47&  1& 301&  1&   *& --& 1047&  1&   *& --&  27&  1&   22.36&   0.88& +39.2\p&     1.2\p&    27&  *& I&   *&    1.9\p&     0.39& 1\cr
20553&  50&  1& 308&  1& 308&  1& 1050&  1& 163&  0&  87&  1&   22.25&   1.52& +37.48&   0.19&   1&  *& H&   C&    2.2\p&     6.15& 1\cr
20557&  48&  1&   *& --&   *& --& 1048&  1& 160&  1&  55&  1&   24.47&   1.06& +38.94&   0.13&   1&  *& *&   *&    6.7\p&     0.52& 1\cr
\noalign {\smallskip}
\noalign{\hrule}
}}}}

\setbox5=\vbox
{
\tabcap{2}{Hyades membership compilation summary (3/4)}
\sevenrm\baselineskip 7pt
\hbox{ 
\def\p{\phantom{0}}
\vbox { 
\halign {
\tabskip 10pt \hfil#\hfil& 
\tabskip 2pt \hfil#& \tabskip 8pt \hfil#& 
\tabskip 2pt \hfil#& \tabskip 8pt \hfil#& 
\tabskip 2pt \hfil#& \tabskip 8pt \hfil#& 
\tabskip 2pt \hfil#& \tabskip 8pt \hfil#& 
\tabskip 2pt \hfil#& \tabskip 8pt \hfil#& 
\tabskip 2pt \hfil#& \tabskip 14pt \hfil#& 
\tabskip 2pt \hfil#& \tabskip 12pt \hfil#& 
\tabskip 2pt \hfil#& \tabskip 4pt \hfil#& \tabskip 15pt \hfil#& 
\tabskip 2pt \hfil#\hfil& \hfil#\hfil& \tabskip 10pt \hfil#\hfil& 
\tabskip 2pt \hfil#& \hfil#& \hfil#\hfil \cr		
\noalign{\hrule}
\noalign {\smallskip}
\tablabels
\noalign {\smallskip}
\noalign{\hrule}
\noalign {\smallskip}
20563& 174&  0& 310&  1& 312&  1&   39&  1& 164&  1&   *& --&   19.35&   1.79& +39.95&   0.16&   1&  *& *&   *&    5.5\p&     1.33& 1\cr
20567&  51&  1& 315&  1& 316&  1& 1051&  1&   *& --&  89&  1&   18.74&   1.17& +40.1\p&     0.6\p&     8&  *& *&   *&    7.1\p&     0.81& 1\cr
20577&  52&  1& 319&  1& 320&  1& 1052&  1& 165&  1&  90&  1&   20.73&   1.29& +38.80&  \#0.08&   1& RV& *&   *&    2.0\p&     0.40& 1\cr
20601& 140&  1&   *& --&   *& --& 1140&  1& 167&  0& 142&  1&   14.97&   1.51& +42.20&   0.12&   1& SB& *&   *&   23.5\p&     8.76& 1\cr
20605&   *& --& 334&  1& 336&  1&    *& --&   *& --&   *& --&   24.41&   6.94& +40.2\p&     0.36&  24&  *& H&   C&    5.4\p&     1.10& 1\cr
20614&  53&  1&   *& --&   *& --& 1053&  1&   *& --&  28&  1&   20.40&   0.74& +36.6\p&     1.2\p&     2&  *& *&   *&    3.4\p&     1.05& 1\cr
20626&  --& --&  --& --&  --& --&   --& --&  --& --&  --& --&   15.92&   1.00& *&        *&      0&  *& *&   *&   18.6\p&    11.59& ?\cr
20635&  54&  1&   *& --&   *& --& 1054&  1&   *& --&  29&  1&   21.27&   0.80& +38.6\p&     1.2\p&    27&  *& I&   *&    4.7\p&     0.11& 1\cr
20641&  55&  1&   *& --&   *& --& 1055&  1&   *& --&  30&  1&   22.65&   0.84& +32.0\p&     2.5\p&     2&  *& I&   *&    4.9\p&     3.27& 1\cr
20648&  56&  1& 355&  1&   *& --& 1056&  1&   *& --&  31&  1&   22.05&   0.77& +38.7\p&     1.3\p&    14&  *& I&   C&    1.5\p&     0.05& 1\cr
\noalign{\vskip 5pt}
20661&  57&  1& 360&  1& 357&  1& 1057&  1&   *& --&  32&  1&   21.47&   0.97& +39.1\p&    \#0.5\p&    15& SB& I&   C&    0.8\p&     0.56& 1\cr
20679& 176&  0& 363&  1& 361&  1& 1176&  1&   *& --&   *& --&   20.79&   1.83& +37.0\p&    \#7.5\p&    21& SB& M&   C&    2.1\p&     0.82& 1\cr
20686&  58&  1&   *& --&   *& --& 1058&  1& 172&  1&  91&  1&   23.08&   1.22& +40.72&  \#0.47&   1& SB& I&   C&    3.5\p&     1.04& 1\cr
20693&  61&  1&   *& --&   *& --& 1061&  1& 177&  0& 181&  0&   22.03&   0.90& +29.67&   0.30&   1&  *& *&   *&    9.3\p&    17.54& 0\cr
20711&  60&  1&   *& --&   *& --& 1060&  1&   *& --&  13&  1&   21.07&   0.80& +35.6\p&     0.6\p&    16&  *& I&   *&    5.2\p&     1.28& 1\cr
20712&  62&  1&   *& --&   *& --& 1062&  1& 178&  1&  56&  1&   21.54&   0.97& +38.77&   0.14&   1& SB& *&   *&    3.9\p&     1.01& 1\cr
20713& 141&  1& 388&  1&   *& --& 1141&  1&   *& --&  93&  1&   20.86&   0.84& +40.8\p&     4.26&  19& SB& *&   *&    1.8\p&     3.65& 1\cr
20719&  63&  1& 389&  1& 382&  1& 1063&  1& 179&  1&  94&  1&   21.76&   1.46& +39.39&  \#0.31&   1& SB& *&   *&    0.5\p&     3.35& 1\cr
20741&  64&  1& 400&  1& 388&  1& 1064&  1& 180&  1&  95&  1&   21.42&   1.54& +40.23&   0.28&   1&  *& *&   *&    0.4\p&     0.24& 1\cr
20745&   *& --& 404&  1& 392&  1&   48&  1& 182&  1&   *& --&   28.27&   3.17& +41.38&   0.18&   1&  *& M&   C&   11.3\p&     1.24& 1\cr
\noalign{\vskip 5pt}
20751&   *& --&   *& --&   *& --&   59&  1& 183&  1&  96&  1&   23.03&   1.66& +41.12&  \#0.20&   1& SB& *&   *&    5.4\p&     1.32& 1\cr
20762&   *& --& 407&  1& 394&  1&   49&  1& 184&  1&   *& --&   21.83&   2.29& +41.22&   0.21&   1&  *& *&   *&    2.9\p&     0.43& 1\cr
20810& 188&  0& 444&  1& 413&  1&    *& --& 190&  0&   *& --&    8.66&   2.61& +60.94&   0.06&   1&  *& *&   *&   69.1\p&    87.60& 0\cr
20815&  65&  1& 446&  1& 415&  1& 1065&  1& 191&  1&  97&  1&   21.83&   1.01& +39.32&   0.24&   1&  *& *&   *&    1.0\p&     0.14& 1\cr
20826&  66&  1&   *& --&   *& --& 1066&  1& 193&  1&  98&  1&   21.18&   1.04& +40.22&   0.21&   1&  *& *&   *&    4.1\p&     0.30& 1\cr
20827& 179&  0& 459&  1& 417&  1&   52&  1& 192&  1&   *& --&   17.29&   2.23& +40.46&   0.07&   1&  *& *&   *&   11.7\p&     1.17& 1\cr
20842&  67&  1&   *& --&   *& --& 1067&  1&   *& --&  33&  1&   20.85&   0.86& +37.5\p&     3.3\p&    11&  *& I&   *&    4.4\p&     0.12& 1\cr
20850& 178&  0& 472&  1& 420&  1&   50&  1& 196&  1& 100&  1&   21.29&   1.91& +40.94&   0.08&   1&  *& *&   *&    2.4\p&     0.27& 1\cr
20873&  68&  1& 485&  1& 429&  1& 1068&  1&   *& --& 101&  1&   18.42&   1.93& +40.6\p&     0.3\p&    24&  *& *&   X&    8.1\p&     2.10& 1\cr
20885&  71&  1& 489&  1&   *& --& 1071&  1& 200&  1&  34&  1&   20.66&   0.85& +40.17&  \#0.08&   1& SB& I&   *&    2.1\p&     3.31& 1\cr
\noalign{\vskip 5pt}
20889&  70&  1&   *& --&   *& --& 1070&  1& 199&  1&   3&  1&   21.04&   0.82& +39.37&   0.06&   1&  *& I&   *&    2.4\p&     0.25& 1\cr
20890&  69&  1&   *& --&   *& --& 1069&  1& 198&  1& 102&  1&   20.09&   1.11& +39.91&   0.08&   1& SB& *&   *&    4.3\p&     0.78& 1\cr
20894&  72&  1& 491&  1&   *& --& 1072&  1&   *& --&  35&  1&   21.89&   0.83& +38.9\p&     0.2\p&    17& SB& I&   *&    0.9\p&     0.05& 1\cr
20899&  73&  1& 495&  1& 439&  1& 1073&  1& 201&  1& 103&  1&   21.09&   1.08& +39.99&   0.16&   1&  *& *&   *&    1.2\p&     0.14& 1\cr
20901&  74&  1& 504&  1&   *& --& 1074&  1&   *& --&  14&  1&   20.33&   0.84& +39.9\p&     4.1\p&    11&  *& *&   *&    4.1\p&     0.26& 1\cr
20916&  75&  1& 511&  1& 448&  0& 1075&  1&   *& --& 104&  1&   20.58&   1.74& +45.0\p&    \#2.5\p&     2& SB& I&   C&    2.3\p&     3.93& 1\cr
20935&  77&  1& 536&  1& 461&  1& 1077&  1& 209&  1& 105&  1&   23.25&   1.04& +39.90&   0.11&   1& SB& *&   O&    3.4\p&     0.77& 1\cr
20948&  78&  1& 544&  1& 469&  1& 1078&  1& 210&  1& 106&  1&   21.59&   1.09& +38.62&   0.24&   1&  *& I&   *&    1.0\p&     0.04& 1\cr
20949&  76&  1&   *& --&   *& --& 1076&  1& 208&  1&  57&  1&   17.08&   1.18& +39.02&   0.17&   1&  *& *&   *&   15.2\p&     0.47& 1\cr
20951&  79&  1& 547&  1& 470&  1& 1079&  1& 211&  1& 107&  1&   24.19&   1.76& +40.70&   0.06&   1&  *& I&   *&    5.1\p&     0.84& 1\cr
\noalign{\vskip 5pt}
20952&   *& --& 550&  0& 474&  0&    *& --&   *& --&   *& --&    7.68&   1.27& +96.3\p&     1.2\p&     2&  *& *&   *&   83.9\p&   512.19& 0\cr
20978& 180&  0& 560&  1& 478&  1&   56&  1& 215&  1& 108&  1&   24.71&   1.27& +40.97&   0.06&   1&  *& *&   *&    5.9\p&     1.69& 1\cr
20995&  80&  1& 569&  1& 481&  0& 1080&  1&   *& --& 171&  0&   22.93&   1.25& +29.3\p&     5.00&  18& SB& I&   C&    2.9\p&     3.30& 1\cr
21008&  81&  1&   *& --&   *& --& 1081&  1&   *& --& 109&  1&   19.94&   0.93& +38.0\p&    \#2.5\p&     2& SB& *&   *&    4.7\p&     1.28& 1\cr
21019&   *& --&   *& --&   *& --&    *& --& 216&  0&   *& --&    3.52&   1.98& +47.89&   0.26&   1&  *& *&   *&  237.8\p&    15.22& 0\cr
21029&  82&  1& 584&  1&   *& --& 1082&  1&   *& --&  36&  1&   22.54&   0.77& +41.0\p&     1.8\p&    13&  *& I&   *&    2.1\p&     0.75& 1\cr
21036&  84&  1& 591&  1&   *& --& 1084&  1&   *& --&  38&  1&   21.84&   0.89& +38.8\p&     1.2\p&     2&  *& I&   *&    2.5\p&     0.14& 1\cr
21039&  83&  1& 589&  1& 493&  0& 1083&  1&   *& --&  37&  1&   22.55&   1.09& +39.56&   0.23&  24& SB& I&   *&    2.2\p&     0.35& 1\cr
21053&  85&  1& 597&  1& 496&  1& 1085&  1&   *& --& 111&  1&   24.28&   0.79& +40.9\p&     1.3\p&     8&  *& I&   *&    5.2\p&     3.77& 1\cr
21066&  86&  1&   *& --&   *& --& 1086&  1& 220&  1& 112&  1&   22.96&   0.99& +41.35&   0.26&   1&  *& *&   *&    5.4\p&     1.03& 1\cr
\noalign{\vskip 5pt}
21092&  --& --&  --& --&  --& --&   --& --&  --& --&  --& --&   19.64&   9.61& *&        *&      0&  *& H&   C&   13.3\p&     0.95& ?\cr
21099&  87&  1&   *& --&   *& --& 1087&  1& 222&  1&  58&  1&   21.81&   1.25& +40.62&   0.08&   1&  *& *&   *&    2.9\p&     0.38& 1\cr
21112&  88&  1& 625&  1& 507&  1& 1088&  1& 224&  1& 172&  0&   19.46&   1.02& +40.98&   0.31&   1&  *& *&   *&    5.6\p&     0.44& 1\cr
21123&   *& --& 627&  1& 509&  1&   63&  1& 225&  1&   *& --&   23.41&   1.65& +40.38&   0.11&   1& SB& *&   O&    3.8\p&     0.41& 1\cr
21137&  89&  1& 644&  1& 516&  1& 1089&  1&   *& --&  39&  1&   22.25&   1.14& +36.0\p&    \#2.5\p&     2& SB& *&   *&    1.7\p&     2.23& 1\cr
21138& 191&  0& 645&  1& 517&  1&   62&  1& 228&  1&   *& --&   15.11&   4.75& +41.28&   0.21&   1&  *& *&   *&   19.9\p&     1.08& 1\cr
21152&  90&  1&   *& --&   *& --& 1090&  1&   *& --& 143&  1&   23.13&   0.92& +39.8\p&     1.0\p&    24&  *& *&   *&    9.4\p&     0.20& 1\cr
21179&   *& --& 677&  1& 532&  1&   60&  1&   *& --&   *& --&   17.55&   2.97& +41.70&  \#1.0\p&    24& SB& *&   *&   11.1\p&     2.01& 1\cr
21194&   *& --& 682&  0& 541&  0&    *& --&   *& --&   *& --&    9.42&   2.76& *&        *&      0&  *& *&   *&   60.1\p&    69.56& 0\cr
21256&   *& --&   *& --&   *& --&   66&  1& 235&  1&   *& --&   24.98&   1.95& +41.39&   0.20&   1&  *& *&   *&    7.2\p&     1.19& 1\cr
\noalign{\vskip 5pt}
21261&   *& --&   *& --&   *& --&   65&  1& 237&  1&   *& --&   21.06&   2.21& +41.43&   0.15&   1&  *& *&   *&    2.5\p&     0.42& 1\cr
21267&  94&  1& 724&  1& 574&  1& 1094&  1&   *& --& 116&  1&   22.80&   0.98& +36.9\p&     0.9\p&     8&  *& *&   *&    3.8\p&     1.38& 1\cr
21273&  95&  1& 725&  1&   *& --& 1095&  1&   *& --&   7&  1&   21.39&   1.24& +37.7\p&     0.9\p&    28& SB& *&   O&    1.9\p&     1.18& 1\cr
21280&  96&  1& 727&  1& 578&  1& 1096&  1&   *& --& 117&  1&   24.02&   1.68& +37.6\p&    \#1.2\p&     2& SB& M&   C&    5.0\p&     2.83& 1\cr
21306&   *& --& 741&  0& 593&  0&    *& --&   *& --&   *& --&   12.62&   1.96& --81.8\p&     6.9\p&    22&  *& *&   *&   33.2\p&   375.20& 0\cr
21317&  97&  1& 748&  1& 598&  1& 1097&  1& 241&  1& 119&  1&   23.19&   1.30& +40.78&   0.16&   1&  *& *&   *&    3.6\p&     0.53& 1\cr
21332&   *& --& 751&  1& 600&  1&    *& --&   *& --&   *& --&    9.87&   1.02& *&        *&      0&  *& *&   *&   55.0\p&    19.28& 0\cr
21353&  98&  1&   *& --&   *& --&    *& --& 242&  0&   *& --&    6.81&   1.34& +28.95&   0.19&   1&  *& *&   *&  101.9\p&    47.17& 0\cr
21395&   *& --& 771&  1& 611&  0&    *& --& 245&  0&   *& --&   13.51&   1.32& +40.37&   0.24&   1& SB& *&   *&   28.1\p&    14.28& 0\cr
21459& 100&  1&   *& --&   *& --& 1100&  1&   *& --&  59&  1&   22.60&   0.76& +43.3\p&     1.2\p&     2&  *& *&   *&    5.9\p&     3.40& 1\cr
\noalign{\vskip 5pt}
21474& 101&  1&   *& --&   *& --& 1101&  1&   *& --& 121&  1&   22.99&   0.95& +33.7\p&    \#1.2\p&     2& SB& *&   *&    3.3\p&     6.02& 1\cr
21475&   *& --&   *& --&   *& --&    *& --&   *& --& 122&  1&   18.93&   1.75& *&        *&      0&  *& I&   *&    7.9\p&    20.92& 0\cr
21482&   *& --&   *& --&   *& --&    *& --& 249&  0&   *& --&   56.02&   1.21& +36.18&   0.08&   1& SB& *&   *&   29.0\p&     1.56& 1\cr
21543& 102&  1&   *& --&   *& --& 1102&  1& 253&  1&  40&  1&   23.54&   1.29& +42.00&  \#0.33&   1& SB& *&   G&    4.4\p&     4.81& 1\cr
21588& 103&  1&   *& --&   *& --& 1103&  1&   *& --&  41&  1&   21.96&   1.04& +38.4\p&     1.2\p&     2&  *& I&   G&    2.2\p&     6.01& 1\cr
21589& 104&  1&   *& --&   *& --& 1104&  1&   *& --&  15&  1&   21.79&   0.79& +44.7\p&    \#5.00&   2& SB& I&   *&    3.9\p&     0.74& 1\cr
21637& 105&  1&   *& --&   *& --& 1105&  1& 259&  1&  42&  1&   22.60&   0.91& +39.86&   0.29&   1&  *& *&   *&    5.9\p&     0.33& 1\cr
21654& 106&  1&   *& --&   *& --& 1106&  1& 262&  1& 123&  1&   20.81&   1.30& +41.86&  \#0.12&   1& SB& *&   *&    3.5\p&     0.74& 1\cr
21670& 107&  1&   *& --&   *& --& 1107&  1&   *& --&  16&  1&   19.44&   0.86& +36.3\p&     1.2\p&     2&  *& I&   *&    9.4\p&     3.34& 1\cr
21673&   *& --&   *& --&   *& --&    *& --&   *& --&   *& --&   21.49&   0.96& +26.1\p&     0.7\p&    11& SB& I&   G&    2.3\p&   124.44& 0\cr
\noalign {\smallskip}
\noalign{\hrule}
}}}}

\setbox6=\vbox
{
\tabcap{2}{Hyades membership compilation summary (4/4)}
\sevenrm\baselineskip 7pt
\hbox{ 
\def\p{\phantom{0}}
\vbox { 
\halign {
\tabskip 10pt \hfil#\hfil& 
\tabskip 2pt \hfil#& \tabskip 8pt \hfil#& 
\tabskip 2pt \hfil#& \tabskip 8pt \hfil#& 
\tabskip 2pt \hfil#& \tabskip 8pt \hfil#& 
\tabskip 2pt \hfil#& \tabskip 8pt \hfil#& 
\tabskip 2pt \hfil#& \tabskip 8pt \hfil#& 
\tabskip 2pt \hfil#& \tabskip 14pt \hfil#& 
\tabskip 2pt \hfil#& \tabskip 12pt \hfil#& 
\tabskip 2pt \hfil#& \tabskip 4pt \hfil#& \tabskip 15pt \hfil#& 
\tabskip 2pt \hfil#\hfil& \hfil#\hfil& \tabskip 10pt \hfil#\hfil& 
\tabskip 2pt \hfil#& \hfil#& \hfil#\hfil \cr		
\noalign{\hrule}
\noalign {\smallskip}
\tablabels
\noalign {\smallskip}
\noalign{\hrule}
\noalign {\smallskip}
21683& 108&  1&   *& --&   *& --& 1108&  1&   *& --&  17&  1&   20.51&   0.82& +35.6\p&     2.5\p&    19&  *& I&   *&    3.4\p&     2.91& 1\cr
21684&   *& --&   *& --&   *& --&    *& --& 264&  0&   *& --&    9.56&   1.73& +30.69&   0.20&   1&  *& *&   *&   58.7\p&    63.17& 0\cr
21723&   *& --&   *& --&   *& --&   80&  1& 266&  1& 124&  1&   23.95&   1.63& +42.50&   0.19&   1&  *& *&   *&    5.9\p&     0.53& 1\cr
21741& 109&  1&   *& --&   *& --& 1109&  1& 267&  1&  60&  1&   15.96&   1.36& +41.34&   0.16&   1&  *& *&   *&   17.7\p&     0.52& 1\cr
21762& 185&  0&   *& --&   *& --&   82&  1& 269&  1& 125&  1&   23.65&   2.53& +40.90&   0.17&   1& SB& M&   C&    4.7\p&     0.76& 1\cr
21788& 110&  1&   *& --&   *& --& 1110&  1& 270&  0& 166&  0&   19.48&   1.26& +35.85&   0.05&   1&  *& *&   *&    8.3\p&     7.85& 1\cr
21829& 163&  0&   *& --&   *& --&    *& --&   *& --&   *& --&    5.88&   0.97& +35.7\p&     1.2\p&     2&  *& *&   *&  126.7\p&    47.56& 0\cr
21923&   *& --&   *& --&   *& --&    *& --& 278&  0&   *& --&   23.23&   1.25& +13.57&   0.20&   1&  *& I&   *&    4.6\p&   211.77& 0\cr
21946&   *& --&   *& --&   *& --&    *& --& 279&  0&   *& --&   21.10&   2.22& +14.49&   0.23&   1& SB& *&   *&    3.5\p&   178.36& 0\cr
21983&   *& --&   *& --&   *& --&   94&  1& 281&  0&   *& --&   21.48&   1.84& +24.47&   0.34&   1&  *& *&   *&    5.4\p&    82.06& 0\cr
\noalign{\vskip 5pt}
22044& 111&  1&   *& --&   *& --& 1111&  1&   *& --&  43&  1&   20.73&   0.88& +39.6\p&     0.5\p&     2&  *& I&   *&    5.9\p&     1.18& 1\cr
22105&   *& --&   *& --&   *& --&    *& --& 282&  0&   *& --&    9.08&   1.79& +25.94&   0.11&   1&  *& *&   *&   64.1\p&    71.72& 0\cr
22157& 112&  1&   *& --&   *& --& 1112&  1&   *& --&  18&  1&   12.24&   0.86& +43.0\p&     1.0\p&    11& SB& I&   *&   36.1\p&     4.32& 1\cr
22176& 164&  0&   *& --&   *& --&    *& --& 283&  0&   *& --&   10.81&   0.94& +44.11&   0.10&   1& SB& *&   *&   46.5\p&    62.75& 0\cr
22177&   *& --&   *& --&   *& --&  119&  1& 285&  1&   *& --&   22.45&   2.32& +43.16&   0.25&   1&  *& *&   *&   11.1\p&     0.36& 1\cr
22203& 142&  1&   *& --&   *& --& 1142&  1& 284&  1& 126&  1&   19.42&   1.09& +42.42&  \#0.71&   1& SB& *&   *&    6.5\p&     0.76& 1\cr
22221& 113&  1&   *& --&   *& --& 1113&  1& 286&  1& 144&  1&   26.26&   1.04& +42.47&  \#0.11&   1& SB& *&   G&   10.5\p&     5.44& 1\cr
22224&   *& --&   *& --&   *& --&   92&  1&   *& --& 127&  1&   24.11&   1.72& +40.32&  \#0.09&   1& SB& *&   *&    6.0\p&     0.38& 1\cr
22253&   *& --&   *& --&   *& --&   93&  1& 290&  1&   *& --&   15.74&   1.98& +41.78&   0.23&   1&  *& *&   *&   18.7\p&     1.09& 1\cr
22265& 114&  1&   *& --&   *& --& 1114&  1&   *& --& 128&  1&   19.81&   1.43& +39.8\p&    \#0.4\p&    15& SB& *&   *&    5.9\p&     1.79& 1\cr
\noalign{\vskip 5pt}
22271&   *& --&   *& --&   *& --&    *& --& 291&  0&   *& --&   22.07&   2.03& +40.30&   0.17&   1&  *& *&   *&    8.5\p&     4.18& 1\cr
22350& 115&  1&   *& --&   *& --& 1115&  1& 296&  1&  61&  1&   19.30&   1.67& +41.84&  \#0.44&   1& SB& *&   G&    7.9\p&     0.72& 1\cr
22380& 116&  1&   *& --&   *& --& 1116&  1& 298&  1& 129&  1&   21.38&   1.46& +41.62&   0.15&   1&  *& *&   *&    4.4\p&     0.97& 1\cr
22394& 117&  0&   *& --&   *& --&    *& --& 299&  1&  62&  1&   18.96&   1.62& +40.60&   0.31&   1& SB& *&   *&   10.4\p&     0.34& 1\cr
22422& 118&  1&   *& --&   *& --& 1118&  1& 300&  1& 130&  1&   19.68&   0.96& +42.04&   0.14&   1&  *& *&   *&    6.3\p&     0.53& 1\cr
22446& 165&  0&   *& --&   *& --&    *& --& 301&  0&   *& --&   13.26&   1.11& +31.84&   0.14&   1&  *& *&   *&   30.3\p&    35.59& 0\cr
22496& 119&  1&   *& --&   *& --& 1119&  1&   *& --& 131&  1&   22.96&   1.17& +41.40&   0.16&  24& SB& *&   G&    5.1\p&     1.31& 1\cr
22505& 120&  1&   *& --&   *& --& 1120&  1& 305&  1& 132&  1&   23.64&   0.99& +42.34&  \#0.33&   1& SB& *&   S&    6.0\p&     1.75& 1\cr
22524& 121&  1&   *& --&   *& --& 1121&  1& 307&  1& 133&  1&   19.30&   0.95& +42.74&   0.17&   1& SB& *&   *&    7.2\p&     0.80& 1\cr
22550& 122&  1&   *& --&   *& --& 1122&  1& 312&  1& 134&  1&   20.15&   1.14& +42.44&  \#0.17&   1& SB& I&   C&    7.4\p&     0.83& 1\cr
\noalign{\vskip 5pt}
22565& 123&  1&   *& --&   *& --& 1123&  1&   *& --&   8&  1&   17.27&   0.82& +36.8\p&     1.2\p&     2&  *& I&   *&   12.8\p&     4.10& 1\cr
22566& 143&  1&   *& --&   *& --& 1143&  1& 313&  1& 135&  1&   17.14&   1.00& +42.92&   0.19&   1&  *& *&   *&   13.1\p&     1.19& 1\cr
22607& 124&  1&   *& --&   *& --& 1124&  1&   *& --& 136&  1&   23.91&   1.04& +39.83&   0.24&   1& SB& I&   C&    6.7\p&     1.87& 1\cr
22654&   *& --&   *& --&   *& --&   98&  1& 318&  1&   *& --&   18.93&   2.02& +42.88&   0.25&   1&  *& *&   *&    8.4\p&     0.58& 1\cr
22684& 145&  1&   *& --&   *& --& 1145&  1& 319&  0& 167&  0&   12.14&   2.22& +48.53&   0.08&   1&  *& I&   C&   36.7\p&    17.34& 0\cr
22751& 125&  1&   *& --&   *& --&    *& --& 325&  0& 168&  0&   11.62&   1.95& +48.72&   0.18&   1&  *& *&   *&   40.9\p&    16.84& 0\cr
22782& 146&  1&   *& --&   *& --& 1146&  1& 327&  0& 174&  0&   14.82&   0.88& +57.10&   0.23&   1&  *& *&   *&   22.0\p&    95.23& 0\cr
22805& 166&  0&   *& --&   *& --&    *& --& 328&  0&   *& --&    5.52&   1.29& +19.14&   0.33&   1& SB& *&   *&  135.2\p&   114.49& 0\cr
22850& 126&  1&   *& --&   *& --& 1126&  1&   *& --& 137&  1&   14.67&   0.95& +38.4\p&     2.0\p&     8&  *& *&   *&   22.9\p&     1.27& 1\cr
22893& 147&  1&   *& --&   *& --&    *& --& 331&  0& 175&  0&    9.66&   1.43& --30.57&   0.26&   1&  *& *&   *&   58.3\p&   993.75& 0\cr
\noalign{\vskip 5pt}
23044& 149&  1&   *& --&   *& --& 1149&  1& 336&  0&   *& --&   12.62&   1.89& +37.96&   0.17&   1&  *& I&   C&   37.6\p&    13.12& 1\cr
23056& 148&  1&   *& --&   *& --&    *& --& 335&  0& 169&  0&   14.29&   1.48& +60.93&   0.17&   1&  *& *&   *&   26.8\p&   109.89& 0\cr
23069& 127&  1&   *& --&   *& --& 1127&  1& 337&  1& 138&  1&   19.66&   1.62& +43.68&   0.16&   1&  *& *&   *&    7.9\p&     0.80& 1\cr
23205&  --& --&  --& --&  --& --&   --& --&  --& --&  --& --&   10.73&   1.66& *&        *&      0&  *& I&   C&   50.5\p&     8.53& ?\cr
23214& 128&  1&   *& --&   *& --& 1128&  1&   *& --&  44&  1&   23.09&   0.83& +42.5\p&     1.5\p&    24&  *& I&   *&    6.7\p&     0.30& 1\cr
23312&   *& --&   *& --&   *& --&    *& --&   *& --& 160&  0&   16.77&   1.79& +42.21&   0.40&  24&  *& *&   *&   18.7\p&     1.18& 1\cr
23409&   *& --&   *& --&   *& --&  105&  1& 341&  0&   *& --&   11.39&   1.66& +78.64&   0.36&   1&  *& *&   *&   45.5\p&   358.75& 0\cr
23497& 129&  1&   *& --&   *& --& 1129&  1&   *& --&   4&  1&   20.01&   0.91& +38.0\p&     1.7\p&     8&  *& *&   *&    8.9\p&     1.29& 1\cr
23498& 187&  0&   *& --&   *& --&  107&  1& 345&  1& 139&  1&   18.44&   1.66& +43.51&   0.19&   1&  *& *&   G&   11.1\p&     0.08& 1\cr
23574& 150&  1&   *& --&   *& --& 1150&  1& 346&  0& 176&  0&    2.26&   1.27& +30.22&   0.19&   1&  *& *&   *&  396.8\p&    31.53& 0\cr
\noalign{\vskip 5pt}
23589&   *& --&   *& --&   *& --&    *& --& 347&  0&   *& --&    5.30&   0.81& +49.35&   0.21&   1&  *& *&   *&  143.2\p&    40.64& 0\cr
23599&   *& --&   *& --&   *& --&    *& --& 348&  0&   *& --&    3.98&   1.61& +111.51&  0.37&   1&  *& *&   *&  205.9\p&   855.75& 0\cr
23662&  --& --&  --& --&  --& --&   --& --&  --& --&  --& --&   16.69&   1.12& *&        *&      0&  *& *&   O&   18.6\p&     4.30& ?\cr
23701& 151&  1&   *& --&   *& --& 1151&  1& 349&  0& 145&  1&   13.78&   2.08& +42.92&  \#0.16&   1& SB& *&   G&   29.7\p&     2.37& 1\cr
23750&   *& --&   *& --&   *& --&    *& --&   *& --& 140&  1&   18.78&   1.40& +42.31&   0.18&  24&  *& *&   *&   10.6\p&     0.15& 1\cr
23772&  --& --&  --& --&  --& --&   --& --&  --& --&  --& --&   12.00&   1.87& +35.38&  \#0.93&  24& RV& H&   C&   39.7\p&    11.47& 1\cr
23983& 130&  1&   *& --&   *& --& 1130&  1&   *& --&  10&  1&   18.54&   0.83& +44.16&   0.14&  24& SB& I&   *&   13.0\p&     0.55& 1\cr
24019& 131&  1&   *& --&   *& --& 1131&  1&   *& --&   9&  1&   18.28&   1.30& +44.90&   0.52&  24&  *& I&   C&   15.5\p&     5.84& 1\cr
24020& 132&  1&   *& --&   *& --& 1132&  1& 353&  1&  64&  1&   18.28&   1.30& +45.00&   0.19&   1&  *& I&   C&   15.5\p&    36.00& 0\cr
24021&  --& --&  --& --&  --& --&   --& --&  --& --&  --& --&   21.39&   1.21& *&        *&      0&  *& *&   *&   11.3\p&     9.56& ?\cr
\noalign{\vskip 5pt}
24035& 152&  1&   *& --&   *& --& 1152&  1& 354&  0& 170&  0&   25.67&   1.53& +16.48&  \#0.39&   1& SB& I&   *&   13.1\p&   175.91& 0\cr
24046&   *& --&   *& --&   *& --&    *& --&   *& --&   *& --&   24.88&   1.06& *&        *&      0&  *& I&   *&   12.6\p&    54.92& 0\cr
24116&  --& --&  --& --&  --& --&   --& --&  --& --&  --& --&   11.56&   1.19& +45.30&   1.20&   2&  *& *&   *&   41.9\p&     1.87& 1\cr
24923&  --& --&  --& --&  --& --&   --& --&  --& --&  --& --&   18.26&   1.58& +43.70&   0.23&  24&  *& *&   *&   14.7\p&     0.17& 1\cr
25141& 167&  0&   *& --&   *& --&    *& --& 361&  0&   *& --&    9.40&   1.48& +34.54&   0.24&   1&  *& *&   G&   63.7\p&    24.65& 0\cr
25639&  --& --&  --& --&  --& --&   --& --&  --& --&  --& --&   11.58&   1.13& +42.1\p&     0.43&  24&  *& *&   *&   46.3\p&    14.11& 1\cr
25694&  --& --&  --& --&  --& --&   --& --&  --& --&  --& --&   11.17&   1.28& *&        *&      0&  *& *&   *&   46.4\p&    11.21& ?\cr
25871&  --& --&  --& --&  --& --&   --& --&  --& --&  --& --&   11.55&   0.91& *&        *&      0&  *& *&   *&   45.0\p&     1.32& ?\cr
25929& 155&  0&   *& --&   *& --&    *& --& 366&  0&   *& --&    6.74&   1.40& +33.22&   0.34&   1&  *& *&   *&  104.9\p&    30.90& 0\cr
26159&  --& --&  --& --&  --& --&   --& --&  --& --&  --& --&   11.13&   1.39& *&        *&      0&  *& *&   S&   47.3\p&    10.82& ?\cr
\noalign{\vskip 5pt}
26227& 156&  0&   *& --&   *& --&    *& --& 367&  0&   *& --&    0.72&   1.90& +46.64&   0.24&   1&  *& *&   *& 1345.6\p&     4.57& 0\cr
26382& 168&  0&   *& --&   *& --&    *& --&   *& --&   *& --&   18.56&   0.86& +41.1\p&     1.2\p&     2&  *& *&   *&   16.2\p&     2.07& 1\cr
26795&   *& --&   *& --&   *& --&    *& --&   *& --& 161&  0&    6.93&   1.13& +27.1\p&     0.27&  23& SB& *&   *&  102.9\p&    66.29& 0\cr
26844&  --& --&  --& --&  --& --&   --& --&  --& --&  --& --&   46.51&   2.35& *&        *&      0&  *& *&   *&   26.7\p&     2.78& ?\cr
27431&  --& --&  --& --&  --& --&   --& --&  --& --&  --& --&   13.11&   0.87& *&        *&      0&  *& *&   *&   37.4\p&     8.89& ?\cr
27502&   *& --&   *& --&   *& --&    *& --&   *& --& 147&  0&    6.15&   1.25& *&        *&      0&  *& *&   G&  120.0\p&    10.01& 0\cr
27791&  --& --&  --& --&  --& --&   --& --&  --& --&  --& --&   11.50&   6.04& *&        *&      0&  *& I& X/S&   49.7\p&     1.46& ?\cr
27933&   *& --&   *& --&   *& --&    *& --&   *& --& 148&  0&   12.76&   0.94& *&        *&      0&  *& *&   *&   39.2\p&    33.71& 0\cr
28356&  --& --&  --& --&  --& --&   --& --&  --& --&  --& --&   14.87&   0.98& +45.00&   2.50&   2&  *& *&   *&   33.1\p&     0.53& 1\cr
28469&  --& --&  --& --&  --& --&   --& --&  --& --&  --& --&   10.52&   0.99& +48.00&   5.00&   2&  *& *&   *&   56.3\p&    10.27& 1\cr
\noalign{\vskip 5pt}
28614& 169&  0&   *& --&   *& --&    *& --&   *& --&   *& --&   21.49&   0.82& +40.9\p&     0.3\p&    20& SB& I&   C&   19.3\p&    26.99& 0\cr
28774&  --& --&  --& --&  --& --&   --& --&  --& --&  --& --&   12.81&  12.80& *&        *&      0&  *& I&   C&   41.6\p&    11.50& ?\cr
\noalign {\smallskip}
\noalign{\hrule}
}}}}

\input psfig		    	

\titlea{Introduction}

The considerable importance of the Hyades cluster in studies of
Galactic structure, in the understanding of the chemical evolution of
the Galaxy, and in the determination of the Population~I distance
scale, is well documented in the literature.  The nearest moderately
rich cluster, with some 300 possible members, a total mass of some
300--400~M$_\odot$, and an age of around 600--800~Myr, it has an
extension in the sky of about 20~degrees.  Although uncertainty in the
distances of individual members has limited the definition of the
cluster's main sequence, and thereby its helium content and
corresponding evolutionary sequence, it has nevertheless been used as
the basic observational material for several fundamental relationships
in astrophysics, including the location of the main sequence in the
Hertzsprung-Russell diagram and the mass-luminosity relationship, as
well as forming the basis for the determination of luminosities of
supergiants, OB stars, and peculiar stars in clusters. Determinations
of the distance to the cluster have provided the zero-point for
distances within our Galaxy and, indirectly through the Cepheids, one
of the foundations on which the extragalactic distance scale
ultimately rests.

At 40--50~pc, the Hyades cluster is somewhat beyond the distance where
the parallaxes of individual stars are easily measured, or generally
considered as fully reliable, from ground-based observations. Over
almost a century, considerable effort using a wide variety of indirect
methods has therefore been brought to bear on the problem of
establishing the distance to the cluster. Distance estimates have been
based on a variety of geometrical manifestations of a cluster of stars
participating in a uniform space motion, while other estimates have
been based on the average trigonometric parallax for a number of
cluster stars, dynamical parallaxes for binaries, and photometric
parallaxes using a variety of photometric systems. Nevertheless, the
details of the HR and mass-luminosity diagrams remain imprecisely
established due to limitations in the accuracy of the parallaxes of
the individual members, while the distance of the cluster is still
open to debate: recent estimates of the distance modulus range from
3.16 based on trigonometric parallaxes (Gatewood et al.\ 1992), 3.40
based on convergent-point analyses using proper motions from the FK5
and PPM Catalogues (Schwan 1991), and 3.42 based on recent Hubble
Space Telescope FGS observations (van Altena et al.\ 1997a).

The present work is inspired by the availability of the final results
of the Hipparcos astrometry mission, which provide a radical
improvement in astrometric data on all stars in the Hipparcos
observing programme, including approximately 240 candidate Hyades
members. The Hipparcos results offer the following principal
improvements: (1) standard errors of the annual proper motions of
typically 1~milliarcsec (mas) with respect to an inertial
(extragalactic) reference frame; (2) absolute trigonometric parallaxes
with standard errors of order 1~mas; (3) systematic errors of the
astrometric parameters below around 0.1~mas (or mas/yr); (4) parameter
determination, or indications, of double or multiple systems for
component separations larger than about 0.1~arcsec and $\Delta
m\la3$~mag; (5) precise photometry and detailed variability indicators
based on the Hipparcos broad-band magnitude $Hp$; (6) homogeneous
$B-V$ and $V-I$ colour indices.

Literature on the Hyades distance determination is considerable: this
paper is intended neither as a comprehensive review nor a critical
evaluation of the previous estimates in the light of the Hipparcos
results. Neither does it aim to answer unambiguously the question
`what is the distance to the Hyades', a somewhat nebulous problem
given the resolution in radial distance provided by the Hipparcos
parallaxes, and the sensitivity of the results to the precise
qualification of distance: if the distance of the centre of mass is
the objective, membership criteria, selection effects, and M/L
relationships become critical. Rather, our objective is to reconcile
previous distance estimates based on the availability of reliable
absolute trigonometric parallaxes, assign improved membership
probabilities, and thereafter probe both the cluster dynamics and the
assumptions on which previous distance determinations have
rested. Finally, we will define the observational main sequence based
on a subset of objects for which membership is secure and
observational data particularly reliable, and compare this with
theoretical determinations of the Hyades zero-age main sequence based
on knowledge of the cluster's metallicity.

In order to establish the complexities of the problem, Sect.~2
provides a summary of (some of) the distance determinations discussed
in the literature to date, with particular reference to the agreement
or disagreement between the results of the various convergent-point
analyses and distance estimates derived by other means. Sect.~3
summarises the data, both from the Hipparcos Catalogue and from the
published literature, used for the present study.  Selection effects
entering the list of candidates for this study are also discussed.

The development of the paper then proceeds as follows. In Sect.~4, we
examine the improvement brought by the Hipparcos proper motions (and
their connection to an inertial frame) which, as we shall demonstrate,
permit a significant advance in the understanding of the systematic
effects entering previous evaluations of the distance to the Hyades
based on convergent-point analyses. New insights and the limitations
of this approach applied to the Hyades are discussed. In Sect.~5 we
use the Hipparcos absolute trigonometric parallaxes to determine a
statistically significant distance estimate for each candidate member,
eventually permitting a provisional mean cluster distance to be
defined.  This is carried out in parallel with a combination of the
parallaxes and proper motions with published and unpublished radial
velocities to determine the position and velocity components of
candidate members with respect to a reference cluster centre.  In
Sect.~6 we discuss the Hipparcos parallaxes: first we combine the
information coming from the Hipparcos parallaxes and proper motions,
and demonstrate their mutual consistency. Then we examine the
differences between ground-based and Hipparcos parallaxes. Finally, we
examine effects (in particular `Lutz-Kelker' type corrections) which
complicate the direct interpretation of the Hipparcos parallaxes.

In Sects~7 and 8 we examine the spatial distribution and dynamics of
the cluster, looking at the question of mass segregation, and
comparing our present results with published $N$-body simulations. We
examine the velocity residuals of each member, which can be fully
explained on the basis of the observational errors. We also examine
the consistency of these results with previous estimates of the
rotation, flattening, and internal velocity dispersion of the system.

Finally, in Sect.~9, we restrict our list of Hyades candidates to
those showing no existing evidence for multiplicity, and construct the
resulting observational main-sequence, comparing it with new models
for the zero-age main sequence. From this, the cluster helium
abundance is estimated, and this information is used to construct
isochrones from which an estimate of the cluster age is determined.

\titlea{Distance determinations to date}   

Under certain conditions which are at least reasonably well
represented in the Hyades, the common proper motion of stars in a
cluster can be used in the well-established, but nonetheless ingenious
`convergent-point' method of distance determination. Since all
distance determinations employed for the Hyades have either been based
on this method, or have been judged according to their agreement or
otherwise with it, a short review of its theoretical basis and a
comparison between its results and those obtained by other methods
over the years is in order. Reviews of the various distance
determination methods include those by van Altena (1974), Hanson
(1980), and Turner et al.\ (1994). We will not discuss details of
distance estimates based on photometric parallaxes (see, e.g., van
Altena 1974, Turner et al.\ 1994). A summary is given in Table~1.

\begtabfull
\tabcap{1} {Distance determinations ordered by date (...
indicates one or more co-authors). The distance modulus (D.M.) is
taken or derived from the original reference, occasionally with some
uncertainty as to a definitive value (and not necessarily in agreement
with values referred to in subsequent compilations).}
\centerline{\box1}
\endtab

>From simple dynamical arguments it can be shown that, for an open
cluster of a few hundred stars within a volume of a few parsecs in
radius, moving together under their mutual gravitation, the internal
velocity dispersion is of the order of 1~km s$^{-1}$ or less, and thus
small compared with the typical linear velocity of the cluster as a
whole relative to the Sun, $\bf v$ (the Hyades cluster has a space
motion of approximately 45~km~s$^{-1}$ with respect to the Sun,
presumably reflecting the velocity of the cloud in which the cluster
formed).  Provided that the cluster is sufficiently nearby to extend
over an area of, say, several degrees, the parallel motions of the
stars in space yield, on the celestial sphere, directions of proper
motions that appear to converge on a unique point -- the direction of
the unit vector $\langle {\bf v}
\rangle$ is known as the convergent point. If $\bf b$ is the
barycentric coordinate vector to a cluster member, and $\langle {\bf
b} \rangle$ its coordinate direction then, neglecting the internal
velocity dispersion, the radial velocity is $\rho=\langle {\bf b}
\rangle ^\prime{\bf v}$ (the prime symbol associated with matrices and
vectors will be used to denote transposition, so that ${\bf
x}^\prime{\bf y}$ denotes the scalar product of the two vectors). With
$\lambda$ denoting the angular distance between the star and the
convergent point, and $\pmb{$\mu$}$ the proper motion vector: 
$$
\langle {\bf b} \rangle ^\prime \langle {\bf v} \rangle = \cos \lambda  \eqno(1)
$$
and:
$$
\vert \pmb{$\mu$} \vert =\pi\, \vert{\bf v}\vert \sin\lambda/A_v  \eqno(2)
$$
where $\pi$ is the parallax of the cluster member.
$A_v=4.74047...$~km~yr~s$^{-1}$ is the astronomical unit expressed in
the appropriate form when $\pi$ and $\mu$ are expressed in mas and
mas~yr$^{-1}$ respectively. For the Hyades, $\lambda\simeq33^\circ$,
the radial velocity $v_{\rm rad} \simeq40$~km~s$^{-1}$ in the cluster
centre, $\vert{\bf v}\vert\simeq45$~km~s$^{-1}$, and $\vert
\pmb{$\mu$} \vert \sim 100$~mas~yr$^{-1}$.

Although $\bf v$ can in principle be determined from the radial
velocity measurements alone, its resulting direction is generally not
well determined because of the limited angular extent of the cluster;
the usual procedure has therefore been to determine $\langle {\bf v}
\rangle$ from proper motions, and $\vert{\bf v}\vert$ from radial
velocities, from which $\lambda$ is obtained from Eq.~(1) and $\pi$
from Eq.~(2).

Although the method is conceptually simple, its application in
practice is not so straightforward. Errors in the individual proper
motions resulting from measurement errors, or defects in the proper
motion system, lead to accidental errors in $\pi$, to an error in
$\langle{\bf v}\rangle$ and, ultimately, to a systematic bias in
$\lambda$ depending on $\langle {\bf b} \rangle $. For the Hyades, the
streaming motion differs by only 60--70~degrees from that of local
field stars towards the solar antapex, so that observational scatter
in the proper motions of member stars, and the random motions of field
stars, complicates membership selection based only on proper motions.

In the basic convergent-point method it is assumed that the cluster is
neither expanding, contracting, or rotating, that the motion of the
cluster with respect to the field is large enough to permit accurate
membership discrimination, and that the system of proper motions is
inertial and without systematic errors. In his review, van Altena
(1974) considered that the first two criteria were adequately
satisfied, but that information on the proper motion system was
incomplete. Hanson (1975) considered the possibility of random motions
contributing significantly to the stars' space velocities, as well as
the effects of expansion, contraction, or rotation, concluding that
any resulting deviations from parallel motion are insignificant at
levels affecting the distance determination by the convergent point
method. Gunn et al.\ (1988) presented weak evidence for rotation at
the levels of $\la 1$~km~s$^{-1}$ rad$^{-1}$ (projected), not
inconsistent with these conclusions.

The convergent point method was first applied to the Hyades cluster by
Boss (1908), using the proper motions of 41 suspected cluster members
supplemented by three radial velocities. The classical
convergent-point method was further developed and discussed by Smart
(1938), Brown (1950), and others.  A systematic regression error
arising from the quadratic form of the proper motion component
coefficients in the normal equations, and leading to an upward
revision of 7~per cent in the distance to the cluster, was identified
by Seares (1944, 1945).

Subsequent distance determinations using the convergent-point method
initially appeared to be in close agreement, although the
correspondence between the van Bueren (1952) and Wayman et al.\ (1965)
results was later attributed in part to the use of the same proper
motion system (van Altena 1974). Hodge
\& Wallerstein (1966) suggested that the cluster was 20 per cent farther than
indicated by the proper motions -- given the previous standard
distance, binary stars in the Hyades would have been overluminous with
respect to their masses, both as compared to normal stars like the
Sun, and as compared to models derived from stellar structure theory
(Wallerstein \& Hodge 1967).

In the classical convergent-point method, the determination of
$\langle {\bf v} \rangle$ depends only on the directions of proper
motions, and not on their absolute values. Upton (1970) derived a
procedure for calculating the distance directly from the proper motion
gradients across the cluster, dispensing with the intermediate step of
locating the convergent point -- the cluster distance is then given by
the ratio of the mean cluster radial velocity to the proper motion
gradient in either coordinate. Use of this method, whose relevant
equations can be derived by differentiating the basic convergent-point
equation (Eq.~2), has the advantage that more complete use is made of
the proper motion data, while the two independently measured gradients
yield two distance estimates whose comparison provides an indication
of the systematic and accidental errors involved.

The accepted distance to the Hyades was revised from about 40~pc to
about 44~pc around 1978 based on models of the chemical composition
(Koester \& Weidemann 1973), and independent astrometric and
photometric results (van Altena 1974, Hanson 1975, Eggen 1982). Hanson
(1975) applied different formulations of the convergent-point method
to new proper motion and cluster membership data, concluding that the
errors due to different formulations of the method appeared to be
quite small, with systematic errors in previous meridian circle proper
motions implicated as the cause of the discrepancies which seemed to
exist between distances derived from earlier proper motion analyses
and those resulting from a broad variety of other observational
methods.

\begfigwid 10.0cm
\vskip-10.0cm
\centerline{\psfig{figure=fig01_dist_mod.ps,width=12cm,angle=270}}
\figure{1}{
Distance modulus (given by $m-M=-5\log\pi-5$, where $\pi$ is the parallax in 
arcsec) for the distance determinations, with errors, since 1980 given in
Table~1. The `Torres' determination refers to Torres et al.\ (1997c).
} 
\endfig

Taking into account systematic magnitude effects in the Hanson proper
motions, McAlister (1977) revised Hanson's distance of 48~pc downward
to 43~pc, close to the value of 43.5~pc given by Corbin et al.\
(1975). Murray \& Harvey (1976) showed how all measurements of proper
motion and radial velocity of the cluster members could be combined
into a general solution for the cluster motion and the parallaxes of
individual stars. A review of astrometric results by Hanson (1980)
concluded that 45.6~pc (distance modulus $3.30\pm0.04$) was indicated
by the best of current data; being a weighted mean of classical
convergent-point methods and trigonometric parallaxes. Meanwhile,
dynamical parallaxes from visual and eclipsing binaries have
traditionally yielded slightly higher values: McClure (1982) found
49~pc, but a reanalysis of that and other data by Peterson \& Solensky
(1987) gave 47~pc (distance modulus $3.36\pm0.05$) still slightly
higher than the astrometric results.

Determination of the convergent point from radial velocities was
applied to the Hyades by Stefanik \& Latham (1985), Detweiler et al.\
(1984), and Gunn et al.\ (1988), based on the methodology applied by
Thackeray (1967) to Sco-Cen.

Schwan (1990, 1991) presented the most recent determinations of the
convergent point based on proper motions of 44 and 145 stars from the
FK5 and FK5/PPM respectively, and derived a distance modulus of
$3.40\pm0.04$.

Individual trigonometric parallaxes have been published for certain
candidate Hyades members, most recently by Patterson \& Ianna (1991),
by Gatewood et al.\ (1992), and in the Fourth Edition of the General
Catalogue of Trigonometric Stellar Parallaxes (van Altena et al.\
1995). Recent determinations of a mean cluster distance have been
given by Turner et al.\ (1994), and van Altena et al.\
(1997b). High-precision orbital parallaxes have recently been
determined for 51~Tau (vB24), 70~Tau (vB57), and 78~Tau (vB72) by
Torres, Stefanik \& Latham (1997a,b,c). In each case, however, their
extrapolation to a mean cluster distance involves the use of relative
ground-based proper motions, so that the high intrinsic accuracy of
their orbital parallax determinations does not propagate through to a
corresponding accuracy on the mean distance.

A weighted mean distance modulus of $3.42\pm0.09$~mag, based on Hubble
Space Telescope FGS observations of seven cluster members, has been
given by van Altena et al.\ (1997a).

These recent determinations are shown, with their published errors, in
Fig.~1. It is evident that the Hyades distance may still not be
considered as a conventional astronomical constant. It is this
uncertainty, and its attendant implications, that this paper seeks to
resolve.

The Hipparcos Catalogue provides parallaxes for all stars in the
observing programme, of sufficient (milliarcsec) accuracy not only to
assign membership probabilities on the basis of the distances alone,
but to resolve the depth of the cluster. The availability of annual
proper motions with standard errors of order 1~mas~yr$^{-1}$ for all
stars, leads to an opportunity to re-discuss membership on the basis
of convergent point analysis, to probe the kinematical assumptions
implicit in such analyses, and to examine, in combination with the
parallaxes, the cluster membership and dynamics independently of any
assumed dynamical model.

\titlea{Observational material} 

\titleb{Data from the Hipparcos Catalogue}

This study makes use of the final data contained in the Hipparcos
Catalogue, which provides barycentric coordinates,
inertially-referenced proper motions, and absolute trigonometric
parallaxes for nearly $120\,000$ stars (ESA 1997). It is based on the
240 candidate Hyades members specifically included in the Hipparcos
Input Catalogue, supplemented by the astrometric and photometric data
for all Hipparcos Catalogue objects within the range $2^{\rm
h}\,15^{\rm m}<\alpha<6^{\rm h}\,5^{\rm m}$ and
$-2^\circ<\delta<+35^\circ$ for independent membership studies of
objects not considered as candidate members in the past (this region
includes all previous candidate cluster members).

We stress from the outset that the Hipparcos Input Catalogue, on which
the final Hipparcos Catalogue contents are based, is not complete to
the observability limit of the Hipparcos observations, although
specific attention was given during its construction to the inclusion
of potentially observable candidate Hyades cluster members. Thus,
although membership analysis can be conducted on previously
unsuspected cluster members contained in the Hipparcos Catalogue, this
catalogue will not contain members fainter than the satellite
observability limit, of around $V\sim12$~mag, nor objects omitted from
the Hipparcos Input Catalogue for other reasons.  Clearly, an
incomplete survey will most likely result in a preferential selection
of stars according to distance, and a biased value of the mean cluster
distance -- although it will not affect the discussion of the main
sequence modelling when using individual distance estimates for each
object.

\begfigwid 10cm
\vskip-10cm
\centerline{\psfig{figure=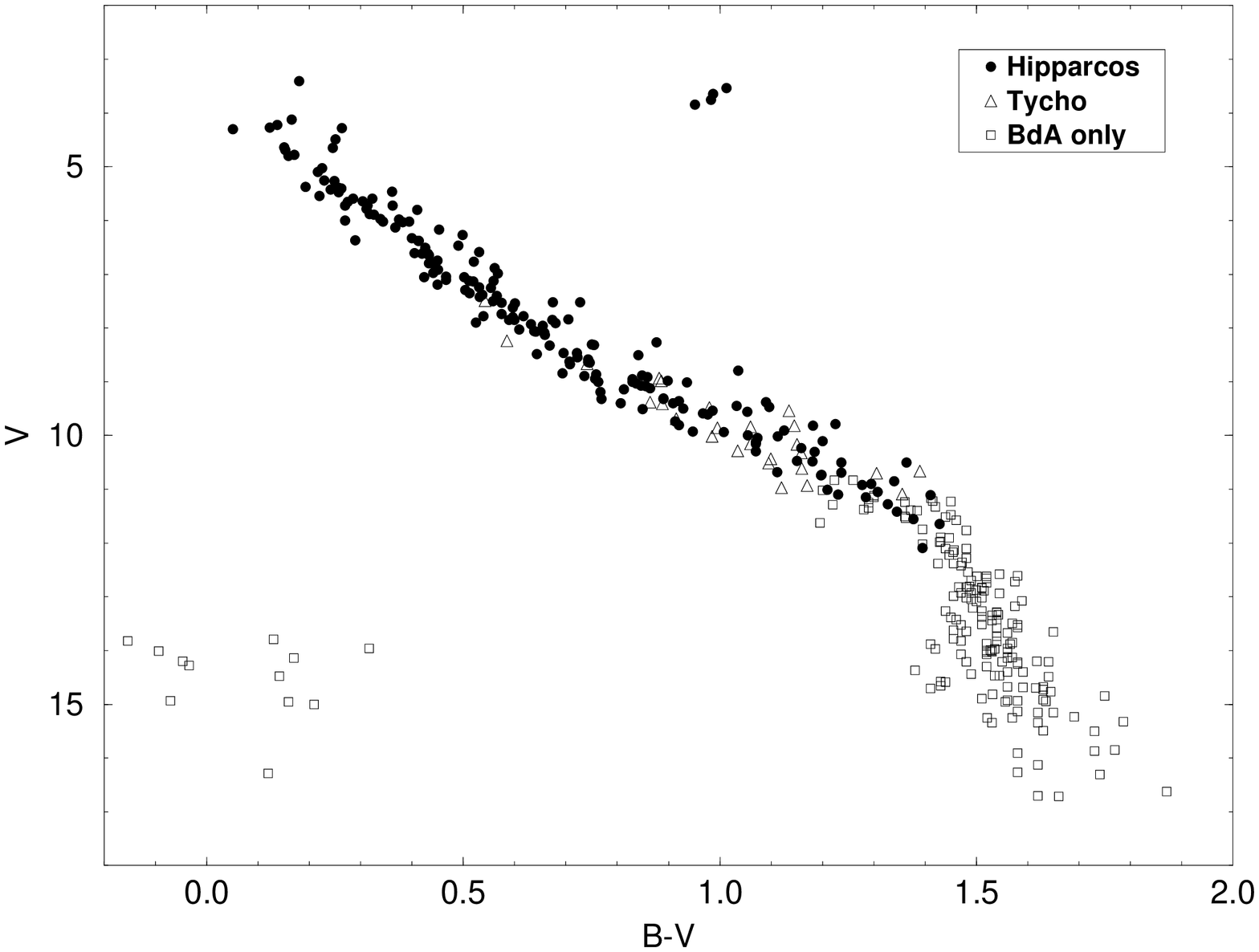,width=14cm,angle=270}}
\figure{2} {Hyades stars from the `Base des Amas' (BDA). Stars contained 
in the Hipparcos Catalogue are displayed as filled circles (190
stars).  Stars not contained in the Hipparcos Catalogue, but appearing
in the Tycho Catalogue, are displayed as open triangles
(27~stars). The remaining 174 stars contained only in the BDA are
displayed as open squares.  }
\endfig

The choice of Hipparcos targets in a given field of the sky was
subject to many operational constraints which were, in some cases, in
conflict with scientific requirements. Fig.~2 illustrates the sample
of Hyades stars contained in the Hipparcos Catalogue with respect to
the global content of Hyades candidate members contained in the data
base for stars in open clusters (`Base des Amas', or BDA, Mermilliod
1995), on the basis of the photometric data, $V$ and $B-V$, contained
in the BDA. Stars from the BDA contained in the Hipparcos Catalogue
are displayed as filled circles (190 stars). Stars not contained in
the Hipparcos Catalogue, but appearing in a second major product of
the Hipparcos mission, the Tycho Catalogue (ESA 1997), are displayed
as open triangles (27 stars). The remaining 174 stars contained only
in the BDA are displayed as open squares. The figure shows, as
expected, the progressive incompleteness of the Hipparcos sample with
increasing magnitude (and is also a useful demonstration of the
completeness of the Tycho Catalogue down to about $V=10.5$~mag).

Up to 114 of the `BDA only' stars lie within 5~pc of the cluster
centre, assuming that all lie at the mean distance of the cluster
centre.  Any possible kinematical bias due to the present sample
selection has not been studied in the present paper, as the distance,
proper motion and radial velocity data for stars other than those
contained in the Hipparcos Catalogue are either insufficiently
accurate, incomplete, or very inhomogeneous.

For objects contained in the Hipparcos Catalogue, those considered to
be members or candidate members by one or more previous workers are
listed in Table~2.  Column~a gives the Hipparcos Catalogue identifier,
while columns~b--m give the designation and membership status
according to a number of previous workers. These are not the only
papers where membership of particular objects have been discussed
(see, for example, references in Griffin et al.\ 1988) although they
represent the most substantial developments of the cluster membership
studies. The membership status listed in columns b--m do not
necessarily reflect fully the membership assignment in the original
papers: in some cases the authors give probabilities for membership or
include some indication of `doubtful membership', which we have
converted (sometimes subjectively) into a 1 or 0. Table~2 thus
reflects our own understanding of previous membership studies
converted to a yes/no status. For full details we refer to the
original papers.  Entries in Table~2 with `--' in columns~b--m
inclusive are new candidates arising from the present study, selected
as described in subsequent sections (we considered it desirable to
list all candidates sequentially in one table, independently of their
history, and will distinguish their historical status by referring to
these as `previous' and `new' members, where appropriate).

\newdimen\oldhsize

\begtabpage
\oldhsize=\hsize
\hsize=18cm
\tabtwocap
\hsize=\oldhsize
\bigskip
\box3
\endtab

\begtabpage
\box4
\endtab

\begtabpage
\box5
\endtab

\begtabpage
\box6
\endtab

The previous candidates compiled in Table~2 form the basic list of
objects for our initial studies. It is noted that this list contains
objects already considered as non-members by some or even all previous
workers (and which we will go on to confirm as non-members), while it
does not include those objects which are considered as possible or
secure Hyades members which are not contained within the Hipparcos
Catalogue (as described above). Later in the paper, having determined
the general spatial and kinematical properties of the cluster on the
basis of the general properties of the previous members, we will
provide our own assignment of membership to this basic list (this
final result is given in the last column of Table~2). We will also
supplement the previous candidates by additional candidates selected
from the Hipparcos Catalogue having spatial and kinematical properties
in common with the general cluster (also included in Table~2).

The Hipparcos Catalogue (ESA 1997) itself describes the details of the
catalogue construction and contents, while recent summaries may be
found in the literature related to the construction of the
intermediate catalogue (Kovalevsky et al.\ 1995), and to the
determination of the trigonometric parallaxes and associated errors
(Perryman et al.\ 1995).

The Hipparcos and Tycho Catalogues have been constructed such that the
Hipparcos reference frame coincides, to within limits set by
observational uncertainties, with the International Celestial
Reference System (ICRS), as recommended by the IAU Working Group on
Reference Frames (Ma et al.\ 1997, see also Lindegren \& Kovalevsky
1995).  The latter system is practically defined by the adopted
positions of several hundred extragalactic radio sources. It
supersedes, although it is consistent with, the optical reference
frame defined by the FK5 catalogue, which was formally based on the
mean equator and dynamical equinox of J2000.  The resulting deviation
from inertial, about all three axes, is considered to be less than
approximately 0.25~mas~yr$^{-1}$. For a discussion of the comparison
of ground-based positions and proper motions with those of Hipparcos,
see Lindegren et al.\ (1995). The Hipparcos Catalogue (ESA 1997,
Volume~1, Section~1.5.7) details the relationship between the
ICRS(Hipparcos) and J2000(FK5) frames. The epoch of the Hipparcos
Catalogues is J1991.25, although the provision of the full covariance
matrix of the astrometric solution for each star permits the
positions, and corresponding standard errors, to be propagated to any
epoch within the same reference system.

The Hipparcos trigonometric parallaxes are absolute, and are
considered to be free from systematic (global) errors at a level of
some 0.1~mas or smaller (Arenou et al.\ 1995).  Studies so far suggest
that the true external parallax and proper motion errors are unlikely
to be underestimated, as compared with the formal standard errors, by
more than about 10--20~per cent. We will demonstrate that the present
results provide further evidence for the reliability of the quoted
astrometric standard errors, and provide independent evidence for the
absence of significant systematic errors in the parallaxes and proper
motions.  Columns~(n--o) of Table~2 provide the Hipparcos parallax and
standard errors (in mas). These values, as well as the remaining
astrometric parameters (and correlations), are as published in the
Hipparcos Catalogue.

The Hipparcos Catalogue also contains detailed photometric data,
including broad-band, high-precision, multi-epoch photometry in the
calibrated Hipparcos-specific photometric system $Hp$. These are
homogeneous magnitudes derived exclusively from the satellite
observations, providing the basis for detailed photometric variability
analyses which are summarised, star-by-star, in the published
catalogue (the $Hp$ magnitudes were not used in the construction of
the HR diagrams in Sect.~9 in view of the absence of appropriate
bolometric corrections). In addition, the catalogue includes $V$
magnitudes, and $B-V$ and $V-I$ colour indices derived on the basis of
satellite and/or ground-based observations.

\titleb{Binary information and radial velocity data}

Information on the binary nature of the stars in the Hyades is
important for a variety of reasons: in addition to the astrophysical
relevance, the confidence which can be placed on the kinematic or
dynamical interpretation of the radial velocities and proper motions
(and hence the space motions) is affected by the (known or unknown)
binary nature of the object. We have therefore attempted to compile
the best available radial velocity and binary information for each
object, and their inter-relationship.

Columns~p--r of Table~2 provide the radial velocity, standard error,
and source of radial velocity, respectively. These columns represent
the result of our literature search, and are supplemented by Coravel
radial velocity results specifically acquired in the context of this
study by one of us (JCM, column~r~=~24). The Coravel velocities
include the `standard' zero-point correction of 0.4~km~s$^{-1}$
(Scarfe et al.\ 1990).  The Griffin et al.\ radial velocities given in
Table~2 are the `uncorrected' values given in their paper: for use in
our subsequent kinematical studies they have been corrected as
described in Eq.~(12) of Gunn et al.\ (1988), but accounting for a
sign error which is present in their equation (and confirmed by the
authors): in their notation, we have added a correction of
$-q(V)-0.5$~km~s$^{-1}$ for stars fainter than $V=6$~mag, and a
correction of $-0.5$~km~s$^{-1}$ for brighter stars.  In the
assignment of errors from the compilation of Griffin et al.\ (1988) we
adopted the `internal error' quoted in their Table~IV for stars with
3~measurements or less.

Column~s indicates whether the object has been classified as a
spectroscopic binary (SB), or as (possibly) variable in measured
radial velocity and therefore indicative of a possible spectroscopic
binary (RV), according to the given source.  Radial velocities are
systemic ($\gamma$) velocities where available. If the radial velocity
has been noted as variable, or if no $\gamma$ velocity is available, a
`\#' precedes the radial velocity error, indicating that it should be
viewed with caution for dynamical studies of the cluster. Of the
previous candidates in Table~2, 71 are classified as spectroscopic
binaries, 37 of which have a $\gamma$ velocity determined; these are
mostly from the work of Griffin et al.\ (1988, and references
therein), some are from refs.~10--11 accompanying Table~2, with others
as referenced individually in the key to Table~2.  [To assist
cross-referencing to the results of Griffin et al., column~j of
Table~2 uses the sequential number of the object in Table~IV of
Griffin et al.; this sequential numbering takes account of two
`blocks' containing~6 objects (rather than the usual~5) in their
Table~IV].

Columns~t--u provide information on the (possible) binary nature of
the star taken from the Hipparcos Catalogue. Each entry in the
Hipparcos Catalogue includes duplicity/multiplicity information
derived from the observations: very broadly, systems are resolved if
their separations are larger than approximately 0.1~arcsec and their
magnitude differences smaller than about 3~mag. For such systems the
catalogue provides detailed information on the components in Part~C of
the Double and Multiple Systems Annex. Additional information, such as
suspected duplicity inferred from the astrometric residuals, or
observed photocentric acceleration implying the presence of
short-period orbital systems, has been derived from the observations,
and relevant catalogue entries are flagged accordingly and assigned to
distinct parts of the Double and Multiple Systems Annex.

Column~t is taken from Field~H56 of the Hipparcos Catalogue and
indicates that a CCDM identifier, denoting entries in the `Catalogue
of Components of Double and Multiple Stars' (Dommanget \& Nys 1994),
has been assigned to the catalogue entry: H~indicates that the system
was determined as double or multiple by Hipparcos (previously
unknown); I~that the system was identified as double in the Hipparcos
Input Catalogue; and M~that the system had been previously identified
as double, but not recorded as such in the Hipparcos Input
Catalogue. Column~u is taken from Field~H59 of Hipparcos Catalogue,
and indicates which part of the Double and Multiple Systems Annex the
entry has been assigned to: `C' indicates components are resolved, `G'
indicates that a non-linear motion of the photocentre has been
detected, `O' that the entry is classified as an orbital system, `V'
that the entry is inferred to be double from a correlation between
photocentric motion and photometric variability, and `X' that the
entry is likely to be an unclassified (close) double or multiple
system. `S' in this column is taken from Field~H61 of the Hipparcos
Catalogue and indicates, somewhat independently, that the entry is a
suspected binary.  Information in column~t may overlap with that given
in column~u, i.e.\ it may simply indicate that detailed information on
components is given in Part~C of the Double and Multiple Systems
Annex, but it may also indicate that the star is a component of a
visual or wide binary (and not included in Part~C).

In summary, `C'~in column~u indicates that components of a double or
multiple system have been resolved by the Hipparcos observations,
while `G', `O', `V', `X', or `S' indicates that the star is, or may
be, a close binary system. The implications for the measured radial
velocities will be taken into account in the discussions of membership
and dynamics of the cluster.

\begfigwid 5.5cm
\vskip-5cm
\centerline{
\hskip -1.0cm       \psfig{figure=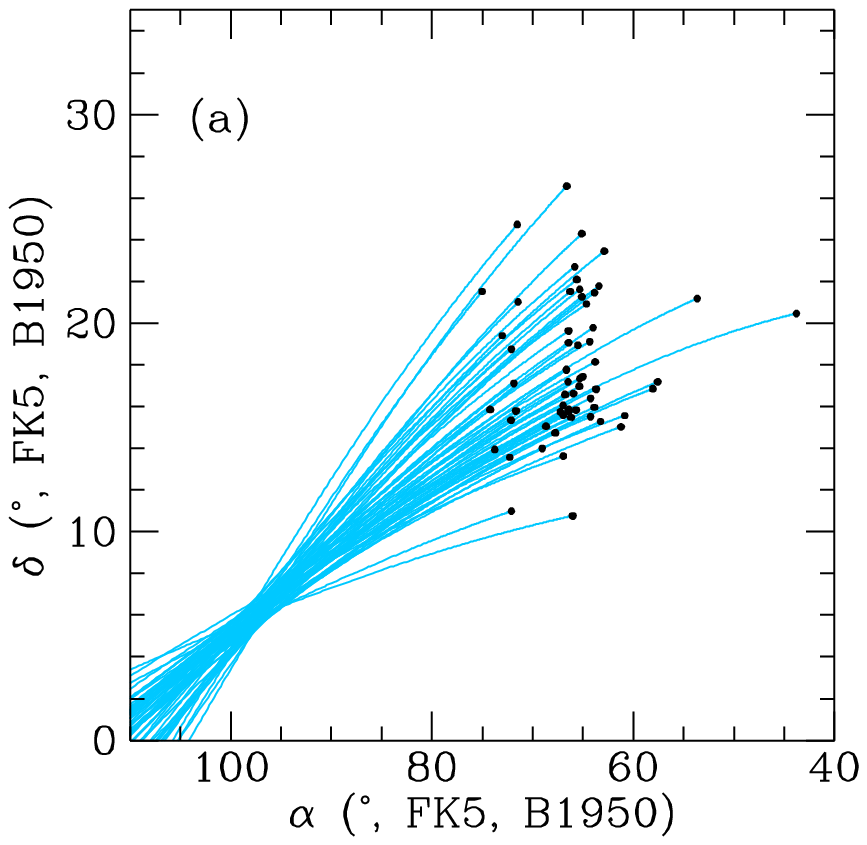,width=6.5cm}
\hskip -1.0cm\hfil  \psfig{figure=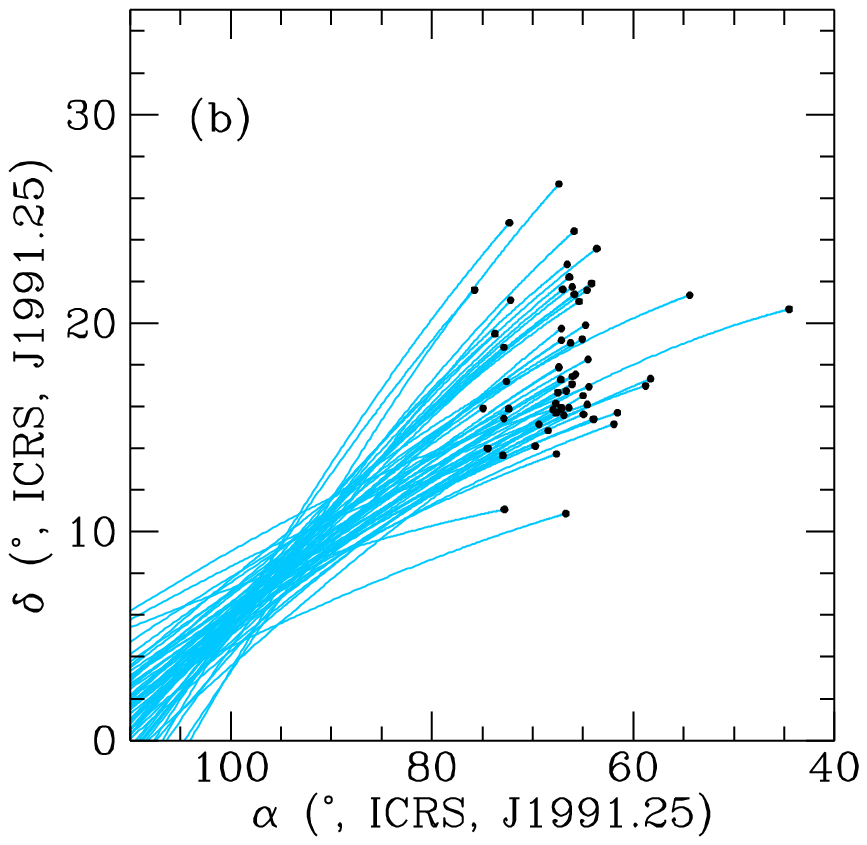,width=6.5cm}
\hskip -1.0cm\hfil  \psfig{figure=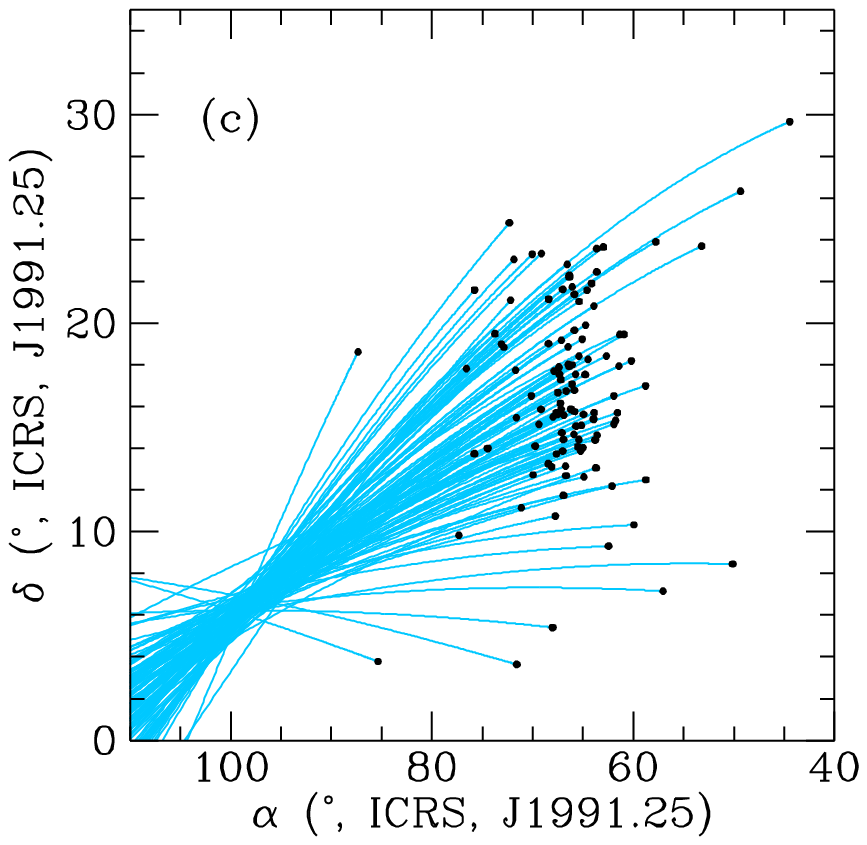,width=6.5cm}
\hskip -1.0cm}
\vskip -10pt
\figure{3} {(a) The set of stars selected by Schwan (1991) for his
convergent point analysis, and according to his adopted membership
criteria, showing the positions (solid circles) and motions of the 
selected stars on their great circles (lines) -- including the
region of the resulting convergent point (intersection of the 
lines) -- based on Schwan's data; (b) shows the Hipparcos Catalogue 
data for precisely the same selection of stars (note the difference 
in reference systems and epochs); (c)
shows the stars selected from the Hipparcos Catalogue data according
to Jones' (1971) method. Notice the much `tighter' distribution of
stellar motions compared with (b).
}
\endfig

\titlea{Proper motions and the convergent point} 

Before developing the analysis of the 6-dimensional (position and
velocity) data set provided by the Hipparcos proper motions and
parallaxes in combination with the ground-based radial velocities, it
is instructive to refer to the most recent determinations of the
convergent point based upon the best-available ground-based data, and
to examine the sensitivity of the resulting analysis to the accuracy
of the available proper motion data. We have not investigated all
numerical implementations of the convergent point method using the
Hipparcos data -- the objective in this section is merely to gain
insight into the performance and consistency of the classical
convergent point methods.

Figs~3(a) and (b) assemble the set of stars selected by Schwan (1991)
for his convergent point analysis, and show the motions of the
selected stars -- including the region of the resulting convergent
point -- based on Schwan's data (a), and for the Hipparcos Catalogue
data for the {\it same\/} selection of stars (b). Note that Schwan's
data are referred to B1950(FK5).

Inspection of the less well-defined convergent point apparent in
Fig.~3(b) compared with that of Fig.~3(a) could lead to the erroneous
conclusion that the Hipparcos proper motions are of a degraded
accuracy compared with those used by Schwan (1991). The correct
explanation is, rather, that for any given set of proper motions and
associated errors, the convergent point analysis selects as candidate
cluster members those having a minimum dispersion about the selected
convergent point.  Evidently, for a given set of candidate members
which have been selected according to a given, but erroneous,
distribution of proper motions, an improvement in the corresponding
proper motion accuracies will not necessarily result in a `tightening'
of the previously-determined convergent point for the same selection
of candidate members. That the Hipparcos Catalogue data result in an
increase in the scatter for the same selection of stars is a direct
consequence of retaining a sub-optimum sample of stars on the basis of
their (imprecise) proper motions. A revised analysis, as we will
demonstrate, leads to a different convergent point, and a
correspondingly different selection of stars. The consequences for the
determination of the individual parallaxes of the candidate cluster
members, and the resulting mean cluster distance, then follow directly
from Eqs.~(1) and (2).

\begfig 8cm
\vskip-8cm
\psfig{figure=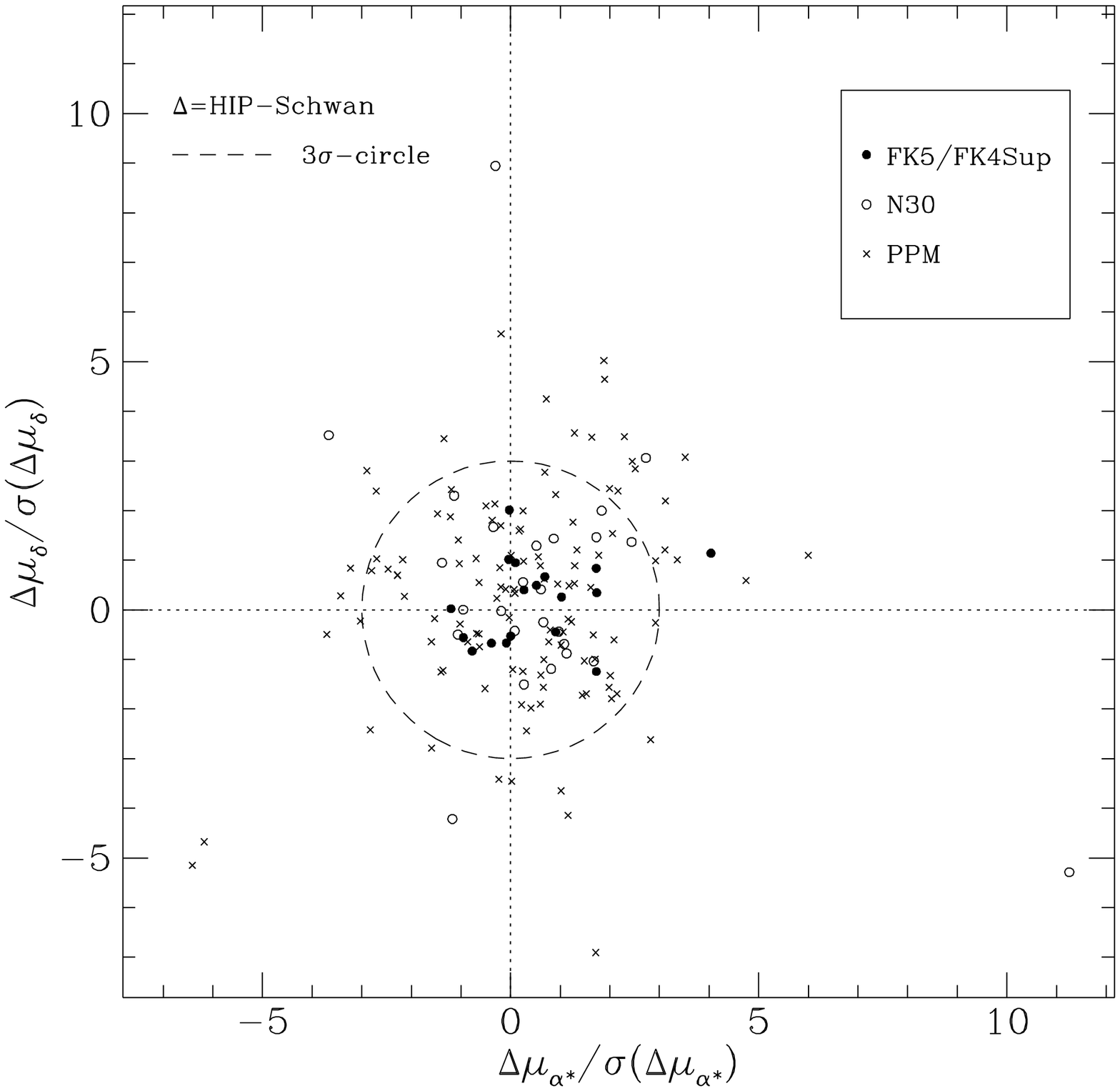,height=8cm}
\vskip -5pt
\figure{4} {The differences in the proper motions components (HIP --
FK5 etc.) for the membership candidates selected by Schwan (1991),
normalised to the combined standard errors in each component. Proper
motion components given by Schwan (1991) in B1950(FK5) coordinates
have been transformed to J2000(FK5) for comparison with the
ICRS(Hipparcos) data. The good agreement between the
FK5 and Hipparcos proper motions, with degraded accuracies from
Schwan's candidates with proper motions only from N30 or PPM, is
evident.
}
\endfig

Our implementation of the convergent point method applied to the
Hipparcos proper motions used the maximum-likelihood technique
described by Jones (1971), which has the merit of locating the
convergent point and, simultaneously, the corresponding cluster
members. In essence, members are searched for amongst the set of stars
showing the clearest converging motions. Applying it to the Hipparcos
proper motions of the previous candidate members listed in Table~2
(those with entries in columns~b--m which, we recall, is our basic
starting point containing a large proportion of possible cluster
members) resulted in about one half of the stars (113) being selected
as cluster members in the first iteration. Two further applications of
the same method to those stars not selected in the first iteration
provided 17 and 18 additional candidate members in the second and
third steps respectively. As shown in Fig.~3(c) the convergence of
stars selected in the first step is now much tighter than seen in
Fig.~3(b), although the successively selected groups have different
convergent points: at $(\alpha,\delta)=(98\ddeg 6,6\ddeg 4), (95\ddeg
1,8\ddeg 3), (96\ddeg 6,5\ddeg 8)$, respectively (ICRS, epoch
J1991.25).

The explanation for the absence of a unique convergent point is that
the methods of Jones and Schwan for selecting members of the cluster
(and convergent point methods in general) rely on finding stars which
show the least deviations of the position angle of their proper
motion, $\theta$, with respect to the direction from the stellar
position to the convergent point, $\theta_{\rm c}$. In reality each
star has its own `convergent point', given by the direction of its
space motion, and the difference $\Delta\theta=\theta -\theta_{\rm c}$
will be small if the stellar convergent point is close to the great
circle connecting the cluster centre to the cluster convergent point.
As a result, the convergent point method tends to select stars with
space motions that lie in the plane defined by the great circle
passing through the cluster centre and the convergent point.

\begfig 8cm  
\vskip-8cm
\psfig{figure=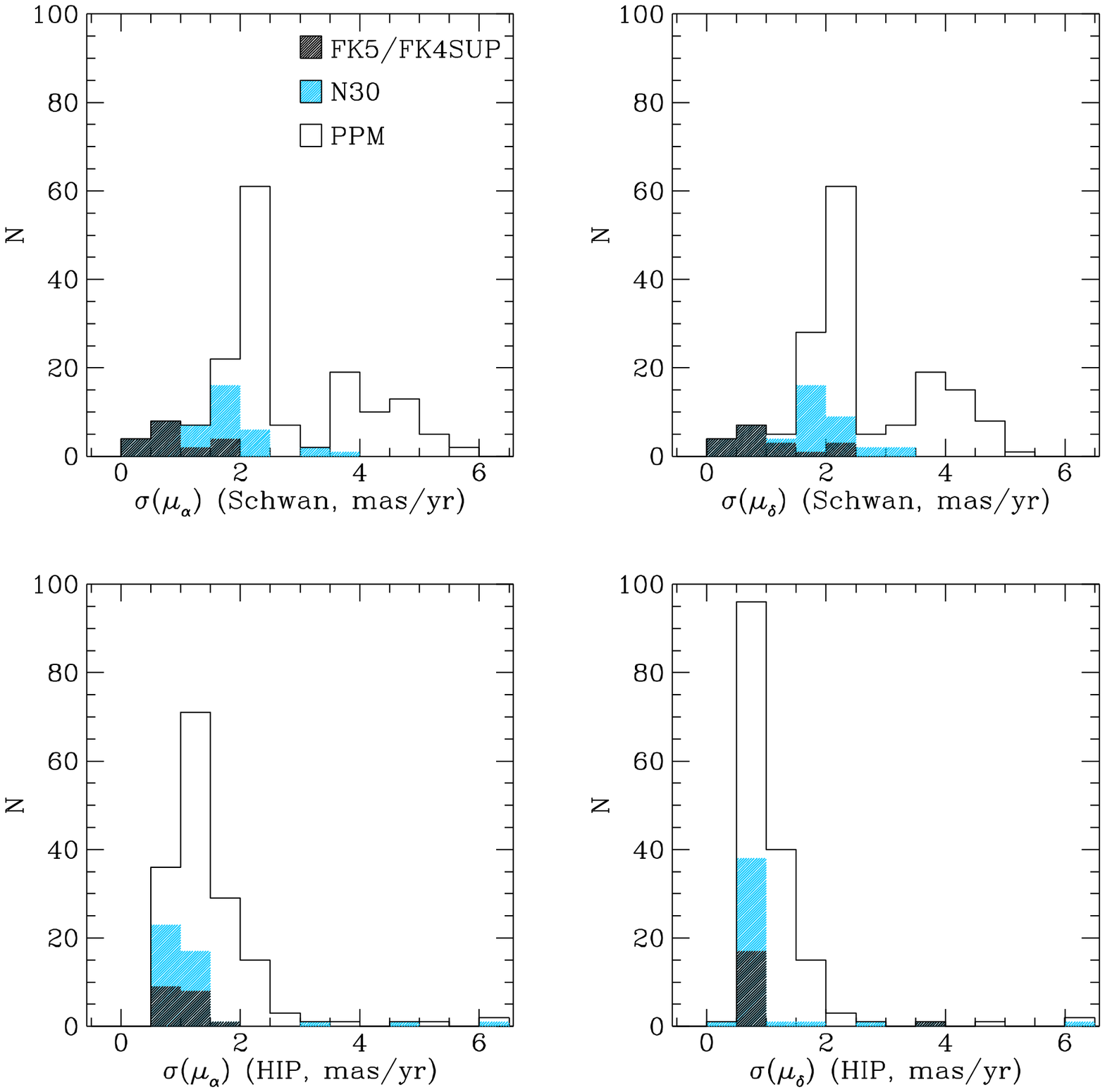,height=8cm}
\vskip -5pt
\figure{5} {The distribution of standard errors in the proper motion
components for the membership candidates selected by Schwan
(1991). Top: as used by Schwan (1991), derived from FK5, N30, and
PPM. Bottom: the distribution of standard errors for the same objects
from the Hipparcos Catalogue, according to their appearance in the
previous source catalogues. The slightly better Hipparcos proper
motion accuracies for the subsets of the FK5 and N30 objects reflects 
their generally brighter apparent magnitudes.
}
\endfig

If there is a significant velocity dispersion in the cluster the
convergent point membership selection will lead to an artificial
flattening of the distribution of candidate members in velocity space,
which may in turn lead, for example, to spurious inferences of
rotation. Conversely if there is significant systematic structure in
the internal velocities of a cluster, the convergent point method may
lead to a spatial bias in the selection of candidates. This would
happen if the cluster possessed a significant component of rotation
with the extreme internal velocities located primarily in a plane
perpendicular to the great circle connecting the cluster centre and
convergent point.

Schwan's proper motions were from mixed sources, drawn from the
FK5/FK4Sup, N30, and PPM catalogues. That the differences in
membership selection resulting from Schwan's values and the present
Hipparcos values are arising from the different quality of the
available proper motions is evident from Fig.~4, which illustrates the
differences in the proper motion components (HIP$-$FK5 etc) for the
membership candidates selected by Schwan (1991), normalised to the
combined standard errors in each component. The corresponding
distributions of the standard errors are shown in Fig.~5. The
generally very good agreement between the FK5 and Hipparcos proper
motions, with degraded accuracies for Schwan's candidates with proper
motions only from N30 or PPM, is evident. It should be borne in mind
that the differences between, for example, FK5 and Hipparcos proper
motions may partially reflect true differences in the measured proper
motions of astrometric binaries where the FK5 proper motions reflect
the long-term photocentric motion, with the Hipparcos measurements
made over a period of only 3.5~years carrying information on orbital
perturbations over these time scales. Indeed, there is some evidence
that the proper motion differences between Hipparcos and FK5 exceed
their combined standard errors, at least in a statistical sense
(Wielen 1997).

We have identified that heterogeneous ground-based proper motions will
affect the determination of the convergent point, the membership
determination, and hence the cluster distance modulus.  Although,
alone, a systematic error of approximately 1~mas~yr$^{-1}$ in the
proper motion system would be needed (Eq.~2) to account for distance
errors of 1~per cent, the $\sin\lambda$ term in Eq.~(2) results in a
greater sensitivity of the distance estimate to a combination of the
proper motions and adopted convergent point.  We will return to a
discussion of the convergent point based on the Hipparcos data, and
the consistency between the distances inferred from the convergent
point (derived from the ground-based and Hipparcos proper motions) and
the Hipparcos-based trigonometric parallaxes, in Sect.~6.1.

\titlea{Membership determination and mean cluster distance}

\titleb{Determination of positions and space motions}

The discussions of Sect.~4 also illustrate the point acutely evident
to previous workers that membership selection based on proper motion
data alone, however accurate, may also lead to erroneous inferences
about cluster membership if significant departures from strictly
parallel motion exist within the cluster.

With the availability of the full 6-dimensional position-velocity
data, based on the 5~astrometric parameters provided by Hipparcos
supplemented by the stellar radial velocity when available, we are in
a position to examine membership based on stricter spatial and
kinematic criteria than has been possible hitherto.

We first assemble the equations used for the transformation between
equatorial and Galactic coordinates, and for the determination of
space velocities based on the observed proper motions, parallaxes and
radial velocities. The transformation between the equatorial and
Galactic systems is given by:
$$
[{\bf x}_{\rm G}~{\bf y}_{\rm G}~{\bf z}_{\rm G}] = 
[{\bf x~y~z}] \, {\bf A}_{\rm G} \eqno{(3)}
$$
where $[{\bf x~y~z}]$ and $[{\bf x}_{\rm G}~{\bf y}_{\rm G}~{\bf
z}_{\rm G}]$ are the basis vectors in the equatorial and Galactic
systems respectively ($\bf x$ is the unit vector towards $(\alpha,
\delta)=(0,0)$, $\bf y$ is the unit vector towards $(+90^\circ,0)$,
and $\bf z$ the unit vector towards $\delta=+90^\circ$), and where the
matrix ${\bf A}_{\rm G}$ relates to the definition of the Galactic
pole and centre in the ICRS system.  Currently no definition of this
relation has been sanctioned by the IAU, and we adopt the following
definition proposed by the Hipparcos project (ESA 1997, Volume~1,
Section~1.5.3), using as celestial coordinates of the north Galactic
pole in the ICRS system:
$$
\eqalign{\alpha_{\rm G} &= 192.\!\!^\circ859\,48  \cr
         \delta_{\rm G} &= +27.\!\!^\circ128\,25  \cr} \eqno{(4)}
$$
with the origin of Galactic longitude defined by the Galactic
longitude of the ascending node of the Galactic plane on the equator
of ICRS, taken to be:
$$
\l_\Omega = 32.\!\!^\circ931\,92  \eqno{(5)}
$$
Eqs~(4) and (5) preserve consistency with the previous B1950
definition of Galactic coordinates (Blaauw et al.\ 1960) to a level
set by the quality of optical reference frames prior to Hipparcos,
accounting for both the transformation to the J2000(FK5) system (cf.\
Eq.~(33) of Murray 1989) and then to the ICRS(Hipparcos) system by
application of the orientation difference between the Hipparcos and
FK5 Catalogues.

These values of the angles $\alpha_{\rm G}$, $\delta_{\rm G}$ and
$\l_\Omega$ are to be regarded as exact quantities.  From them, the
transformation matrix ${\bf A}_{\rm G}$ may be computed to any desired
accuracy.  To 8~decimals the result is:
$$
{\bf A}_{\rm G} = \kern-3pt \pmatrix{
-0.054\,875\,56& +0.494\,109\,43& -0.867\,666\,15 \cr 
-0.873\,437\,09& -0.444\,829\,63& -0.198\,076\,37 \cr
-0.483\,835\,02& +0.746\,982\,24& +0.455\,983\,78 \cr
} \eqno{(6)}
$$
If ${\bf b}$ denotes the barycentric position of the star, measured in parsec, 
and ${\bf v}$ its barycentric space velocity, measured in km~s$^{-1}$, then:
$$
{\bf b} = A_p\,{\bf u}/\pi \eqno{(7)}
$$
and:
$$
{\bf v} = ({\bf p}\mu_{\alpha *}A_v/\pi + {\bf q}\mu_\delta A_v/\pi
+ {\bf r}V_R)k \eqno{(8)}
$$
where $\bf u$ is the unit vector in the barycentric direction, $[{\bf
p~q~r}]$ is the normal triad defined below, and $A_p=1000$~mas~pc and
$A_v=4.74047...$~km~yr~s$^{-1}$ designate the astronomical unit
expressed in the appropriate form; $\pi$ is the parallax expressed in
mas, and $\mu_{\alpha *}=\mu_\alpha\cos\delta$ and $\mu_\delta$ are
the proper motion components expressed in mas~yr$^{-1}$.  The Doppler
factor, $k=(1-V_R/c)^{-1}$, is required to account rigorously for
light-time effects in the calculation of the space velocity in terms
of the observed proper motion and radial velocity.

In the equatorial system the components of the normal triad $[{\bf
p}$, ${\bf q}$, ${\bf r}]$ are given by the matrix:
$$
\eqalignno 
  {{\bf R} &= \pmatrix{p_x&q_x&r_x\cr p_y&q_y&r_y\cr p_z&q_z&r_z\cr }\cr
		   &=
\pmatrix{ -\sin\alpha & -\sin\delta\cos\alpha & \cos\delta\cos\alpha \cr
\phantom{-}\cos\alpha & -\sin\delta\sin\alpha & \cos\delta\sin\alpha \cr
 0 & \cos\delta & \sin\delta \cr}    &(9) \cr}
$$
The equatorial components of ${\bf b}$ and ${\bf v}$ may thus be
written:
$$
\pmatrix{b_x\cr b_y\cr b_z\cr} =
{\bf R}\pmatrix{0 \cr 0 \cr A_p/\pi \cr}\eqno{(10)}
$$
and:
$$
\pmatrix{v_x\cr v_y\cr v_z\cr} =
{\bf R}\pmatrix{k\mu_{\alpha *}A_v/\pi \cr k\mu_\delta A_v/\pi \cr kV_R \cr}
\eqno{(11)}
$$
The Galactic components of ${\bf b}$ and ${\bf v}$ are obtained
through pre-multiplication by ${\bf A}_{\rm
G}^{\scriptscriptstyle\prime}$.  In the following we ignore the
Doppler correction factor, and set $k=1$.

\titleb{Preliminary membership selection}

At this point, our notion of cluster `membership' is intentionally
vague, based only on some general preconceptions about the uniformity
of the space velocities in the central region.  As we shall see,
realistic $N$-body simulations predict, or reflect, dynamical
properties such as mass segregation and cluster evaporation as members
diffuse beyond the cluster tidal radius, or are ejected in dynamical
interactions closer to the cluster core. Thus we might expect an
increasing dispersion of the space velocities with increasing distance
from the cluster centre.

Our approach will therefore involve the following steps: (i) assign
preliminary membership based on rather non-rigorous spatial and
kinematical criteria; (ii) estimate a preliminary centre of mass and
centre of mass motion; (iii) examine the displacements and velocity
residuals of each candidate member with respect to these preliminary
reference values; and finally (iv) refine the membership criteria
accordingly, once the preliminary spatial and velocity structure
becomes more evident.  Unlike previous implementations of the moving
cluster method, we need not assume anything about the degree to which
the cluster members participate in uniform parallel motion in space;
rather, we will be able to examine directly the assumptions on which
these methods have been invoked and, in particular, whether there is
evidence for cluster rotation, expansion, or shear.

Our starting list of possible candidate members is given in Table~2,
which includes all objects in the Hipparcos Catalogue which have, in
the past, been considered as (possible) Hyades members or assigned a
reference number in the quoted sources.  Preliminary membership was
assigned to a subset of these previous candidate members based on
approximate limits placed on the parallax, the radial velocity, the
object's position in the proper motion vector point diagram, and
distributions of the Galactic components of ${\bf b}$ and ${\bf
v}$. This resulted in the elimination of the most obvious non-members
and a list of 188 preliminary members. The subsequent stages of the
membership selection are insensitive to this preliminary membership,
so neither the details of this selection process, nor the list of
preliminary members, are provided.

\titleb{Preliminary determination of the centre of mass}

The preliminary members show a projected spatial distribution
extending over 10--20~pc in each coordinate on the sky; thus for a
distance of $\simeq45$~pc, and for median Hipparcos parallax errors of
$\sigma_\pi\simeq1$~mas ($\simeq1.5$~pc at this distance), the depth
of the cluster is clearly resolved by the Hipparcos parallaxes. Our
next step is to determine an approximate centre of mass for the
cluster, not with the ultimate goal of determining a mean cluster
distance, but rather in order to establish a well-defined cluster
reference point and mean space motion to which more careful membership
assignment may be referred.

In order to determine the centre of mass of the cluster, masses of
single stars have been determined using a reference isochrone from the
models of Schaller et al.\ (1992). The point on the isochrone nearest
to the observed values of $M_V$ and $B-V$ was determined for each
star, and a mass assigned corresponding to that point on the
isochrone. The masses of spectroscopic binaries were taken from the
literature or, if no published mass was available, an estimate of the
minimum mass was made based on the luminosity of the primary. For
resolved binaries, masses of the secondary were assigned according to
the mass-luminosity calibration given by Henry \& McCarthy (1993). For
components of binaries with separations larger than about 20~arcsec,
the number density criterion given in Brosche et al.\ (1992) was used
to infer whether they are physically associated with the primary or
not. Optical companions cross-referenced in the Hipparcos Input
Catalogue through their CCDM identifier (see Sect.~3.2), mostly at
separations of more than 100~arcsec, were not considered to be members
of the Hyades, and were not included in the determination of the mass
of the system -- although the proper motions of these components may
be listed in the Hipparcos Input Catalogue as identical to those of
the `primary', it is more probable that most of these are background
objects. For an estimation of the errors on the position of the centre
of mass, and for the dynamical investigations in Sect.~8, associated
standard errors were arbitrarily assigned to be 0.1~M$_\odot$ for
single stars and 0.5~M$_\odot$ for double stars.

The centre of mass was then determined as $\Sigma m_i {\bf b}_i/\Sigma
m_i$.  To avoid outliers in the space positions affecting the centre
of mass, stars located in the central regions of the cluster, defined
(in parsecs and Galactic coordinates) by:
$$\eqalign{
-50\le &b_x \le -30 \cr
-10\le &b_y \le +10 \cr
-25\le &b_z \le -10 \cr}  \eqno{(12)}
$$
were selected for the determination of the centre of mass. Of the 188
preliminary members 142 lie in this central region.  The resulting
(preliminary) centre of mass in Galactic coordinates (in pc) is shown
in the first line of Table~3.

The same 142 stars (of which 141 have a measured radial velocity) were
used to derive the centre of mass motion. The derived velocity
components in Galactic coordinates, and total space motion, are also
given in the first line of Table~3. Assigning the binaries half the
weight of single stars, to account for the larger uncertainties in
their space motions, or using the inverse of the standard errors as
weights, results in coordinates of the centre of mass which differ by
no more than 0.3~pc in each component from the unweighted results, and
in a mean velocity within a few tenths of km~s$^{-1}$. We conclude
that these results are rather insensitive to the weighting scheme
adopted.

\titleb{Final membership selection}

The space velocity derived in the previous subsection can now be used
to refine the membership criteria, and to determine additional
candidate Hyades members (not previously considered as members
according to columns~b--m of Table~2) from the field around the
cluster according to kinematic criteria. As stated in Sect.~2 the
expected intrinsic velocity dispersion around the mean cluster motion
is less than 1~km~s$^{-1}$; a dispersion of around 0.2~km~s$^{-1}$
would be expected for a Plummer potential with a core radius of
4$^\circ$ (Gunn et al.\ 1988 derived a corresponding core radius of
3.15~pc) and a mass of about 400~M$_\odot$. For more realistic, simple
stellar systems, a larger dispersion, of order 0.4~km~s$^{-1}$ for a
half-mass radius of 5~pc (see Sect.~8) may apply (Binney \& Tremaine
1987, Eq.~4--80b). In selecting cluster members, at least this
intrinsic dispersion must be considered. But numerous other effects
may be responsible for a dispersion of the velocities around the mean
cluster motion.  In addition to observational errors, including the
contribution of different zero points for different radial velocity
sources, the presence of (undetected) binaries will inflate the
intrinsic velocity dispersion due to their orbital motion, affecting
both the observed proper motion and radial velocity distribution.

The combined effects are clearly observed in the velocity
distributions of the preliminary members. For the 142 candidate
members located in the central region of the cluster (of which 141
have a measured radial velocity compiled in Table~2), the standard
deviation in the Galactic components of $\bf v$ are $2.3$, $1.9$ and
$2.1$~km s$^{-1}$. The median errors in the three components are
$0.59$, $1.04$, and $0.84$~km s$^{-1}$, suggesting that the internal
dispersion is resolved. However, taking only the 51 single stars that
have a radial velocity measured by Griffin et al.\ 1988 (essentially
providing a single source of radial velocities with a carefully
defined origin) the standard deviations are found to be $0.4$, $0.8$,
and $1.0$~km s$^{-1}$ (in this case we have chosen half the
inter-quartile range, which provides a more robust estimator for small
samples of data), with median errors of $0.48$, $1.29$, and $0.92$~km
s$^{-1}$. Thus the internal velocity dispersion is not resolved
significantly in any of the components of~$\bf v$.

Hence we assume at this stage that all cluster members move with the
same velocity vector even though, due to the use of different radial
velocity sources and the presence of binaries, the velocity data do
show considerable dispersion. This has been accounted for by assigning
an uncertainty to the relative velocity of each star with respect to
the mean cluster space motion, which is the sum of the standard errors
and covariances in the centre of mass motion and the observed standard
errors and covariances in the individual data. We stress that this
uncertainty does not reflect the physical internal dispersion of the
velocities in the Hyades, but rather the quality of the data presently
available.

The membership selection then proceeds as follows. For each star we
can calculate, from Eq.~(11), the expected values of the transverse
and radial velocities for a star moving with the common cluster
motion:
$$ 
\pmatrix{V_{\alpha*}\cr V_\delta\cr V_{\rm R}\cr}_0 = 
{\bf R^{\prime}}\pmatrix{v_x\cr v_y\cr v_z\cr}_{\rm C} \eqno{(13)}
$$
where $V_{\alpha*}=\mu_{\alpha*}A_v/\pi$, $V_\delta=\mu_\delta
A_v/\pi$, and $V_{\rm R}$ are the velocity components in equatorial
coordinates, and the subscript~C refers to the centre of mass motion.
Similarly, from the observed values of $\pi, \mu_{\alpha*},
\mu_\delta,$ and $V_{\rm R}$ we can calculate the values 
of the observed transverse and radial motions $(V_{\alpha*}, V_\delta,
V_{\rm R})$.

A comparison between the expected and observed velocities requires an
evaluation of the associated covariance matrices. If the covariance
matrix associated with vector $\bf x$ is $\bf C_x$, then the
covariance matrix of ${\bf y}={\bf F}({\bf x})$ is given by ${\bf
C_y}={\bf JC_xJ}^{\prime}$ where $\bf J$ is the Jacobian matrix
associated with the transformation from $\bf x$ to $\bf y$. The
Jacobian matrices for the transformations from ${\bf v}_{\rm C}$ to
$(V_{\alpha*}, V_\delta, V_{\rm R})_0$, and from $(\pi, \mu_{\alpha*},
\mu_\delta, V_{\rm R})$ to $(V_{\alpha*}, V_\delta, V_{\rm R})$ are: 
$$
{\bf J}_0= {\bf R}^{\prime} \eqno{(14)}
$$
and:
$$
{\bf J}= \pmatrix{
 -\mu_{\alpha*}A_v/\pi^2 & A_v/\pi & 0 & 0 \cr
 -\mu_{\delta}A_v/\pi^2  & 0 & A_v/\pi & 0 \cr
 0 & 0 & 0 & 1 } 	\eqno{(15)}
$$
respectively. We will then consider a star to be a candidate Hyades
member based on these kinematic criteria if the difference between the
expected and observed velocities lies within a certain combined
confidence region of the two calculated vectors. Assuming the two are
statistically independent, the combined confidence region is described
by the sum of the two covariance matrices, $\pmb{$\Sigma$}$, as:
$$
c={\bf z}^{\scriptscriptstyle \prime}\,\pmb{$\Sigma$}^{-1} \,{\bf z} \eqno{(16)}
$$
where $\bf z$ is the difference vector, and $c$ is dimensionless.  In
order to account more easily for objects for which the radial velocity
was unavailable, the membership selection was carried out in
equatorial coordinates, with ${\bf v}_C = (-6.22, 44.97, 5.36)$ km
s$^{-1}$, this value corresponding to the velocity in Galactic
coordinates given in the first line of Table~3. The uncertainty in the
centre of mass motion corresponding to the 142 candidate members
derived in Sect.~5.2 is:
$$
\pmatrix{
	+2.40 & -0.18 & +0.04 \cr
	-0.18 & +2.45 & +0.17 \cr
	+0.04 & +0.17 & +1.26 } \eqno{(17)}
$$
where the diagonal elements are the standard errors in km s$^{-1}$ and
the off-diagonal elements are the associated correlation coefficients.
For stars with unknown radial velocity the same procedure, excluding
the component $V_R$, can be applied. In this case, we use information
restricted to the tangential velocity components, using only the
$2\times3$ sub-matrix of Eq.~(15). The quantity $c$ is distributed
according to a $\chi^2$ distribution with 2 or 3 degrees of freedom,
depending on the dimensions of $\bf z$. An adopted 99.73~per cent
confidence region (corresponding, somewhat arbitrarily, to
$\pm3\sigma$ for a one-dimensional Gaussian) corresponds to a value of
$c=14.16$ for the full 3-dimensional difference vector ($P=0.9973$ for
$\chi^2=14.16$ and $\nu=3$), or to $c=11.83$ for the 2-dimensional
difference vector if the radial velocity is unknown ($P=0.9973$ for
$\chi^2=11.83$ and $\nu=2$).  We note that the uncertainties on the
mean cluster motion are correlated, and return to this point in our
discussion of the velocity distribution of the cluster members.

\begfig 8cm
\vskip-8cm
\psfig{figure=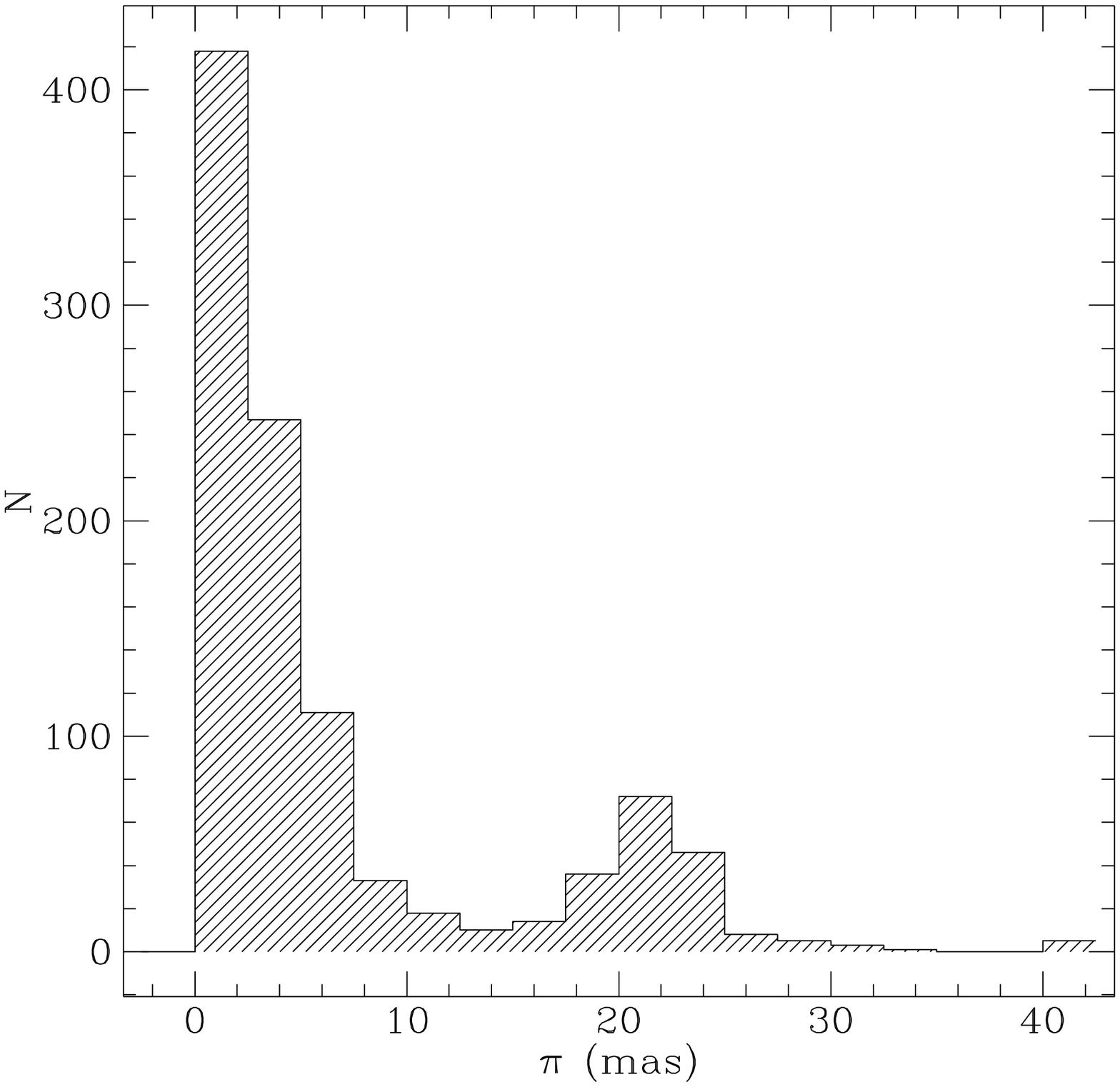,height=7.4cm}
\vskip -5pt
\figure{6} {The distribution of $\pi$ for the 1027 stars described
in the text. A subsequent membership selection criterion of $\pi\ge10$
mas was applied to these candidates.
}
\endfig

Applying the selection procedure to the 5499 stars in the Hipparcos
Catalogue located in the area of sky noted in Sect.~3, and using
$V_{\rm R}$ from Table~2 when available, or from the Hipparcos Input
Catalogue compilation, when available, results in a list of 1027
candidate members. Fig.~6 shows the distribution of $\pi$ for these
stars -- the Hyades cluster corresponds to the peak in the parallax
distribution between 15 and 27~mas. Most of the `new candidates' from
this list of 1027 stars have small parallaxes with relatively large
$\sigma_\pi/\pi$, and small expected proper motions, due to a
combination of their large distances and/or their location close to
the cluster's convergent point. Radial velocity information is largely
absent for these additional objects, which are generally likely to be
unassociated with the Hyades cluster, and identified here as
`possible' candidates (at least we cannot exclude them on the basis of
our data) simply due to the large uncertainties in the observational
material.

The distribution of $\pi$ offers no unambiguous criteria for further
constraining possible membership. In the following we simply exclude
objects with $\pi<10$~mas from further consideration, leaving a list
of 218 candidate Hyades members; adopting a membership threshold of
$\pi\ge8$~mas would result in 246 candidate members. Of the 218, 179
are in the list of previous candidate members from Table~2, with the
39 `new' candidates also listed in Table~2 (`--' in all of
columns~b--m). The remaining 64 out of the 282 objects in Table~2 are
considered now as non-members.

\begfig 8cm
\vskip-8cm
\psfig{figure=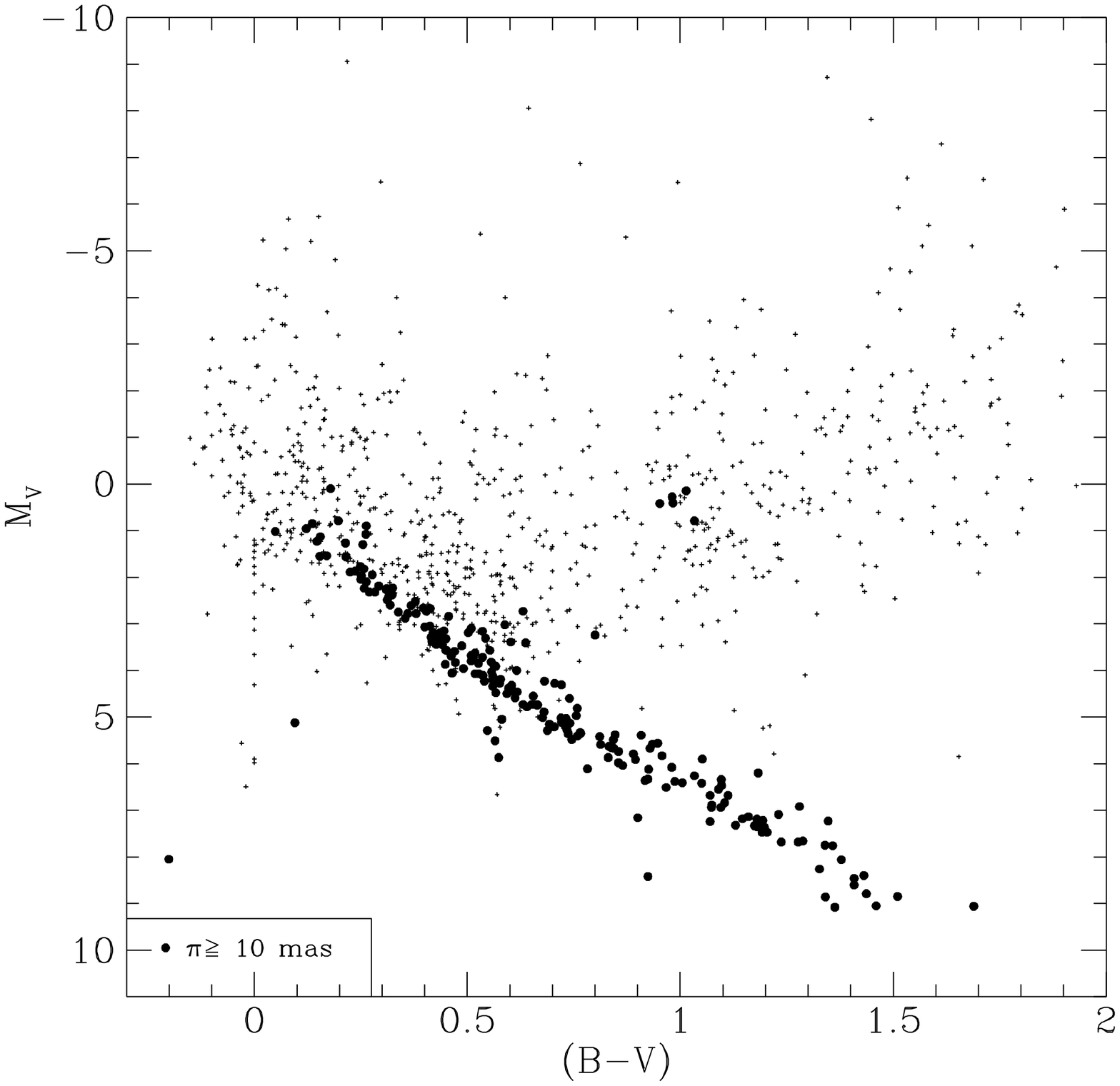,height=7.4cm}
\vskip -5pt
\figure{7} {The HR diagram represented by the 1027 stars selected
according to the kinematical criteria (all points), and those retained
according to the additional criterion $\pi\ge10$ mas ($\bullet$).
}
\endfig

In compiling the list of new candidates in Table~2 we have already
made use of additional radial velocities acquired by one of us (JCM)
using Coravel, which were valuable in excluding some of the stars
located at large distances as members -- while Griffin et al.\ (1988)
estimated a 1~per cent probability that a field star will exhibit a
radial velocity indistinguishable from that of a cluster member, the
combination of radial velocity and proper motion data provides an
almost unambiguous membership criterion for objects participating in
the same overall space motion. These new radial velocities have
already been introduced into Table~2, as indicated in the notes to the
table. New velocities acquired during late 1996 were already used to
suppress intermediate candidates: this is the case for HIP 10920,
11815, 15288, 15406, 19757, 22802, 22809, 23810, 25419, while a
further 23 stars with $\pi<10$ mas could also be excluded as candidate
members based on the new radial velocities. The final status is that
all of the previous 179 candidate members in Table~2 now have a known
radial velocity, while 18 of the 39 `new' candidates in Table~2 have a
known radial velocity.  More definitive membership assignment would
clearly benefit from the acquisition of radial velocities for those
candidates for which the full 3-d space motion is presently unknown.

\begfigwid 11cm
\vskip-11cm
\centerline{
\hskip -1.0cm           \psfig{figure=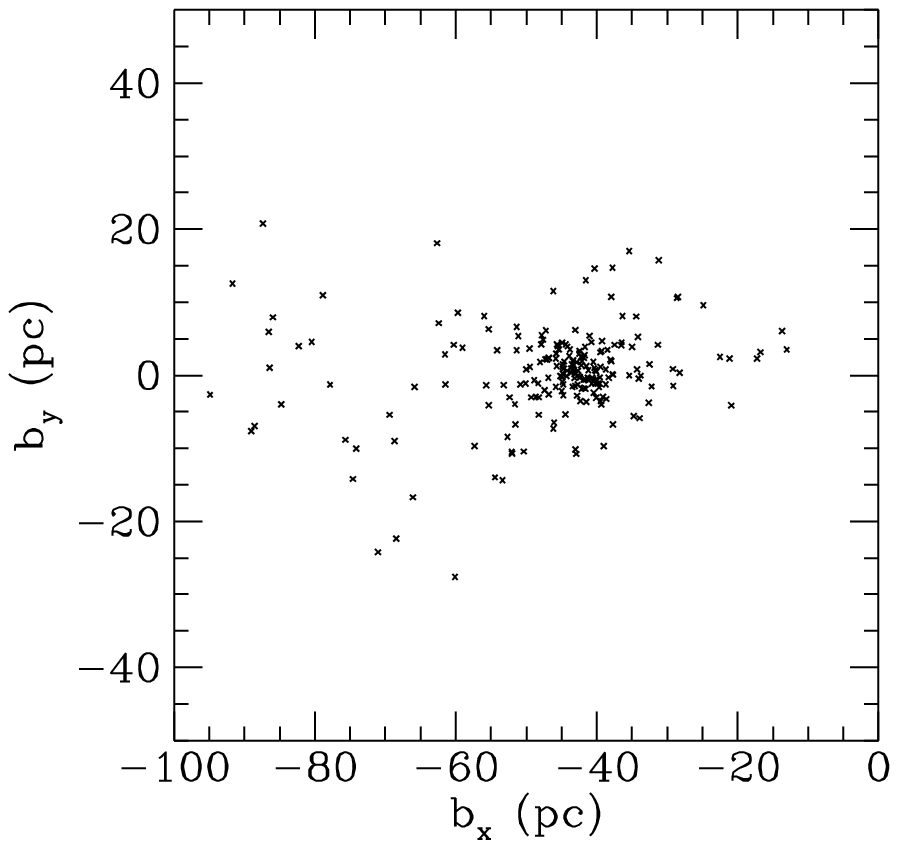,width=6.0cm}
\hskip -1.0cm\hfil      \psfig{figure=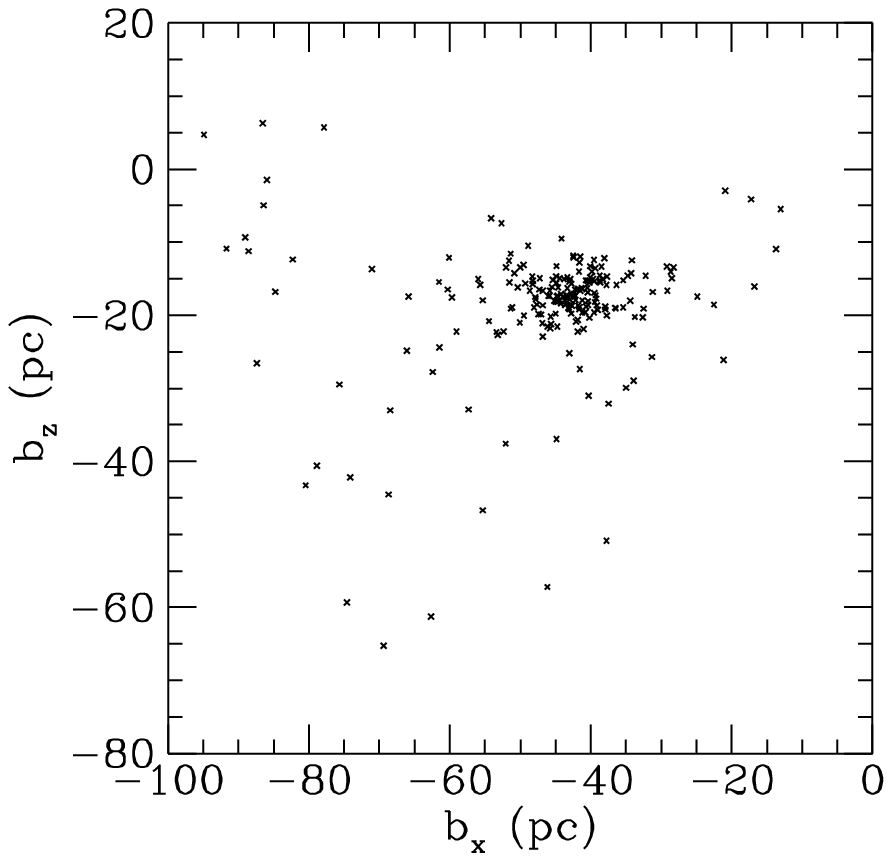,width=6.0cm}
\hskip -1.0cm\hfil      \psfig{figure=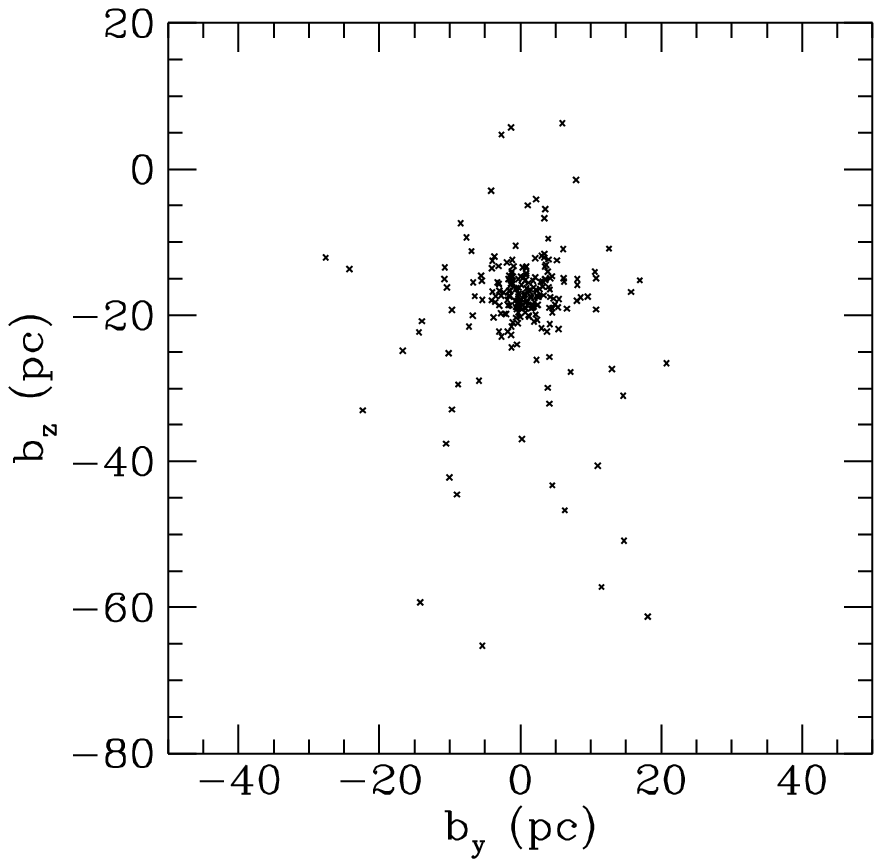,width=6.0cm}
\hskip -1.0cm}
\vskip -0.5cm
\centerline{
\hskip -1.0cm           \psfig{figure=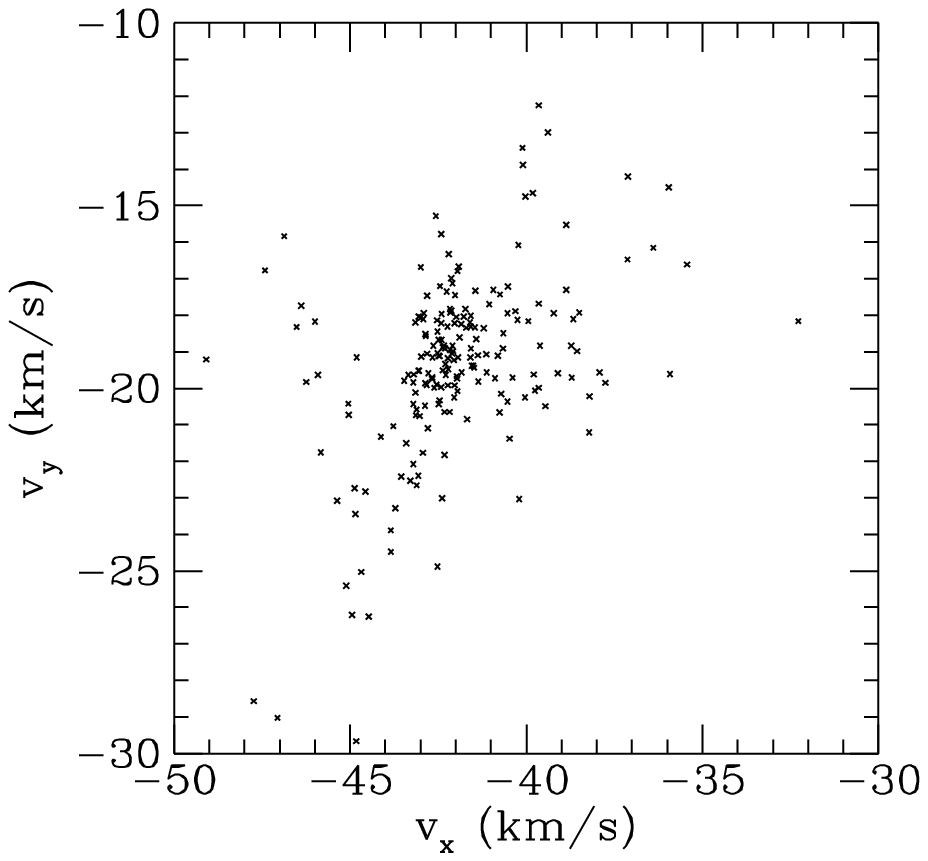,width=6.0cm}
\hskip -1.0cm\hfil      \psfig{figure=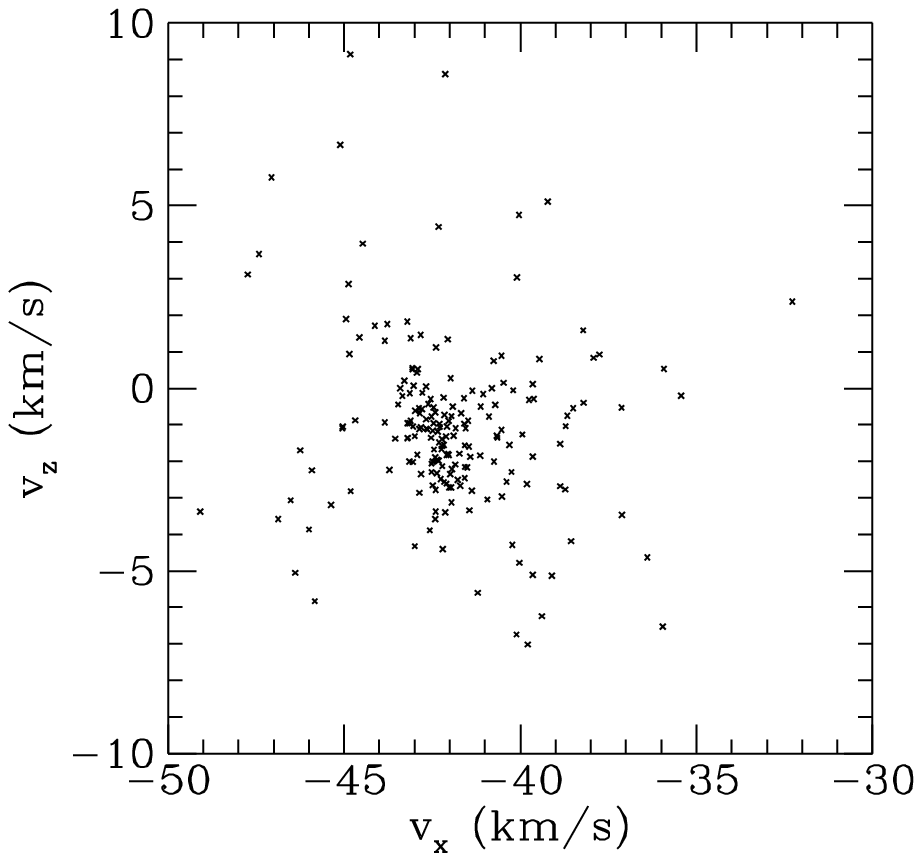,width=6.0cm}
\hskip -1.0cm\hfil      \psfig{figure=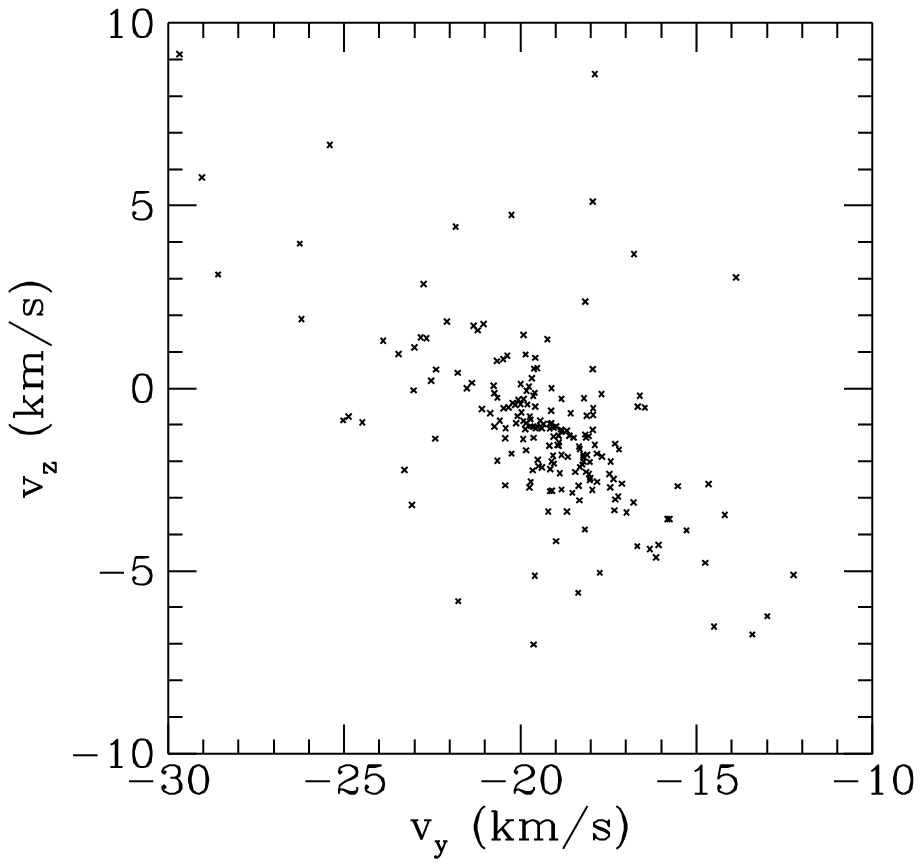,width=6.0cm}
\hskip -1.0cm}
\vskip -10pt
\figure{8} {(a) Projected positions for the 218 candidate members,
in Galactic coordinates (top); (b) projected velocity distributions,
in Galactic coordinates (bottom).
}
\vskip -0.5cm
\endfig

\begfigwid 5.5cm
\vskip-5.5cm
\centerline{
\hskip -1.0cm           \psfig{figure=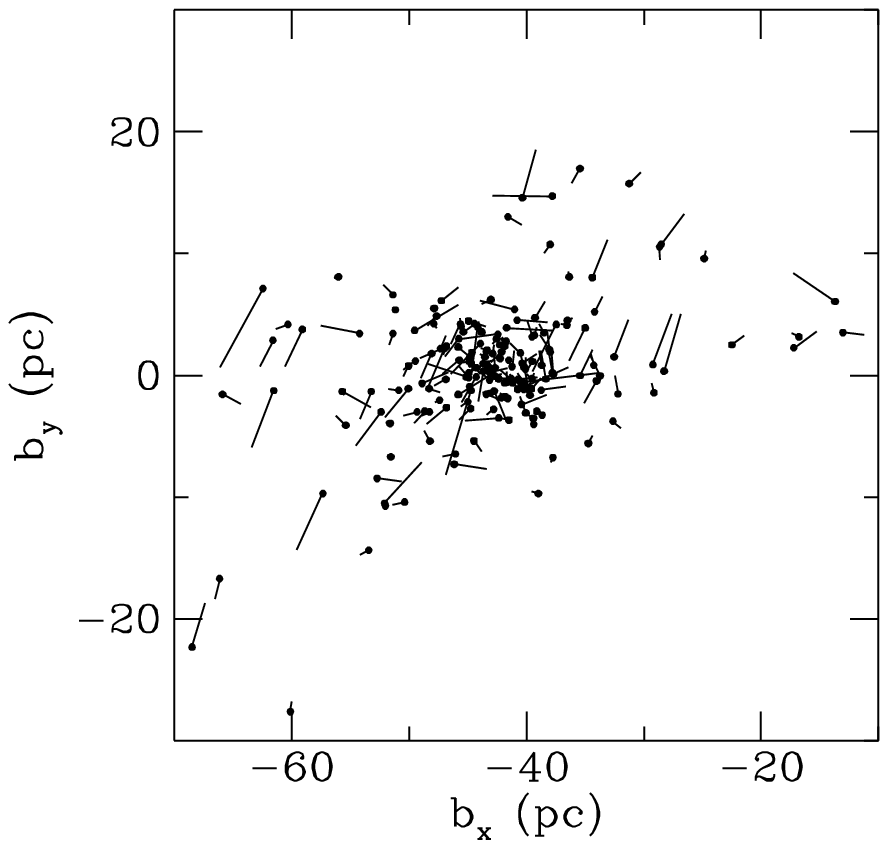,width=6.0cm}
\hskip -1.0cm\hfil      \psfig{figure=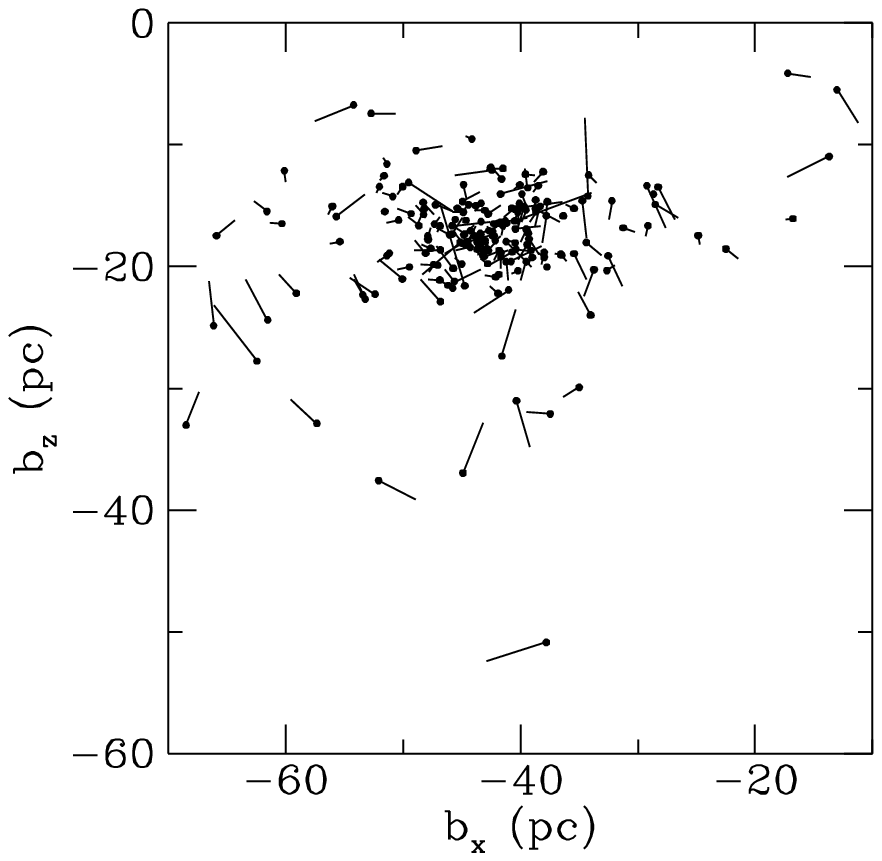,width=6.0cm}
\hskip -1.0cm\hfil      \psfig{figure=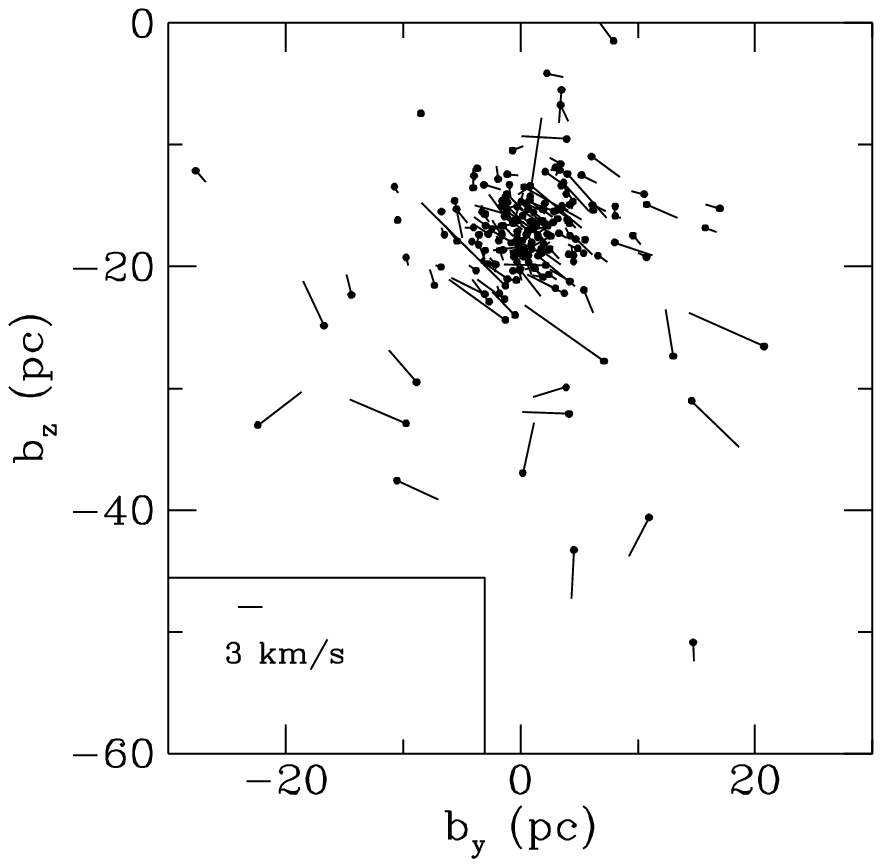,width=6.0cm}
\hskip -1.0cm}
\vskip -10pt
\figure{9} {Projected velocities as a function of position for the
197 candidate members with available radial velocities. The residuals
are given with respect to the velocity of the cluster centre in
Galactic coordinates. HIP~13117 is omitted due to large errors 
in its space velocity.
}
\endfig

The process of recalculating the centre of mass after each membership
selection step, and further refining it once new radial velocities
were acquired and putative members rejected, leads to velocity changes
by only $\sim0.1$~km~s$^{-1}$ or so in each component. The resulting
sensitivity of membership selection to changes in the resulting
systemic space motion is also small: varying all three velocity
components by $\pm0.5$~km~s$^{-1}$ (over a $3\times3\times3$ grid),
the number of selected stars in the inner 10~pc only changes by 3 or 4
at most (out of 134).  Taking steps of 1~km~s$^{-1}$ the number of
stars changes by $\pm5$. So we may be confident that the precise value
of the adopted space motion will not significantly affect the
determination of members.

After membership selection a redetermination of the centre of mass
could be made for the stars within, say, $r<10$~pc (roughly
corresponding to the tidal radius of the cluster) or $r<20$~pc of the
Hyades centre. Starting from the preliminary centre of mass only two
iterations were needed to converge to corresponding determinations of
the centre of mass and mass motion, and these are very robust in terms
of the initial estimate. The results are given in the second and third
lines of Table~3. We adopt a reference distance of $46.34\pm0.27$~pc,
corresponding to a distance modulus $m-M=3.33\pm0.01$~mag, for the
objects contained in the Hipparcos Catalogue within 10~pc of the
cluster centre (roughly corresponding to the tidal radius).

The resulting value of the space velocity in equatorial coordinates
(ICRS) is $(-6.28, +45.19, +5.31)$ km~s$^{-1}$ with a corresponding
convergent point of $(\alpha,\delta)=(97\ddeg 91, 6\ddeg 66)$ for the
inner 10~pc (134~stars), and $(-6.32, +45.24, +5.30)$ km~s$^{-1}$ with
a corresponding convergent point of $(\alpha,\delta)=(97\ddeg 96,
6\ddeg 61)$ for the inner 20~pc (180~stars).  The resulting motion of
the Hyades with respect to the LSR is derived from a solar motion of
16.5 km s$^{-1}$ in the direction $(\ell,b)=(53^\circ,25^\circ)$
(Binney \& Tremaine 1987), and is approximately $(-32.7, -7.3, +5.9)$
km~s$^{-1}$ in Galactic coordinates.

\begtabfullwid
\tabcap{3} {Distance and velocity of the inferred centre of mass
of the Hyades, as described in the text. The first line corresponds to
the preliminary determination based on 142 stars in the central
region.  The last two lines correspond to the `final' determination
for the 134 stars within $r=10$~pc of the cluster centre, and for the
180 stars within $r=20$~pc of the cluster centre, respectively.}
\centerline{
\vbox{
\def\spm{$\scriptstyle \pm$}
{\sevenrm \baselineskip 8pt
\tabskip 6pt plus 5pt minus 3pt
\halign {#\hfil& \quad \hfil#& 
\hfil#\hfil& \hfil#\hfil& \hfil#\hfil& \hfil#\hfil& \hfil#\hfil& 
\hfil#\hfil& \hfil#\hfil& \hfil#\hfil& \hfil#\hfil& \hfil#\hfil \cr
\noalign{\hrule}
\noalign {\smallskip}
Selection&  $\scriptstyle N$& 
$\scriptstyle {\bf b}_{\rm C}$ (pc) \span\omit\span\omit&&
$\scriptstyle {\bf v}_{\rm C}$ (km s$\scriptstyle ^{-1}$)\span\omit\span\omit&& 
   $\scriptstyle D$&   $\scriptstyle V$ \cr
&& $\scriptstyle x$& $\scriptstyle y$& $\scriptstyle z$& \quad&  
   $\scriptstyle u$& $\scriptstyle v$& $\scriptstyle w$& \quad& (pc)& 
 (km s$\scriptstyle ^{-1}$) \cr
\noalign {\smallskip}
\noalign{\hrule}
\noalign {\smallskip}
Preliminary& 
142& --42.23\spm0.24&  +0.15\spm0.06& --17.09\spm0.10&&
      --41.53\spm0.16& --19.07\spm0.11& --1.06\spm0.11&&  45.56\spm0.27& 45.72\spm0.22\cr
\noalign {\smallskip}
$\scriptstyle r<10$~pc&
134&   --43.08\spm0.25& +0.33\spm0.06& --17.09\spm0.11&& 
      --41.70\spm0.16& --19.23\spm0.11& --1.08\spm0.11&&  46.34\spm0.27& 45.93\spm0.23\cr
\noalign {\smallskip}
$\scriptstyle r<20$~pc& 
180&  --43.37\spm0.26& +0.40\spm0.09& --17.46\spm0.13&& 
      --41.73\spm0.14& --19.29\spm0.11& --1.06\spm0.10&&  46.75\spm0.31& 45.98\spm0.20\cr
\noalign {\smallskip}
\noalign {\smallskip}
\noalign{\hrule}
}}}
}
\endtab

Although the results for the $r<10$~pc and $r<20$~pc samples are
reasonably consistent, it should be evident that we are not in a
position to provide an unambiguous value for the `mean distance' of
the Hyades, since the centre of mass is sensitive to the subset of
stars used to calculate it, which in turn depends on the selection of
stars contained in the Hipparcos Catalogue, as well as on the
contribution of faint stars, white dwarfs, and secondary components of
unresolved double systems.

In principle the consistency of the results can be improved by using
the redetermined centre of mass to carry out a new iteration of the
membership selection. This process should ultimately lead to a
consistent set of members and centre of mass.  We used the centre of
mass velocity for the stars in the inner 10~pc and a redetermination
of the matrix in Eq.~(17) from their space velocities, to carry out a
second iteration of the membership selection. The result is that ten
objects listed as (possible) members in Table~2 drop out as
non-members. All these are located beyond 10 pc from the cluster
centre listed in line 3 of Table~3. Hence, a redetermination of the
centre of mass would lead to the same result for the stars in the
inner 10 pc and to convergence of the membership selection
process. This illustrates the robustness of our membership selection
procedure.

The positions of the resulting candidates in the Hertzsprung-Russell
diagram are shown in Fig.~7. Those candidates retained according to
the additional criterion $\pi\ge10$~mas are indicated
separately. Fig.~8(a) shows the positions of the 218 candidate members
in Galactic coordinates. Fig.~8(b) shows the projected velocity
distributions. Fig.~9 shows the projected velocities as a function of
position for the 197 candidate members with available radial
velocity. We return to a discussion of these distributions in Sect.~7.

In the final three columns of Table~2, we provide the distance, $d$
(in pc), of each object with respect to the adopted cluster centre
defined by 134 stars within $r<10$~pc (see Table~3) (column~v); the
statistic ${\bf z'}\pmb{$\Sigma$}^{-1}{\bf z}$, constructed with
respect to the velocity of the preliminary centre of mass (column~w);
and the assignments `0', `1', or `?' (column~x) as our final
membership indicator. A `1' is assigned on the basis of the ${\bf
z'}\pmb{$\Sigma$}^{-1}{\bf z}$ statistic alone (with the limits of
$c=14.16$, or $c=11.83$ if the radial velocity is unknown,
corresponding to the $3\sigma$ confidence interval defined
previously), and independent of distance from the cluster centre; `?'
is assigned in column~w to objects with appropriate values of the
${\bf z'}\pmb{$\Sigma$}^{-1}{\bf z}$ statistic but for which we have
no radial velocity, and are thus unable to rule on the overall stellar
space motion for that object. Future assignments can be made on the
basis of new or improved radial velocities, or $\gamma$ velocities as
new binary orbits are determined. In interpreting Table~2 it should be
noted that objects with column~x~=~0 but without known radial velocity
are unlikely to be members, irrespective of their radial velocity.
Objects with apparently `reasonable' values of $\pi$ and $V_{\rm
rad}$, but with large values of $c$ (and hence column~x~=~0) will have
discrepant proper motions (not evident from the table).

\titlea{Discussion of the Hipparcos parallaxes}

\titleb{Parallaxes determined from the convergent point}

In Sect.~4 we demonstrated that the proper motions used by Schwan
(1991) result in a different convergent point from that derived from
the Hipparcos proper motion data, and that the Hipparcos proper motion
data themselves do not lead to a unique convergent point, but to a
successive selection of objects at each iteration of the convergent
point method occupying a different location in velocity space (a
closer study of the stars listed in Table~2 and our final list of
members reveals that no corresponding spatial bias is introduced).

Before proceeding with a discussion of the resulting space and
velocity distributions of the candidate members, we will now examine
to what extent the distances inferred from the convergent point
analysis using ground-based and Hipparcos proper motions respectively,
are consistent with the Hipparcos trigonometric parallaxes. In
principle, this should give additional confidence in the quality of
the Hipparcos parallaxes and proper motions, and an insight into
whether our explanations for the distance discrepancies resulting from
previous ground-based proper motion investigations are correct.

Fig.~10(a) shows the parallaxes inferred by Schwan (1991) from his
proper motion analysis (and based on his published convergent point)
compared with the Hipparcos trigonometric parallaxes for the same
stars. Fig.~10(b) is again constructed using Schwan's proper motions,
but using the space motion derived in Sect.~5.3 to calculate the
parallaxes from his values of the proper motions. The lower panels
show corresponding results but now based entirely on the Hipparcos
proper motions, using the convergent point determined by Schwan~(c),
and finally the space motion from Sect.~5.3~(d).  The important
feature of these diagrams is that Schwan's inferred parallaxes are
systematically smaller than the Hipparcos trigonometric values, a
trend which is visible in Fig.~10(b), and still not completely
eliminated in Fig.~10(c), i.e.\ when the individual Hipparcos proper
motions are used but taken in combination with Schwan's convergent
point.  In contrast, as seen in Fig.~10(d), there is no systematic
difference between the parallaxes derived from the Hipparcos proper
motions and our present determination of the convergent point on the
one hand, and the Hipparcos trigonometric parallaxes on the other.

\begfig 8cm
\vskip-8cm
\psfig{figure=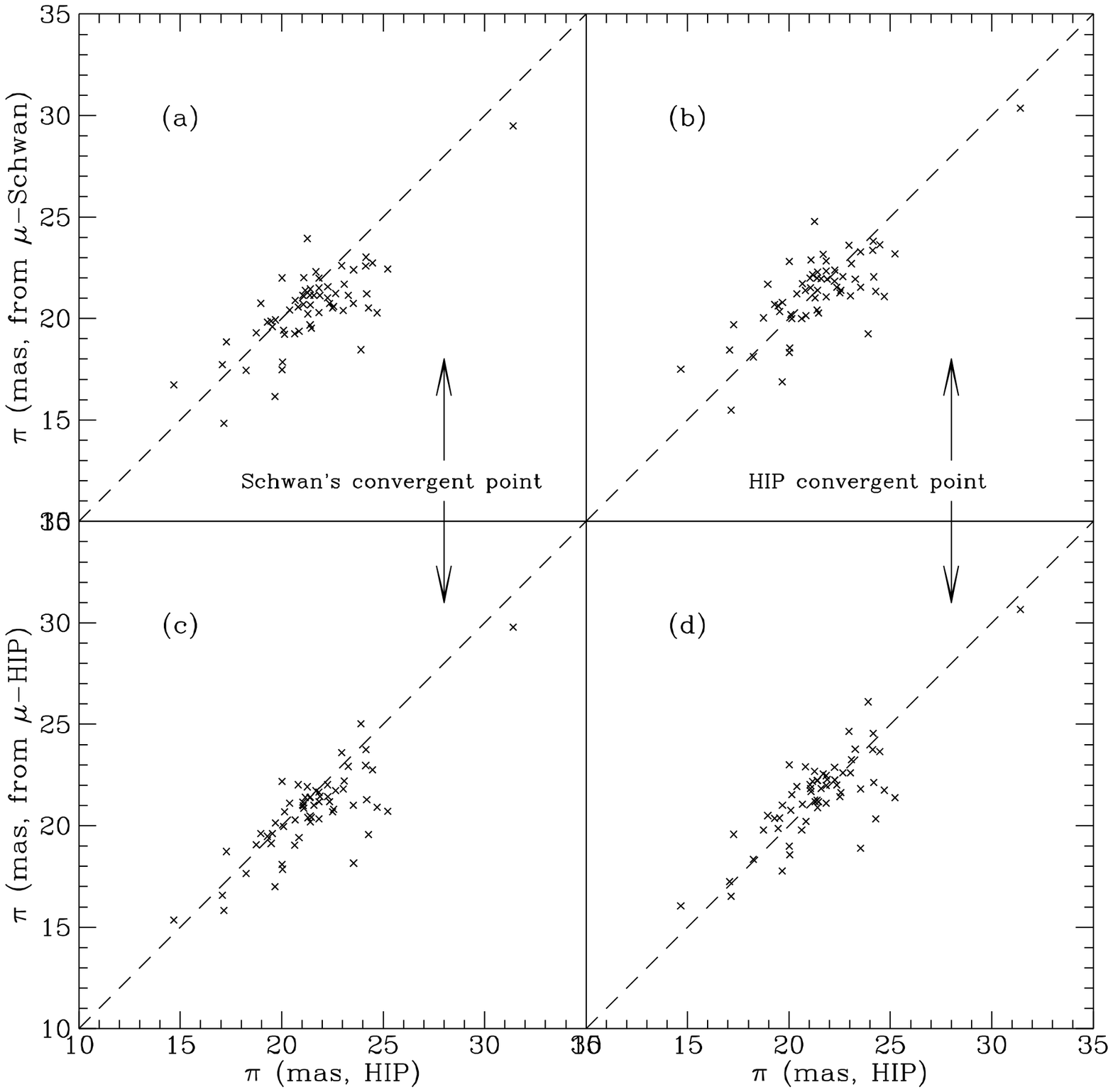,height=8cm}
\vskip -10pt
\figure{10} {This figure shows, for the stars used by Schwan (1991),
the parallaxes inferred from the convergent point, the space motion
and the proper motions, versus the the Hipparcos parallax. The two
left panels show the parallaxes inferred from Schwan's convergent
point and space motion, using Schwan's proper motions (top) and the
Hipparcos proper motions (bottom). The two right panels show the same,
but now the space motion derived in Sect.~5.3 is used to calculate the
parallaxes from the respective proper motions.  }
\endfig

Fig.~11 shows the parallaxes inferred from the Hipparcos proper
motions and the present determination of the space motion of the
cluster centre of mass (ordinate) versus the Hipparcos parallaxes
(abscissa) for all the kinematically selected members discussed in
Sect.~5.4. For the subset of $\pi\ge10$~mas objects, the excellent
correlation between the two quantities indicates that the space motion
and the trigonometric parallaxes are fully consistent. The outliers
include stars with large errors in their astrometry, objects located
close to the convergent point, or objects located at almost the same
declination as the convergent point with $\mu_{\alpha*}\gg\mu_\delta$
(which will lead to a greater probability of erroneously accepting the
object as a member).  >From Fig.~4 the proper motions from Hipparcos
appear systematically larger than those used by Schwan in both right
ascension and declination. Applying a sign-test to the observed
differences shows that this trend is statistically significant at the
95~per cent confidence level.

\begfig 8cm
\vskip-8cm
\psfig{figure=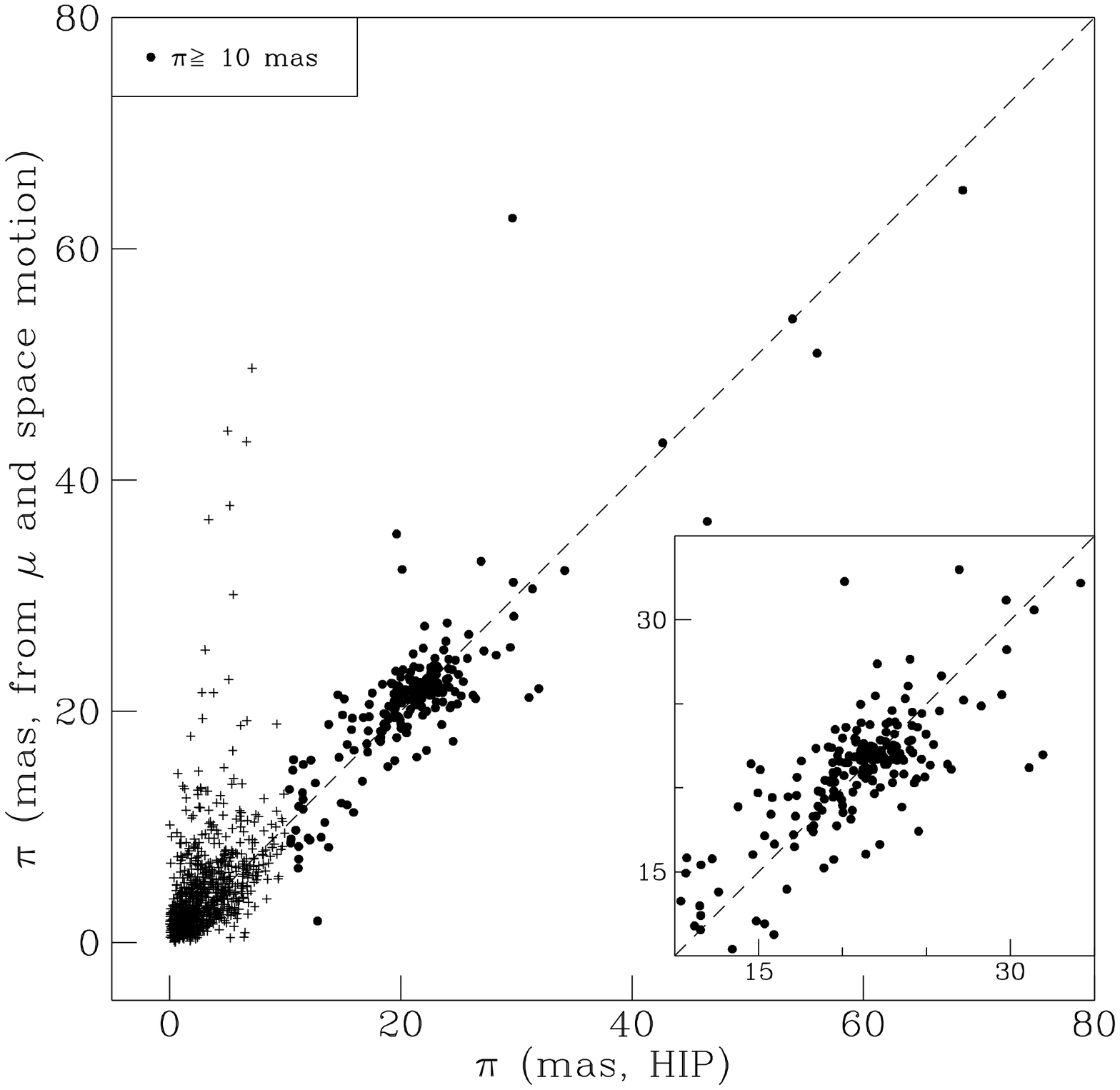,height=8cm}
\vskip -10pt
\figure{11} {The parallax inferred from the Hipparcos proper motions
and the present determination of the space motion of the cluster
centre of mass versus the Hipparcos parallaxes, for all the
kinematically selected members discussed in Sect.~5.4. The outliers
are stars with large errors in their astrometry, objects located close
to the convergent point, or objects located at almost the same
declination as the convergent point with $\mu_{\alpha*}\gg\mu_\delta$
(two other outliers fall off the top of the plot). The inset shows the
relevant cluster region in more detail.  }
\endfig

We can reconcile these results as follows.  The discrepancy between
our reference distance modulus of $m-M=3.33$, and that derived by
Schwan (1991) of $m-M=3.40$, corresponds to 0.07~mag in distance
modulus, or a ratio of $47.9/46.34=1.034$ in distance.  >From Eq.~(2),
$d_{\rm S}/d_{\rm H}= |\mu_{\rm H}|/|\mu_{\rm S}| \times |V_{\rm
S}|/|V_{\rm H}| \times \sin(\lambda_{\rm S})/\sin(\lambda_{\rm H})$,
where H and S refer to parameters from Hipparcos and Schwan (1991)
respectively. Using $|V_{\rm S}|/|V_{\rm H}|=46.60/45.72=1.02$, where
$V_{\rm H}$ corresponds to the velocity of the 142 stars used for the
membership selection (Table~3); and median values of
$\sin(\lambda_{\rm S})/\sin(\lambda_{\rm H})=1.003$ and $|\mu_{\rm
H}|/|\mu_{\rm S}|=1.007$ corresponding to the objects in common
between the two determinations (and after transforming Schwan's data
from B1950 to J2000) also leads to a combined difference of 3~per
cent.

While the median differences between the two sets of proper motions
are below 1~per cent, the derived cluster distance is also sensitive
to the cluster's space velocity -- Schwan's larger radial velocity at
the cluster centre (39.1~km~s$^{-1}$ compared to 38.6~km~s$^{-1}$),
and larger angular distance from cluster centre to the convergent
point, both lead to a larger space velocity, and to larger values of
$\sin(\lambda)$ for all stars.

\titleb{Comparison with previous parallax determinations} 

In demonstrating that the Hipparcos parallaxes and proper motions
together provide a consistent picture of the Hyades structure, space
velocity, and dynamics, our results provide independent evidence (in
addition to that provided by the catalogue construction) that the
trigonometric parallaxes and their standard errors may be taken at
face value. Since recent ground-based determinations of trigonometric
parallaxes for candidate Hyades members have reached formal standard
errors of a few milliarcsec, a comparison between these and the
Hipparcos Catalogue values should therefore permit further insight
into discrepancies between parallaxes (and resulting cluster distance
modulus) determined by different ground-based observatories. In this
section we undertake a first examination of these differences. It is
important to recall that the Hipparcos parallaxes are to be considered
as absolute, while ground-based determinations require corrections to
convert the measurements to absolute values, taking into account the
parallax distribution of the reference stars.

\begtabfull
\tabcap{4} {Parallaxes for the objects in common between 
the Hipparcos Catalogue and Upgren et al.\ (1990).}
\centerline{
\vbox{
\def\spm{$\scriptstyle \pm$}
{\sevenrm \baselineskip 8pt
\vbox {\tabskip 1em plus 2em minus 0.5em
\halign {\hfil#\hfil& \hfil#\hfil& \hfil#\hfil& \hfil#\hfil \cr
\noalign{\hrule}
\noalign {\smallskip}
HIP&	van Altena&	$\scriptstyle \pi_{\rm abs}$ (HIP)&  
$\scriptstyle \pi_{\rm abs}$ (U 90) \cr
No.& No.& (mas)& (mas)\cr
\noalign {\smallskip}
\noalign{\hrule}
\noalign {\smallskip}
20485& 276&   21.08\spm2.69&  17.4\spm3.8\cr  
20527& 294&   22.57\spm2.78&  18.5\spm4.8\cr  
20563& 310&   19.35\spm1.79&  24.2\spm3.2\cr  
20679& 363&   20.79\spm1.83&  23.4\spm4.4\cr  
20827& 459&   17.29\spm2.23&  22.4\spm5.6\cr  
20850& 472&   21.29\spm1.91&  29.5\spm5.2\cr  
20978& 560&   24.71\spm1.27&  29.0\spm5.8\cr  
21138& 645&   15.11\spm4.75&  22.0\spm5.0\cr 
\noalign {\smallskip}
\noalign{\hrule}
}}}
}}
\endtab

Eight of the 23 Hyades parallax stars observed by Upgren et al.\
(1990), three of the stars from the list of Ianna et al.\ (1990), six
of the 10~Hyades parallax stars observed by Patterson \& Ianna (1991),
the spectroscopic interferometric binary 51~Tauri observed by Gatewood
et al.\ (1992) and Torres et al.\ (1997a), and the spectroscopic
interferometric binaries 70~Tauri and 78~Tau observed by Torres et
al.\ (1997b,c), are also contained in the Hipparcos Catalogue. In
addition, 60~of our candidate members are contained in the Fourth
Edition of the General Catalogue of Trigonometric Parallaxes (van
Altena et al.\ 1995).

Mean magnitudes of ground-based parallax stars are typically about
$V=11$~mag, significantly fainter than the median of the Hipparcos
programme, which leads to relatively large values of $\sigma_{\pi,
{\rm HIP}}\sim2-4$~mas for these comparison objects compared with the
median value, of $\sigma_{\pi, {\rm HIP}}\sim1$~mas, for the Hipparcos
Catalogue as a whole.

Fig.~12 shows the relationship between the Hipparcos parallaxes and
the determinations derived from the Van Vleck observations by Upgren
et al.\ (these are listed in Table~4, which also gives the
correspondence between the HIP number and the van Altena number used
by Upgren et al.).  We have eliminated HIP~20605 (vA~334) from the
comparisons because of its inaccurate Hipparcos parallax, largely as a
result of its faint magnitude ($V=11.7$~mag).  Upgren et al.\ used the
corrections from relative to absolute parallaxes according to the
precepts of van Altena et al.\ (1991). The ground-based and Hipparcos
parallaxes are in reasonable overall agreement, with $\langle\pi_{\rm
HIP}-\pi_{\rm U90}\rangle =-2.5\pm4.6$~mas, to be compared with the
average correction from relative to absolute parallaxes, applied by
Upgren et al., of 2.6~mas.

\begfig 6cm
\vskip-6cm
\psfig{figure=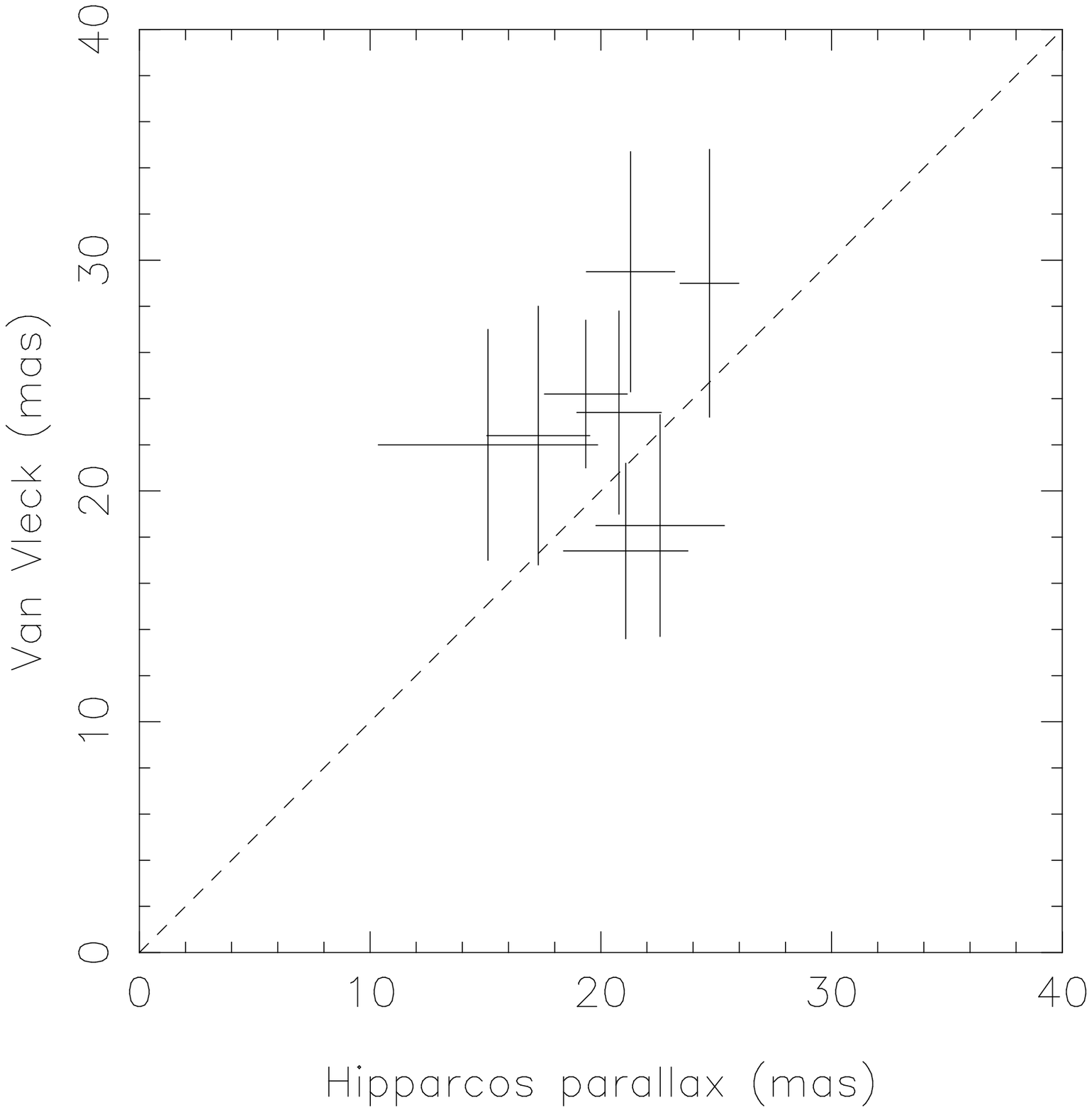,width=8.8cm,angle=270}
\figure{12}{
The differences between the Hipparcos parallaxes, and $\pi_{\rm abs}$ 
determined by Upgren et al.\ (1990) from observations made at 
the Van Vleck Observatory for the eight stars in common between the two 
programmes.} 
\endfig

\begfig 6cm
\vskip-6cm
\psfig{figure=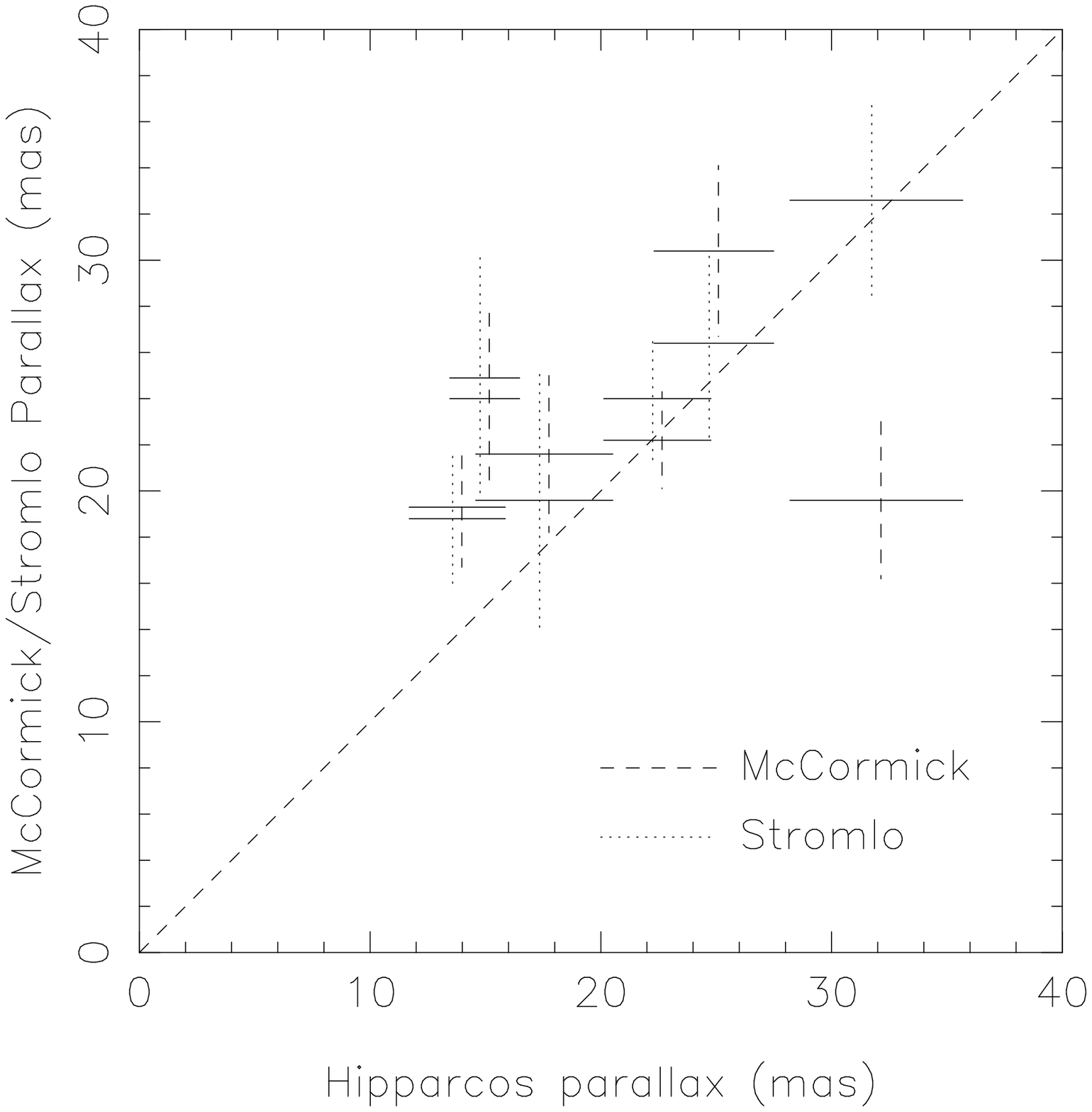,width=8.8cm,angle=270}
\figure{13}{
The differences between the Hipparcos parallaxes, and $\pi_{\rm abs,\
GCTP}$ determined by Patterson \& Ianna (1991) from observations made
at the McCormick and Stromlo Observatories for the six stars in common
between the two programmes. To avoid overlap of the error bars, the
Hipparcos values have been shifted by $+0.2$~mas for the McCormick
comparisons, and by $-0.2$~mas for the Stromlo comparisons.  }
\endfig

Three Hyades objects observed by Ianna et al.\ (1990),
and referred to there by their Johnson number (Johnson et al.\ 1962), 
were also observed by Hipparcos, and these yield 
$\pi_{\rm HIP}=19.17\pm1.93$~mas and $\pi_{\rm I90}=18.9\pm4.7$~mas 
	for HIP~20419 (HY~259);
$\pi_{\rm HIP}=21.10\pm2.22$~mas and $\pi_{\rm I90}=11.4\pm4.8$~mas 
	for HIP~21946 (HY~318); and
$\pi_{\rm HIP}=18.44\pm1.66$~mas and $\pi_{\rm I90}=35.7\pm5.3$~mas 
	for HIP~23498 (HY~351, more commonly called vB187); 
the significant discrepancy for the latter may be partly attributable to the 
acceleration of the photocentric motion observed by Hipparcos.

\begtabfullwid
\tabcap{5} {Parallaxes from the Hipparcos Catalogue and from 
Patterson \& Ianna (1991).}
\centerline{
\vbox{
\def\spm{$\scriptstyle \pm$}
{\sevenrm \baselineskip 8pt
\vbox {\tabskip 1em plus 2em minus 0.5em
\halign {\hfil#\hfil& \hfil#\hfil& \hfil#\hfil& \hfil#\hfil& \hfil#\hfil& 
\hfil#\hfil& \hfil#\hfil& \hfil#\hfil& \hfil#\hfil& \hfil#\hfil& \hfil#\hfil \cr
\noalign{\hrule}
\noalign {\smallskip}
HIP&	van Bueren&   Johnson& \quad&	$\scriptstyle \pi_{\rm abs}$ (HIP)&  
\quad& $\scriptstyle \pi_{\rm abs}$ (PI 91)\span\omit& 
\quad& $\scriptstyle \pi_{\rm abs,\ GCTP}$ (PI 91)\span\omit \cr  
No.&	No.&	No.&&	&&	McCor.&	Stromlo&&	McCor.& Stromlo \cr 
&&&& (mas)&&  (mas)&  (mas)&&  (mas)&  (mas)\cr
\noalign {\smallskip}
\noalign{\hrule}
\noalign {\smallskip}
19316&  --& 233&& 24.90\spm2.59&& 31.6\spm3.7& 27.6\spm4.1&& 30.4\spm3.7& 26.4\spm4.1 \cr
19834&  --& 245&& 31.94\spm3.74&& 20.0\spm3.4& 33.0\spm4.4&& 19.6\spm3.4& 32.6\spm4.4 \cr
20601& 140&  --&& 14.97\spm1.51&& 24.5\spm3.7& 25.4\spm5.2&& 24.0\spm3.7& 24.9\spm5.2 \cr
21179&  --& 288&& 17.55\spm2.97&& 23.2\spm3.4& 21.1\spm5.5&& 21.6\spm3.4& 19.6\spm5.5 \cr
22177&  --& 326&& 22.45\spm2.32&& 22.7\spm2.1& 24.5\spm2.8&& 22.2\spm2.1& 24.0\spm2.8 \cr
23701& 151&  --&& 13.78\spm2.08&& 20.7\spm2.6& 20.2\spm2.8&& 19.3\spm2.6& 18.8\spm2.8 \cr
\noalign {\smallskip}
\noalign{\hrule}
}}}
}}
\endtab

Fig.~13 shows the relationship between the Hipparcos parallaxes and
the determinations derived from the McCormick and Stromlo observations
by Patterson \& Ianna, $\pi_{\rm abs,\ GCTP}$ (these are listed in
Table~5; note that their Tables~1 and 3 refer to van Bueren numbers
for the first four stars, and Johnson numbers for the last six stars).
The comparison uses their reductions from relative to absolute
parallaxes based on the statistical corrections determined from the
General Catalogue of Trigonometric Parallaxes (van Altena et al.\
1991), with an average correction of 2.8~mas. The corrections to
absolute derived by Patterson \& Ianna (1991) on the basis of
spectroscopic parallaxes of field stars were slightly larger and
result in an increasing discrepancy with the Hipparcos values for all
but the McCormick observations of HIP~19834.  On the basis of this
small number of stars, the ground-based trigonometric parallaxes (and
especially those from the Stromlo observations) are seen to be
reasonably consistent with those from Hipparcos, with reliable
estimates of the standard errors, and $\langle\pi_{\rm HIP}-\pi_{\rm
PI91}\rangle =-3.5\pm3.5$~mas for the GCTP-corrected Stromlo values.

Gatewood et al.\ (1992) have used the Multichannel Astrometric
Photometer at the Allegheny Observatory to determine the parallax of
the spectroscopic-interferometric binary 51~Tauri (HIP~20087, vB24) of
$19.4\pm1.1$~mas (employing a correction from relative to absolute
parallaxes of 1.7~mas), compared with $\pi_{\rm
HIP}=18.25\pm0.82$~mas.  A more recent orbital parallax for the same
object has been derived by Torres et al.\ (1997a), $\pi_{\rm
orb}=17.9\pm0.6$~mas, a value which puts these two fundamental
distance determinations of this object in excellent agreement.
Determinations for the binary 70~Tau (HIP~20661, vB57) by Hipparcos,
$\pi_{\rm HIP}=21.47\pm0.97$~mas, and by Torres et al.\ (1997b),
$\pi_{\rm orb}=21.44\pm0.67$~mas, as well as for the binary 78~Tau
(HIP~20894, vB72, $\theta^2$~Tau) by Hipparcos, $\pi_{\rm
HIP}=21.89\pm0.83$~mas, and by Torres et al.\ (1997c), $\pi_{\rm
orb}=21.22\pm0.76$~mas, are also in particularly good agreement, and
may be taken as further evidence for the reliability of these separate
distance determinations, as well as for the robustness of the
Hipparcos parallax determinations for binary systems.

It is noted that the subsequent determinations of the mean cluster
distance by Torres et al.\ (1997a,b,c), based on these three specific
objects, is sensitive to the proper motions adopted for the individual
cluster members. While they derive a resulting distance modulus of
$3.40\pm0.07$ based on 51~Tau, of $3.38\pm0.11$ based on 70~Tau, and
of $3.39\pm0.08$ based on 78~Tau, we can infer that any revised
estimate of the mean cluster distance based on the use of the
inertially referenced Hipparcos proper motions would yield essentially
the same distance modulus as that presented here. For example, using
the Hipparcos parallax for 51~Tau, our convergent point for the inner
20~pc region, and the Hipparcos proper motions, we find a mean cluster
distance of 46.14~pc for the 53 stars used by Torres et al.  (1997a),
a value very close to our centre of mass value. The formula used by
Torres et al.\ to derive distances ($d_i = d_0 (\mu_0/\mu_i)
(\sin(\lambda_i)/\sin(\lambda_0))$, where the subscript zero refers to
the reference object), utilises the PPM proper motion for 51~Tau
(about 5~per cent larger than the Hipparcos value), and the PPM proper
motions of the additional cluster stars, which are almost all smaller
than those of Hipparcos (as discussed in Sect.~6.1). These effects
together cause the systematically larger distances derived by Torres
et al.\ for the cluster.

McClure (1982) derived a dynamical parallax for the Hyades binary
HD~27130 (vB~22, HIP~20019) leading to a distance modulus of
$3.47\pm0.05$~mag, considerably larger than most astrometric
determinations. The Hipparcos parallax, $\pi_{\rm
HIP}=21.40\pm1.24$~mas (distance modulus $3.35\pm0.12$~mag) suggests
that the distance modulus inferred by McClure was overestimated.  The
resulting M-L relationship for binaries (including vB~22) has been
discussed by Torres et al.\ (1997a).

As discussed by van Altena et al.\ (1993, 1994), and in the
Introduction to the Fourth Edition of the General Catalogue of
Trigonometric Stellar Parallaxes (van Altena et al.\ 1995), the
heterogeneous nature of ground-based parallaxes makes any comparisons
between them and the corresponding Hipparcos parallaxes difficult to
interpret in any unified manner. Thus, in establishing the system of
the Fourth Edition of the GCTSP (van Altena et al.\ 1995) three
distinctly different problems were addressed: (1) the correction from
relative parallax to absolute parallax; (2) the relative accuracy of
parallaxes determined at different observatories; and (3) systematic
differences, or zero-point differences between observatories. Whether
the present results, and in particular Figs~12 and 13, suggest that
the corrections from relative to absolute parallaxes, or the relative
parallaxes themselves, have typically been slightly overestimated from
ground-based observations, remains to be understood.

A comparison for the 60 candidate Hyades members contained in the
Fourth Edition of the GCTSP is shown in Fig.~14. Determination of the
distance modulus based on 104 Hyades members from the GCTSP yielded
$m-M=3.32\pm0.06$~mag (van Altena et al.\ 1997b), in good agreement
with our present determination, suggesting that the GCTSP and
Hipparcos systems are indistinguishable to within the limits set by
the ground-based parallax accuracies, at least for the Hyades region.

The are significant differences between the Hipparcos parallaxes (and
proper motion components) and the individual Hubble Space Telescope
Fine Guidance Sensor observations reported by van Altena et al.\
(1997a). For the three brightest stars out of the four in common
between the two sets of observations (HIP~20563/vA310,
HIP~20850/vA472, HIP~21123/vA627) the Hipparcos parallaxes are between
26--42~per cent larger than the corresponding HST
values. Discrepancies between the proper motion components reach
10--20~mas~yr$^{-1}$. We offer no convincing explanation for these
differences but, given the consistency of the Hipparcos measurements
presented elsewhere in this paper, presently favour the Hipparcos
values. Additional investigations will be required in order to
substantiate these claims.

\begfig 7.5cm
\vskip-7.5cm
\psfig{figure=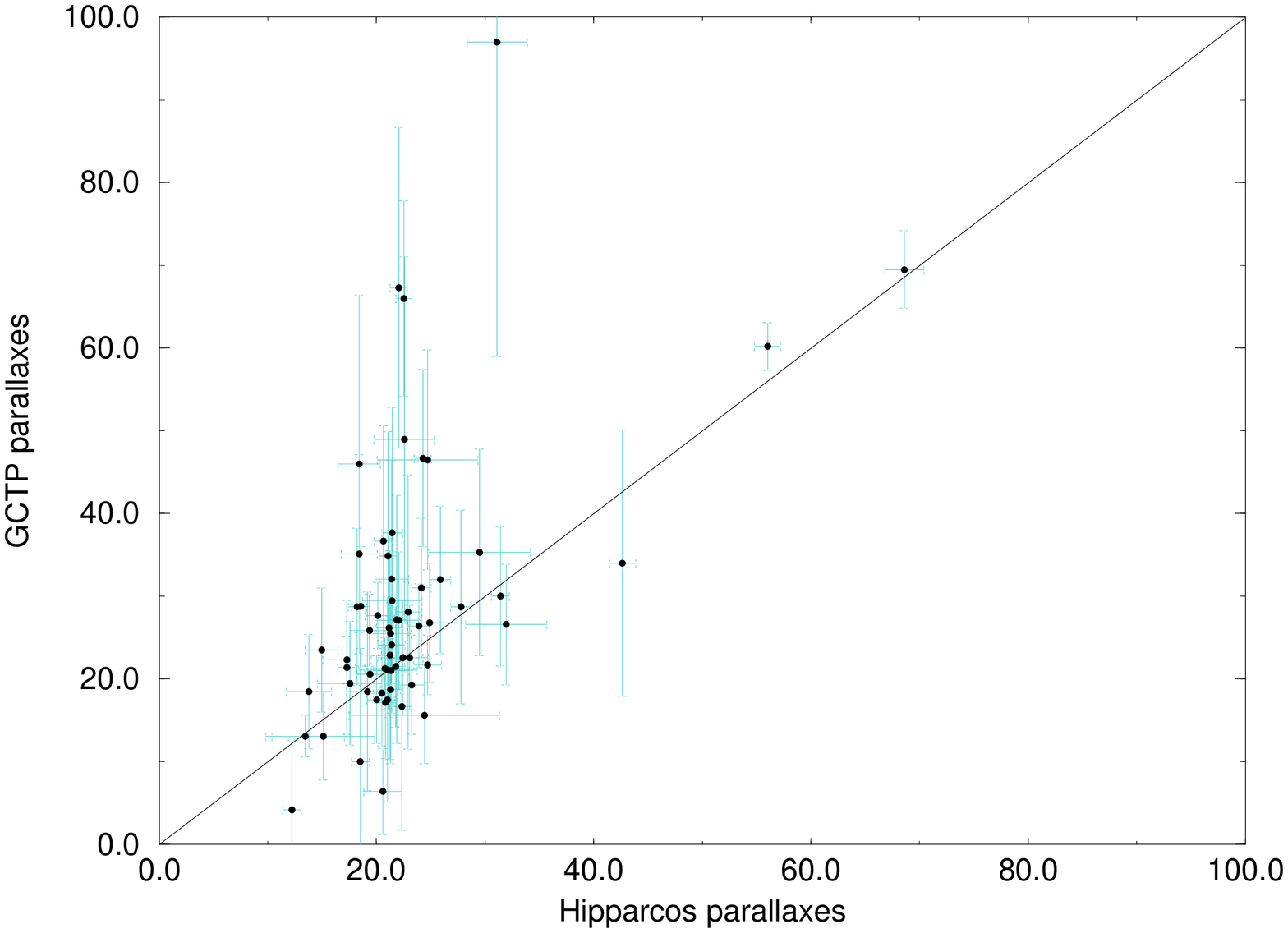,width=8.5cm,angle=270}
\figure{14}{
A comparison between the GCTSP parallax (van Altena et al.\ 1995) and the
Hipparcos parallax (both in mas) for the 60 candidate members in common 
between the two catalogues.}
\endfig

\titleb{Lutz-Kelker corrections, and other effects}

Problems with the use of trigonometric parallaxes have long been
recognised (Eddington 1913, Lutz \& Kelker 1973, Smith \& Eichhorn
1996). One particular difficulty of interpretation arises because the
distance estimate $d=1/\pi$ is biased, with the error distribution of
this estimate obtained by multiplying the probability density function
of $\pi$ by the Jacobian of the transformation from $\pi$ to $d$ (Luri
\& Arenou 1997). As a consequence, with $d=1/\pi$ used to estimate
distances, an {\it a posteriori\/} correction is required -- this
statistical correction is applicable when determined for and applied
to a particular sample population. `Lutz-Kelker' type luminosity
corrections arise correspondingly, and may be understood qualitatively
as arising from the fact that the error volume beyond the distance
corresponding to the measured parallax is larger than the associated
error volume at smaller distances; accounting for this effect makes
all derived luminosities brighter by an amount depending on the
relative parallax error, $\sigma_\pi/\pi$. Applicable corrections are
not the same for a uniform space distribution of stars compared with a
sample drawn from a concentrated population.

The bias on the distance estimates for individual objects in the
Hyades cluster based on the Hipparcos parallaxes can be calculated
analytically (Smith \& Eichhorn 1996), depending on the relative error
in the parallax, which is less than 0.1 for 78 per cent of the Hyades
members (94 per cent have $\sigma_\pi/\pi\le 0.2$). Hence for most of
the members the expectation value of the bias in the distance will be
less than 1 per cent. Since the same bias enters in the tangential
velocities $V_{\alpha*}$ and $V_\delta$, the process of selecting
members based on kinematics should not be affected.  However, for
individual objects the bias in $v_x$ and $v_z$ (derived from
$V_{\alpha*}, V_\delta$ and $V_R$) can be quite large (up to 20 per
cent overestimate for $v_x$ and 8 per cent underestimate for
$v_z$). Thus particular care must be taken when interpreting space
velocities for individual objects with a large relative error on the
parallax.

\break
Smith \& Eichhorn (1996) demonstrated that although one can calculate
analytically the bias in the distance derived from the parallax of an
individual star, the variance in the expectation value for the
distance is infinite, so that a correction of the measured distance is
not possible. However, if the distribution of true parallaxes were
known, the parallax of each individual object could be corrected, in a
statistical sense, {\it a posteriori}. Given the observed parallax and
the associated error, the most likely value of the true parallax, and
hence the true distance, could be estimated.  In principle one could
deconvolve the distribution of measured parallaxes (e.g.\ Lindegren
1995) to derive the error-free distribution of parallaxes. However,
this necessarily assumes that the observed sample of stars is
statistically representative of the underlying parent sample. In
reality, the parallaxes are measured for a sample of stars subject to
specific selection effects.

In the case of Hipparcos the selection effects are complicated, and
the parent sample of stars is known with only limited accuracy.  In
view of the uncertainties in the knowledge of both the parent sample
of stars and the selection effects, and given that for the relevant
values of $\sigma_\pi/\pi<0.1$ individual magnitudes are likely to be
biased at levels at or below about 0.01~mag (cf.\ Smith \& Eichhorn,
Fig.~6) no statistical corrections of the individual parallaxes have
been attempted in the present work. Any resulting effect on the HR
diagram, including the slope of the main sequence (the fainter stars,
having larger errors, will have a larger bias) is expected to be
small, and has not been considered further in this work.

However, an assessment of the possible resulting bias in the {\it
mean\/} Hyades distance and velocity, corresponding to a realistic
space distribution of Hyades member stars, has been investigated by
Monte Carlo simulation based on: (i) a synthetic cluster at the
position of the Hyades, generated with an (albeit simplified) Plummer
density distribution with a core radius of 2.85~pc (see Sect.~8.1);
(ii) the luminosity function for the Hyades derived by Reid (1993);
(iii) a crude approximation to the Hipparcos selection of stars as a
function of magnitude; (iv) errors on the parallax which increase with
the apparent magnitude of the star.

For 1000 realizations of the cluster, 220 stars were generated and
realistic errors reflecting the Hipparcos observations (with a median
increasing with apparent magnitude) were added to the parallaxes. For
each star the corresponding (ICRS) values of $b_x$, $b_y$ and $b_z$
were calculated, and the space velocity components were calculated
from the true proper motions and radial velocities combined with the
parallax. From these quantities the mean values of the components of
$\bf b$ and $\bf v$ were calculated, as well as the distance and
velocity of the cluster.

These results suggest that our mean distance to the cluster and
overall velocity are each overestimated by about $0.09$ per cent.  The
coordinate transformations imply that the components of $\bf b$ are
then overestimated by $0.2, 0.06$ and $0.03$ per cent respectively,
while the velocity component $v_x$ is overestimated by 0.8 per cent,
$v_y$ by $0.06$ per cent and $v_z$ is underestimated by $0.29$ per
cent. These biases are typically small in comparison with the errors
quoted for these quantities in Table~3.

To these problems of working with trigonometric parallaxes, we note
finally one effect resulting from the specific observational
configuration of Hipparcos, which results in a correlation length of
1--2$^\circ$ over which some small residual correlations will probably
exist in the derived astrometric parameters (Lindegren 1989). Such an
effect results in the error in the mean decreasing not as $n^{0.5}$
but probably more as $n^{0.35}$, leading to a small underestimate in
the final error on the estimated distance of the centre of mass.

\titlea{Structure and kinematics of the Hyades cluster} 

\titleb{Spatial distribution}

Fig.~8 shows clear evidence for a centrally concentrated group,
possibly extended in $b_x$ (in the direction of the Galactic centre),
with an evident correlation between the velocities in the Galactic
coordinate directions $y$ and $z$. Fig.~9 indicates velocity residuals
which increase with distance from the cluster centre, and which may be
suggestive of systematic motions. In this section we compare these
results with the expected space and velocity distributions predicted
from $N$-body simulations, and address the question of whether the
cluster members display velocity residuals consistent with a co-moving
system with constant space velocity, or whether there is evidence for
systematic expansion (or contraction), rotation, or shearing motion
due, for example, to the effects of the Galactic tidal field or
passing interstellar clouds, or to the shearing effect of differential
Galactic rotation. These effects, as well as the evaporation of stars
through relaxation by stellar encounters, are considered to represent
primary mechanisms responsible for the disruption of open clusters.

Realistic $N$-body simulations of the dynamical evolution of open
clusters have been made by Terlevich (1987), de la Fuente Marcos
(1995), Kroupa (1995) and others, aiming to reproduce features such as
the observed distribution of cluster ages, mass density, binary
distribution, and mass loss. Comparison with the models of Terlevich
prove to be of particular interest since her $N$-body interactions not
only took into account specific forms for the initial mass function
and mass loss due to stellar evolution, but were also supplemented by
the effects of the Galactic tidal field and transient tidal shocks
produced by passing interstellar clouds. Since it is known that the
age distribution of Galactic open clusters barely extends beyond about
1~Gyr, and that disruptive encounters are likely to be responsible for
cluster break-up, we may hope to investigate whether the space and
velocity distribution of objects beyond the tidal radius provides
evidence for such a disruptive encounter. For example, Terlevich finds
that a cluster loses 90~per cent of its stars in 100~million years
after a collision with a giant molecular cloud, in which the stars
would acquire rather high velocities.

Kroupa (1995) starts with a very large proportion of primordial
binaries, and traces the stellar luminosity function as a result of
mass segregation, evaporation, and changing proportion of binary
systems, assuming models with $\langle M \rangle=0.32$ and
$M_C=128$~M$_\odot$, near to the peak of the mass function of Galactic
clusters. He takes the disintegration time to be the time taken for
the number density to reach 0.1 stars pc$^{-3}$, characteristic of the
Galactic disc in the proximity of the Sun and, finding that these
disintegration times are significantly longer than the lifetimes of
real clusters (Battinelli \& Capuzzo-Dolcetta 1991), suggests that
mechanisms other than internal dynamical evolution must be responsible
for cluster disintegration, such as impacts with giant molecular
clouds. These encounters have been further modelled by Theuns (1992a,
b), and confirm the systematic velocity signatures in the outer
cluster regions, depending on the nature of the encounter, predicted
previously (e.g.\ Binney \& Tremaine 1987, Sect.~7.2).

The equipotential cluster surface becomes open, due to the effects of
the Galactic tidal potential, at distances from the cluster centre
referred to as the tidal radius $r_t$ (King 1962, Wielen 1974). These
openings provide an escape route for stars to evaporate from the
system.  In the case of the Hyades, we find $r_t\simeq10$~pc (see
Sect.~8.3).  The existence of an extended halo formed by stars outside
this tidal radius, but sharing the proper motion characteristics of
the rest of the cluster members was originally noted by Pels et al.\
(1975), and has since been reported for other clusters.

A striking feature of Fig.~8 is that, although the tidal radius is of
the order of 10~pc, about 45 stars are nevertheless found between
10--20~pc, a result consistent with the simulations by Terlevich
(1987) -- these demonstrated that since the openings of the
equipotential surface are on the $x$-axis, stars can spend some
considerable time within the cluster before they find the windows on
the surface to escape through. Such $N$-body simulation models
consistently show a halo formed by 50--80 stars in the region between
1--2 tidal radii -- some of these stars, despite having energies
larger than that corresponding to the Jacobi limit, are still linked
to the cluster after 300--400~Myr.\fonote{ Another mechanism which can
allow a cluster to hold on to stars that are beyond the tidal limit is
binding by an angular-momentum-like non-classical integral, cf.\
H\'enon 1970, Innanen et al.\ 1983.}

These results appear to be weakly dependent on the slope of the IMF:
Terlevich used $\alpha=2.75$, although de la Fuente Marcos (1995) has
argued that for systems with large numbers of members the different
models have very similar behaviour for the escape rate. [In any
detailed interpretation of the spatial distribution of our candidate
members in Fig.~8, it should be noted that certain selection effects
operate in restricting the region of these diagrams which may be
populated, due to the restricted region of $\alpha$, $\delta$ used for
the study (Sect.~3) combined with the projection of the equatorial
coordinates into Galactic coordinates.]

It would seem natural to identify escaping stars, evident both from
numerical simulations and from our observations, with the extended
Hyades stellar group. The existence of such a system -- stars having
the same average motion as the cluster but a very much larger space
and velocity distribution -- was first suggested by Hertzsprung
(1909), and followed up by Str\"omberg (1922, 1923), Eggen (1960, 1982
and references therein) and others. Recent models have been discussed
by Casertano et al.\ (1993). This group is characterised by proper
motions coinciding in direction with that of the Hyades, but
considerably smaller in size. We will examine the relationship between
this group and the central cluster in Sect.~7.2.

A possible flattening of the Hyades cluster has long been debated.
Flattening of the equipotential surfaces, perpendicular to and
directed toward the Galactic plane, was predicted by Wielen (1967) and
in the $N$-body models of Aarseth (1973). Van Bueren (1952) noted that
the cluster appeared to be flattened along the Galactic plane.
Although this was not evident in the observational studies of Pels et
al.\ (1975), the suggestion re-emerged with a fairly clear indication
of flattening in the outer region noted by Oort (1979), and
subsequently by Schwan (1991). The effect is clearly evident in the
$N$-body simulations by Terlevich (1987, Fig.~8).

A principal components analysis of the distribution of space positions
for our Hyades members within 20~pc from the cluster centre shows that
the cluster has a prolate shape. The major axis lies almost along
$b_x$ in Galactic coordinates, making an angle of $\sim16^\circ$ with
the positive $x$-axis. The intermediate axis lies almost along $b_y$,
making an angle of $\sim16^\circ$ with the positive $y$-axis. The
short axis lies along $b_z$. The axis ratios are $1.6:1.2:1$, where
these values have been derived from the standard deviations in
position along the three axes, corrected for the median error in
position along the corresponding axes (similar axis ratios of
$1.5:1.2:1$ are derived by multiplying the quartiles of the
distribution by the factor converting the quartile to the standard
deviation for a normal distribution).  The inner 10~pc region of the
Hyades is more nearly spherical. The fact that the shape of the outer
parts of the cluster is prolate suggests that it is primarily extended
in $b_x$, although it is possibly also slightly compressed in $b_z$,
as shown in Sect.~8.2. This is consistent with the extension being
caused by stars slowly escaping through the Lagrangian points on the
$x$-axis.

\begfig 8.8cm
\vskip-8.8cm
\psfig{figure=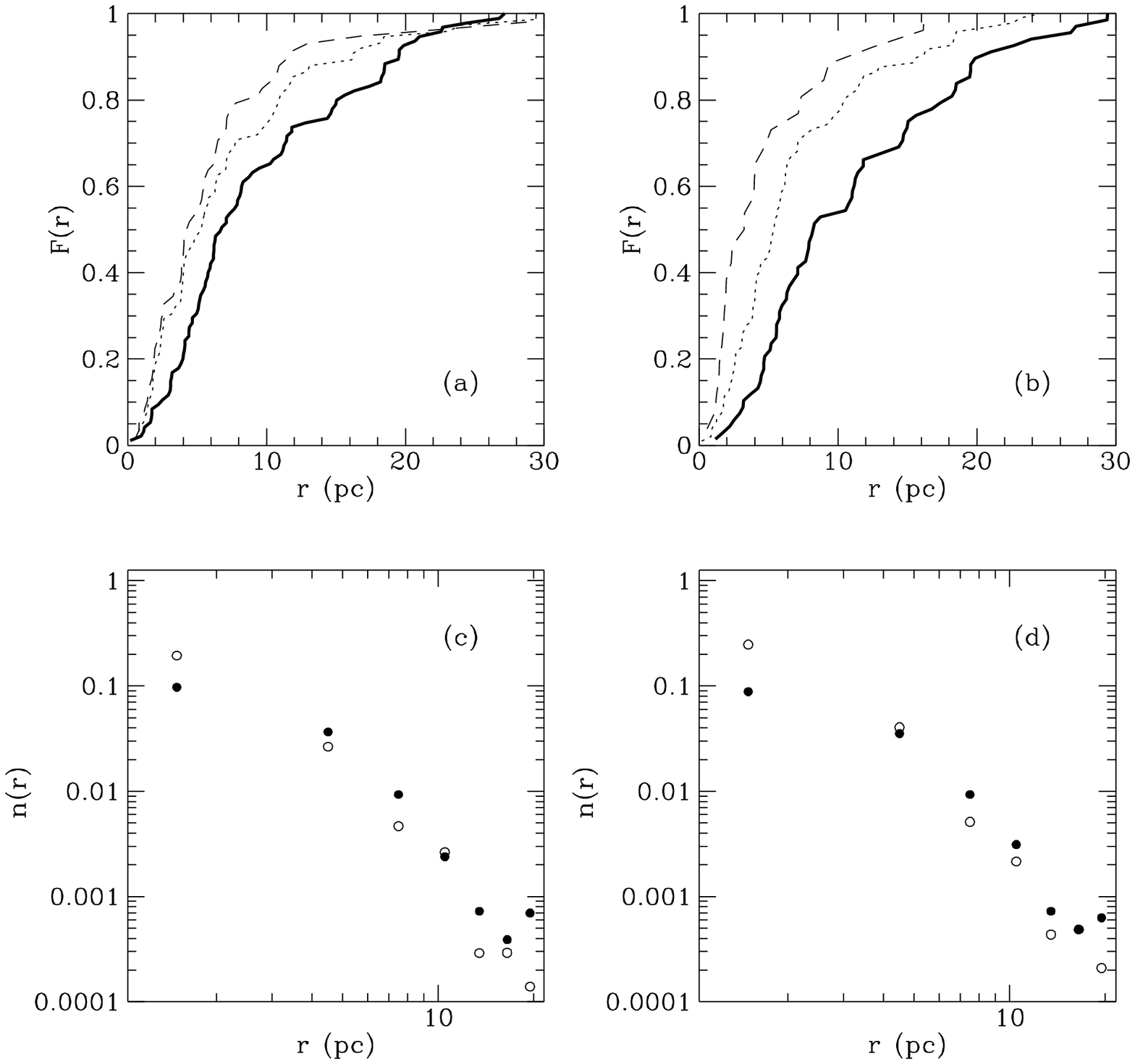,width=8.8cm}
\figure{15} {In (a) cumulative distributions versus distance from
the cluster centre are shown for single stars (solid line, 95
objects); binaries (resolved or spectroscopic; dotted line, 75
objects); and spectroscopic binaries only (dashed line, 58
objects). In (b) cumulative distributions are shown for $M\le1{\rm
M}_\odot$ (solid line 68 objects); $1\le M\le2{\rm M}_\odot$ (dotted
line, 97 objects); and $M>2{\rm M}_\odot$ (dashed line, 26
objects). In (c) the corresponding number densities are shown for
single stars ($\bullet$, 89 objects) or binaries ($\circ$, 71
objects). In (d) corresponding number densities are shown for
$M\le1.2{\rm M}_\odot$ ($\bullet$, 90 objects and $M>1.2{\rm M}_\odot$
($\circ$, 91 objects).  }
\endfig

Fig.~15 shows the cumulative distributions and number densities versus
distance from the adopted cluster centre, for single stars and
binaries for various mass groups. In the central 2~pc region, only
stars more massive than about 1~M$_\odot$ are found (right-hand
figures), and most of these are binaries (left-hand figures). This
general effect was already noted by van Bueren (1952) and Pels et al.\
(1975), and is precisely as found in the simulations by Terlevich
(1987). We note in passing that the division between `hard' and `soft'
binaries, taken as the separation at which the (circular) orbital
velocity ($v=\sqrt{GM/r}$) is equal to the rms random motion is, for a
system mass of 1~M$_\odot$ and $v=0.25$~km~s$^{-1}$ (see Sect.~7.2) is
about 0.07~pc, or $\sim5$~arcmin at the distance of the Hyades.
Numerical and statistical theories of star clusters predict the
formation of a halo where the density falls as $r^{-q}$, with $q$
between 3 and 3.5 (King 1966, Spitzer 1975). Although for clusters
with low central concentration such a power-law density distribution
over the whole cluster is an inadequate representation, from Fig.~15
we find $q\simeq3.3$ for single stars in the range $r=4-10$~pc,
$q\simeq2.9$ for $M<1.2{\rm M}_\odot$ and $q\simeq3.5$ for $M>1.2{\rm
M}_\odot$. Thus less massive stars are more spatially extended than
the more massive ones, with a lower density than for the high mass
stars in the central regions, consistent with results from numerical
simulations. Binaries follow the radial distribution according to
their mass, implying that mass is the predominant segregation factor,
rather than whether the system is binary or not.

This mass segregation in turn implies that cluster luminosity
functions derived from the central regions of the cluster will include
an artificial flattening due to such mass segregation. The
overabundance of bright stars is well established observationally,
with Reid (1993) attributing this to mass segregation and stellar
evaporation, while Eggen (1993) also shows that the luminosity
function is depleted at the faint end compared with the observed
field-star luminosity function.

In the numerical models, massive stars sink to the central region,
forming binaries, which become harder through the interaction with
lighter stars, which in turn acquire enough energy to reach the outer
parts of the cluster. There is a strong preference for energetic
binaries to be formed among the heaviest members, which tend to
segregate towards the dense central part of the cluster. Binaries
themselves may not play a significant role in the evolution of the
cluster, unless there is a significant population of primordial
binaries (e.g.\ as concluded by Kroupa 1995).  Griffin et al.\ (1988)
estimate that 30 per cent of the cluster members with $2.6<M_V<10.6$
are radial velocity binaries.  For systems brighter than $M_V\simeq13$
Eggen (1993) finds a photometric binary proportion of 0.4. Kroupa
interprets his own simulation results as suggesting that the total
proportion of systems in the central 2~pc sphere that are binary stars
may be as high as 65~per cent. Such models have been extrapolated by
Kroupa (1995), Reid (1993), Weidemann et al.\ (1992) and others to
derive a mass of around 1300~M$_\odot$ (and around 3000 stars in
total) at birth.

Counting the stars with an SB or RV indication in column~s of Table~2
and those with an indication C, G, O, V, X, or S in column~u as {\it
bona fide\/} binaries, we find in our sample of members a binary
fraction of 40~per cent. This fraction increases to 61~per cent for
the stars located within 2~pc from the cluster centre, with almost all
binaries in the central region being spectroscopic.

\titleb{Velocity distribution}

We now turn to an examination of the velocity distribution within the
cluster. As stated in Sect.~5.2 the internal velocity dispersion in
the centre of the cluster is not resolved with the accuracy of our
present velocity data. The intrinsic dispersion expected for a cluster
like the Hyades, in dynamical equilibrium, is $\sim 0.2$ km~s$^{-1}$
(van Bueren 1952, Gunn et al.\ 1988), below the upper limit of the
observed dispersion.

Investigation of the possible systematic effects due to rotation of
the cluster was investigated by Wayman et al.\ (1965), while Wayman
(1967) estimated a limit on the contraction of $K=-0.013\pm0.015$~km
s$^{-1}$ pc$^{-1}$. Hanson (1975) considered the possibility of
expansion, contraction, or rotation contributing significantly to the
star's space velocities, and inferred that the convergent point
solution was sufficiently insensitive to rotation that observable
effects would be seen before impacting on the convergent point
distance. Hanson also discounted significant shear due to differential
Galactic rotation. Gunn et al.\ devoted particular attention to the
assessment of the velocity dispersion and the possible rotational
flattening of the system.

The residual velocities shown in Fig.~9 suggest that a number of stars
in the outer regions of the Hyades show very substantial deviations
from the mean cluster motion. These deviations are, by definition,
within `3$\sigma$' of our mean cluster motion, but a systematic
pattern suggestive of a rotation or shearing motion in the cluster
does seem to exist, although it is noted that the interpretation of a
projection into two-dimensions of systematic structures in three
dimensions is not necessarily clearly evident to the eye. Although
other interpretations have been considered (see below), the
correlations between the residual velocities in Fig.~8 turn out to be
fully consistent with the observational errors, as we shall
demonstrate.

In deriving space velocities for the cluster stars we make use of the
observed vector $(\pi,\mu_{\alpha*},\mu_\delta,V_{\rm R})$. This
vector is transformed to a space velocity, implicitly invoking a
transformation to $(V_{\alpha*},V_\delta,V_{\rm R})$.  On the
assumption that the astrometric errors are uncorrelated the
transformation of the observables to the vector
$(\pi,V_{\alpha*},V_\delta,V_{\rm R})$ yields the covariance matrix:
$$
\pmatrix {
\omit & \omit & \omit & 0 \cr
\omit & {\bf S} & \omit & 0\cr
\omit & \omit & \omit & 0 \cr
0 & 0 & 0 & \sigma_{V_{\rm R}}^2 } \eqno{(18)}
$$
With $a=A_v/\pi^2$, ${\bf S}$ is given by:
$$
\pmatrix {
\sigma_\pi^2 & -\mu_{\alpha*}a\sigma_\pi^2 &
               -\mu_\delta a\sigma_\pi^2 \cr
-\mu_{\alpha*}a\sigma_\pi^2 &
a^2 \mu_{\alpha*}^2 \sigma_\pi^2 +
      A_v a \sigma_{\mu_{\alpha*}}^2 &
a^2 \mu_{\alpha*} \mu_\delta \sigma_\pi^2 \cr
-\mu_\delta a\sigma_\pi^2 &
a^2 \mu_{\alpha*} \mu_\delta \sigma_\pi^2 &
a^2 \mu_{\delta}^2 \sigma_\pi^2 +
      A_v a \sigma_{\mu_\delta}^2 }
$$
Hence, even in the absence of correlations between astrometric errors,
the parallaxes and velocity components $V_{\alpha*}$ and $V_\delta$
will in general be correlated. Moreover, because of the position of
the convergent point of the system with respect to the cluster centre,
$\mu_{\alpha*}$ is positive and $\mu_\delta$ is negative for most
cluster members, and hence the product ${\mu_{\alpha*}}\mu_\delta$ is
negative. Thus for most stars the uncertainties in $\pi$ and
$V_{\alpha*}$ are anti-correlated, the uncertainties in $\pi$ and
$V_\delta$ are correlated, and the uncertainties in $V_{\alpha*}$ and
$V_\delta$ are anti-correlated, which will lead to systematic
behaviour of the uncertainties in the sample as a whole. These
systematics will be transferred to the space velocities.

\begfigwid 5.5cm
\vskip-5.5cm
\centerline{
\hskip -1.0cm           \psfig{figure=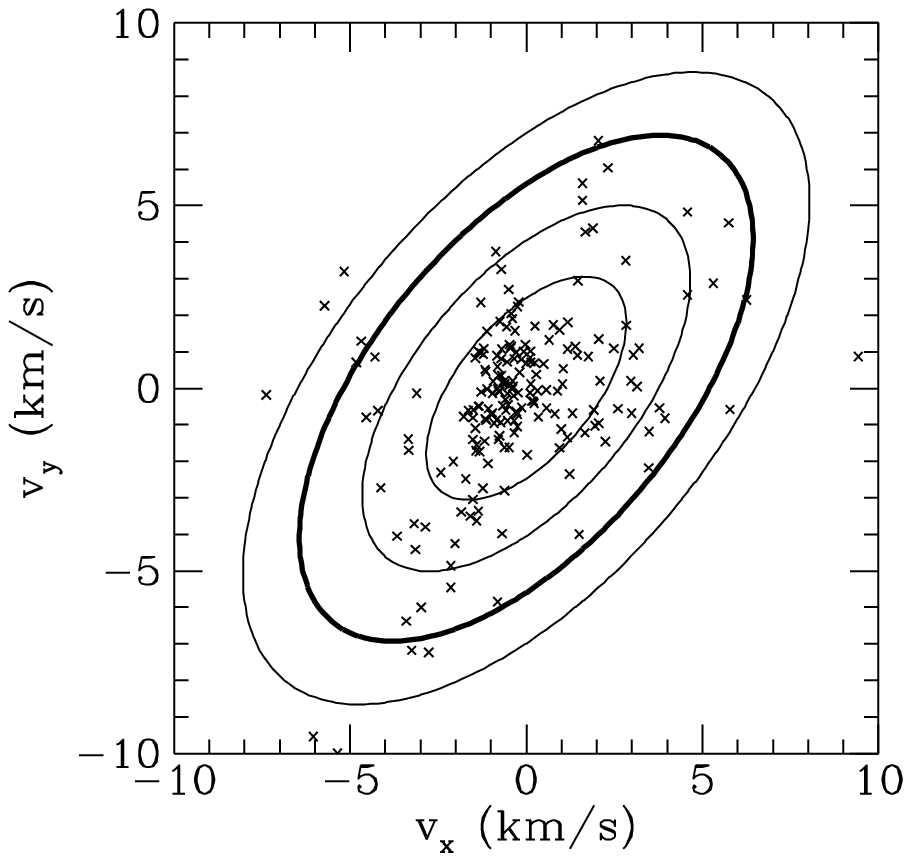,width=6.5cm}
\hskip -1.0cm\hfil      \psfig{figure=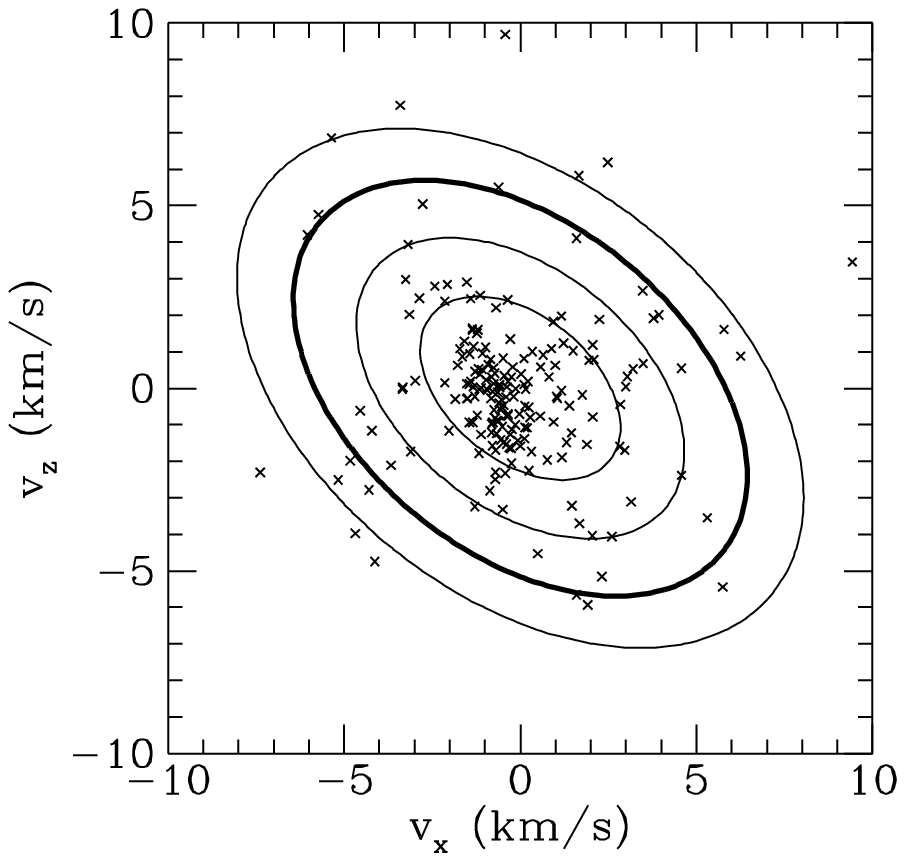,width=6.5cm}
\hskip -1.0cm\hfil      \psfig{figure=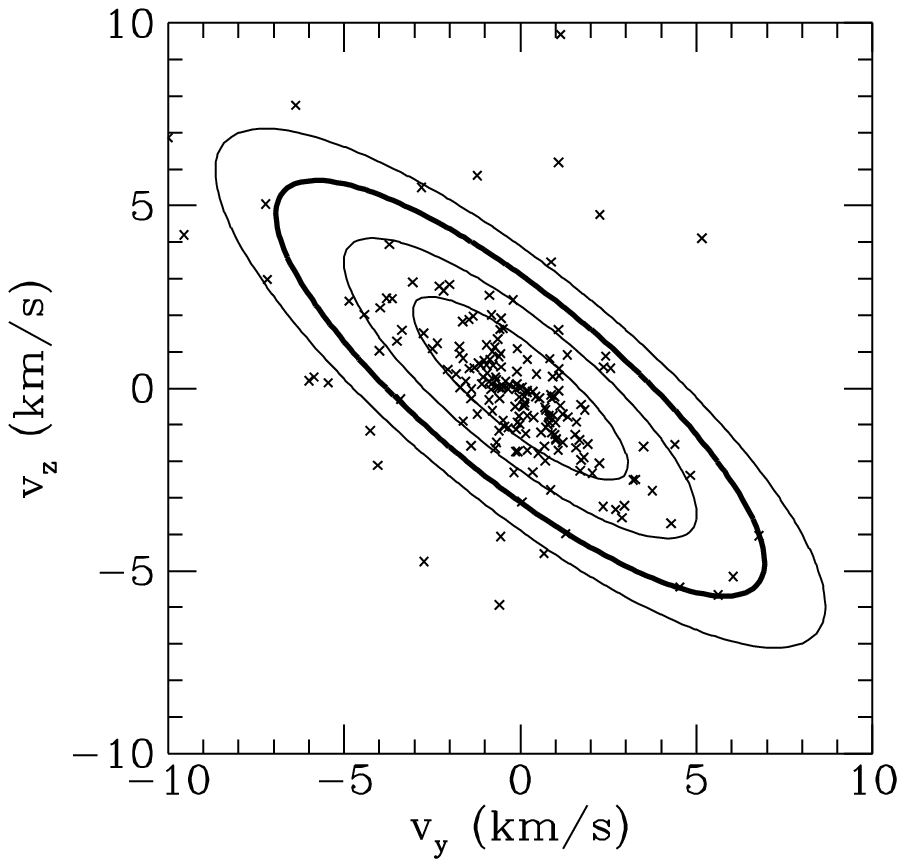,width=6.5cm}
\hskip -1.0cm} 
\figure{16} {
Contour diagrams delineating the (projected) confidence regions
corresponding to the mean covariance matrix of the space velocities
for all candidate members with available radial velocities
(cf.~Fig.~8b).  The confidence regions indicate the expected
distributions of residual velocities in the absence of intrinsic
dispersion. The contours correspond to the $68.3$\%, $95.4$\%,
$99.73$\%, and $99.99$\% confidence levels. The thick line, at
99.73\%, is the $c=14.16$ contour (see Eq.~16).  The crosses
correspond to the residuals between the observed space motions, and
the mean (centre of mass) cluster motion.  }
\endfig

To confirm the prediction of correlations in the velocity residuals we
proceed as follows. If we assume that the motions of the cluster
members are only due to the mean motion of the cluster, then the
observations of the velocities of the cluster members will all have
the same expectation value. We can then average all measured
velocities to obtain a mean motion, with the uncertainty in the mean
given by the mean of the covariance matrices of the individual
members. This mean covariance matrix can then be used to construct the
confidence region within which all residual velocities should
lie. This is illustrated in Fig.~16 for all the Hyades members with a
known radial velocity. The contours delineate the confidence region at
confidence levels $68.3$\%, $95.4$\%, $99.73\%$, and $99.99$\%. From
an eigenvector analysis of the mean covariance matrix, the minor axis
of the distribution of velocity residuals is found to point in the
direction $\ell=105^\circ, b=46^\circ$, which explains the flattened
appearance of the distribution of residuals in Fig.~8b (rightmost
diagram).

An examination of Fig.~9 reveals a correlation between the velocity
residuals (magnitude and direction) and the distances (parallaxes) of
the stars. This effect, especially evident in the leftmost diagram,
can be understood as follows. The difference between the observed and
true stellar parallaxes ($\Delta\pi=\pi_{\rm obs}-\pi_{\rm true}$) is
not correlated with the true parallaxes. However, adding $\Delta\pi$
to $\pi_{\rm true}$ implies that, on average, the stars with the
largest observed parallaxes will have a positive $\Delta\pi$ (and vice
versa for the stars with the smallest observed parallaxes). So the
sign of the parallax error is correlated with the observed
parallax. The correlation between $\Delta\pi$ and $V_{\alpha*}$ and
$V_\delta$, discussed above, will then lead to a correlation between
the observed distances of the stars and the velocity residuals.  Thus
both the overall distribution of the velocity residuals, as well as
the correlation of the direction of the residuals with spatial
position, can be attributed to observational errors.

Nevertheless, a large number of stars (32 out of 197) are located
outside the `$3\sigma$' contour. About half of these lie beyond 10~pc
from the centre of the cluster, with around one half of the 32 stars
being binaries. This indicates that many of these stars may have
suspect velocities. But even for the 165 stars inside the `$3\sigma$'
contour, ${\bf z}'\pmb{$\Sigma$}^{-1}{\bf z}$ is not distributed
according to the expected $\chi^2$ distribution with 3 degrees of
freedom. A large fraction of the stars shows larger deviations from
the mean motion, suggesting that the model `mean motion plus error in
the mean' is insufficient to fully describe the residual velocities.

To investigate the distribution of ${\bf z}'\pmb{$\Sigma$}^{-1}{\bf
z}$ further we restrict our attention to a `high-precision subset' of
the cluster members: stars without any indications of multiplicity,
with a radial velocity determined by Griffin et al.\ (1988), and with
standard errors on the Hipparcos parallax and proper motions of less
than 2 mas and 2 mas yr$^{-1}$ respectively.  This selection results
in a subset of 40~stars from which HIP~18962 is suppressed because it
is located at a very large distance from the cluster centre (45~pc),
and HIP~21788 is suppressed because of its large residual $v_x$ (5~km
s$^{-1}$). Fig.~17 shows the cumulative distribution of ${\bf
z}'\pmb{$\Sigma$}^{-1}{\bf z}$ for these stars (lower solid line),
where we have used the appropriate mean velocity of this subset ${\bf
v}_C = (-42.31, -19.08, -1.43)$ km s$^{-1}$ (in Galactic coordinates)
to characterise the mean cluster motion. The dotted line shows the
expected distribution in the absence of any intrinsic dispersion in
the velocities, from which it is clear that there is an extra
dispersion unaccounted for in our model, or present in our
observational errors.

This extra dispersion may originate from underestimates of the
standard errors in the radial velocities, from the Hipparcos
astrometry, from the presence of undetected binaries (Mathieu 1985),
from systematic motions (e.g.\ rotation), or simply from the intrinsic
velocity dispersion of the cluster. If the cause of the extra
dispersion is attributed solely to the quoted standard errors being
underestimates of the true external errors, then the standard errors
in the astrometry would have to be increased by 50~per cent (factor
1.5), the errors on the radial velocities would have to be increased
by a factor of~3, or both would have to be increased by 40~per
cent. We consider such explanations unlikely.

Given the fact that the observed binary fraction provides a firm lower
limit on the actual binary population, and that many workers consider
that the true binary fraction may approach unity, the cumulative
distribution of ${\bf z}'\pmb{$\Sigma$}^{-1}{\bf z}$ has been
constructed as before, but with a standard error of $0.3$~km s$^{-1}$
added quadratically to each of the components $v_x, v_y$, and $v_z$
(shown by the thick solid line in Fig.~17). The resulting agreement is
now excellent.  For a cluster mass of $\simeq$~460~M$_\odot$, an
estimated core radius of $\sim 2.5$--3 pc, and a half-mass radius of
roughly 4--5~pc, the expected mean cluster velocity dispersion is
$0.2$--$0.4$ km s$^{-1}$.  Our `high-precision subset' contains mainly
stars in the central region of the cluster where the velocity
dispersion will be higher than the mean.  Thus we postulate that the
$0.3$ km s$^{-1}$ added to the standard errors of the space velocity
components can be ascribed to a combination of the internal motion of
the cluster, possibly supplemented by the presence of undetected
binaries in the high-precision sample which would contribute further
to the observed dispersion. This is in good agreement with a recent
estimate of the internal cluster velocity dispersion of $0.25\pm0.04$
km s$^{-1}$ derived by Dravins et al.\ (1997) on the basis of a
maximum-likelihood determination of the cluster's astrometric radial
velocities.

\begfig 8cm
\vskip-8cm
\psfig{figure=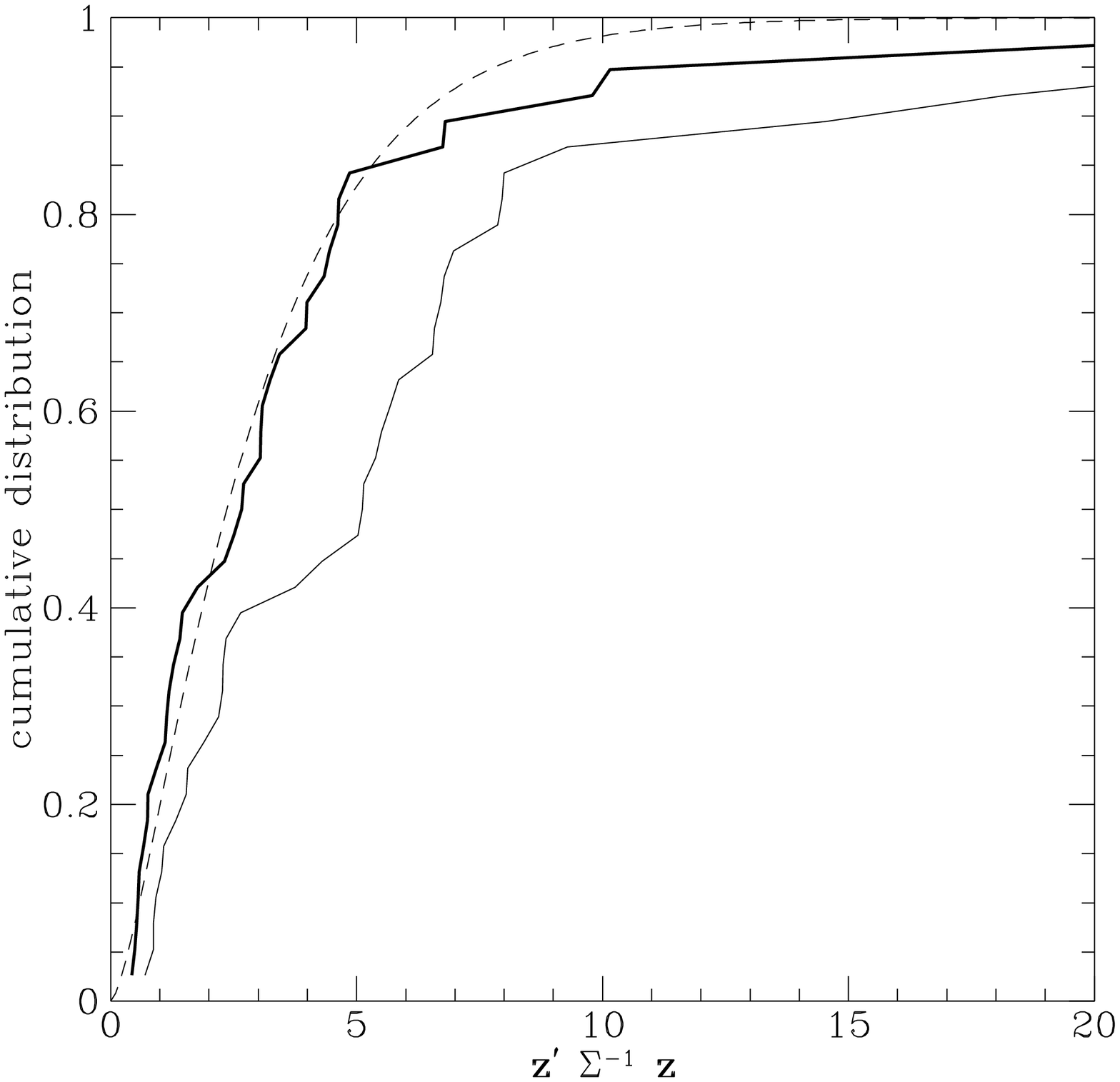,width=8cm}
\figure{17} {The cumulative distribution of ${\bf z}'\pmb{$\Sigma$}^{-1}{\bf z}$ 
for the high-precision subset of objects described in the text (lower
solid line) compared to the expected distribution in the absence of
intrinsic dispersion given by the $\chi^2$ distribution with 3 degrees
of freedom (dashed line).  The thick solid line has been constructed
by adding a standard deviation of $0.3$~km s$^{-1}$ to each of the
components $v_x$, $v_y$, and $v_z$.  }
\endfig

In principle the errors in the space velocities originating only from
the radial velocities can be decoupled from those in the (more
homogeneous) proper motions by selecting objects in a thin `parallax
slice' in order to assess whether the cluster dispersion can be
resolved exclusively in the proper motions. Restricting the data set
to the parallax slice 21--21.5~mas, and using only single stars within
5~pc from the cluster centre, each proper motion is transformed into a
component parallel to the direction of the convergent point (which can
be corrected for the angular size of the cluster), and a perpendicular
component which should reflect only observational or intrinsic
dispersions (the classical $\upsilon$ and $\tau$ components, e.g.\
Smart 1938). In the perpendicular direction the spread in the proper
motions as measured by the inter-quartile range is smaller than the
median error. Although the spread in the parallel component is about
twice as large as the median error, the parallel motion is very large
(about 110~mas~yr$^{-1}$) so that even the 2~per cent spread in
distance caused by the narrow parallax range selected leads to a
considerable artificial spread in the proper motions, of the order of
1--2~mas~yr$^{-1}$. As a result, the intrinsic dispersion in the
proper motions remains unresolved.

Although our observations therefore appear fully consistent with
uniform space motion, we also examined four possible causes of
systematic structure in the space velocities in an attempt to obtain a
better fit to the residuals seen in Fig.~9 without resorting to the
assumption of an unmodelled internal velocity dispersion:

(i) systematic errors in the Hipparcos parallaxes or proper motions,
or in the ground-based radial velocities, will affect the inferred
velocity field. Geometrical considerations demonstrate that a
spatially extended system participating in uniform space motion does
not yield constant space velocities if the radial velocities, or
$\mu_\alpha$ and/or $\mu_\delta$, are subjected to offsets of the type
expected for our particular observational quantities. For example, a
constant offset in the radial velocity zero-point does not transform
to a constant displacement in the space velocity because of the
angular extension of the cluster on the sky.  Similarly, a
non-inertial `spin' in the Hipparcos proper motion system (an effect
whose elimination was the subject of considerable effort during the
finalisation of the Hipparcos Catalogue) would lead to proper motion
displacements of the form:
$$\eqalign{
\Delta\mu_\alpha\cos\delta&=-\omega_x\sin\delta\cos\alpha 
	-\omega_y\sin\delta\sin\alpha + \omega_z\cos\delta \cr
\Delta\mu_\delta&=+\omega_x\sin\alpha - \omega_y\cos\alpha\cr} \eqno(19)$$
where $(\omega_x, \omega_y, \omega_z)$ represents the non-inertial
spin components of the proper motion system, and
$\Delta\mu_\alpha\cos\delta$ and $\Delta\mu_\delta$ are the resulting
offsets in the individual proper motions. We were not able to model
both the magnitude and structure of the residuals as due to a
combination of these zero-point errors, and large radial velocity
offsets alone would lead to a very different structure in the residual
velocities from that seen in Fig.~9. Taking only the magnitude of the
residuals into account $\vert\pmb{$\omega$}\vert$ would have to be of
the order of 10~mas yr$^{-1}$, more than an order of magnitude above
the limits on the non-inertial spin-components of the Hipparcos proper
motions noted in Sect.~3.1;

(ii) objects beyond the cluster tidal radius will be subjected to
systematic velocity perturbations from the Galactic tidal field, which
will lead to a systematic pattern of residual velocities with
increasing distance from the cluster centre, and periodic with
time. The effects can be calculated on the basis of the epicyclic
approximation (Binney \& Tremaine 1987, Sect.~3.2). It is therefore
expected that the velocity residuals in the regions of the cluster
beyond the tidal radius will deviate from a pattern of uniform space
motion. However, the resulting velocity perturbations depend on the
escape velocity and the time of escape, making it difficult to model
them on the basis of only a few escaping members;

(iii) the velocity residuals increase with distance from the centre,
and superficially appear consistent with a gradient of $\Delta v\simeq
0.3$ km s$^{-1}$ pc$^{-1}$ out to distances of about 10~pc, a gradient
necessary to explain the largest velocity residuals in Fig.~9. Such a
high value would exclude rotation as the source of the largest
velocity deviations seen in Fig.~9, since this would require a mass
orders of magnitude larger than the observed mass if the cluster is
not to be disrupted on a short time scale.  A lower rotation within
the central 5~pc region is not required to explain the velocity
residuals which, we re-iterate, are consistent with a non-rotating
system and our observational errors;

(iv) we have examined the possibility that the cluster recently
experienced an encounter with a massive object causing a tidal shear
in the outer regions of the cluster. The magnitude of space motion of
the Hyades with respect to the LSR is rather large (34~km s$^{-1}$),
and since most known massive objects in the vicinity of the Hyades are
molecular or atomic interstellar clouds having a much smaller motion
with respect to the LSR, the consequent large relative velocity
between the Hyades and any object encountered make the interacting
system well suited to treatment using the impulsive approximation
(e.g.\ Binney \& Tremaine 1987, Chapter~7, Theuns 1992a, b). In this
approximation the relative velocity of the two colliding objects is
perpendicular to the velocity disturbances generated in the less
massive system. We can use the direction of the minimum velocity
dispersion given by the results of the eigenvector analysis noted
above to derive the direction of the relative velocity of a postulated
encounter. Since the space motion of the Hyades is known, we can infer
the minimum velocity with which the perturbing object moves with
respect to the LSR. This turns out to be about 30~km s$^{-1}$, thus
tending to exclude the possibility of a recent high-speed encounter
with a nearby giant molecular cloud. Moreover, since the velocity
increments imparted to the perturbing object scale as $2{\rm
G}M/b^2V$, where $M$ is the mass of the perturbing object, $V$ the
relative impact velocity, and $b$ the impact parameter, for any
plausible values of $b$ and $V$ a very high mass, of the order of
$10^6$~M$_\odot$, is required for the perturbing cloud, uncomfortably
large compared with estimates for nearby giant molecular clouds (e.g.,
Dame et al.\ 1987).

We conclude that systematic effects in our data, or external
perturbations of the velocity field, are not evident in our results,
for which the parallaxes, proper motions, and radial velocities are
all consistent with uniform space motion.  In contrast, our results do
not permit an unambiguous or definitive assignment of cluster
membership, at least beyond the cluster tidal radius, or as the
velocity discrepancies between the individual 3-d space velocity and
the mean cluster motion increase.

Table~2 includes objects with low values of ${\bf
z'}\pmb{$\Sigma$}^{-1}{\bf z}$, but with large distances (small
parallaxes) for which the distance from the cluster centre is large.
Candidate `escapers' can also be identified, on the basis of their
small $d$, but discrepant motions.  Of the list of 97 Hyades group
stars listed by Eggen (1982) 92 are contained in the Hipparcos
Catalogue and, of these, 8 are in the region of sky covered by the
present study. Table~6 lists these objects, together with their
distance from the cluster centre, the velocity deviation, and the
value of ${\bf z'}\pmb{$\Sigma$}^{-1}{\bf z}$.

HIP~18692 and 26382 are among the 218 members already contained in
Table~2. HIP~22697 has a velocity consistent with membership, but with
a parallax (9.27~mas) just below our adopted threshold, while HIP
18975 is only just outside our $3\sigma$ membership limit.  All of the
stars in Table~6 are located beyond the tidal radius of 10~pc;
HIP~18975 may be especially interesting in this context since it
appears to be an example of an object emerging from the Hyades, still
close to the tidal radius.  HIP~21788 (vB~110), rejected from our
high-precision subset, may fall into the same category -- while still
within our $3\sigma$ kinematical contour, it evidently has a somewhat
discrepant motion, and was rejected as a plausible member on the basis
of its radial velocity by Griffin et al.\ (1988).

Breger (1968) used Str\"omgren photometry to classify false members of
Eggen's Hyades-group list on the basis of their discrepant metallicity
indices. Breger modelled the metallicity distribution of the moving
group as a 50:50 mixture of Hyades stars with field stars of lower
metallicity. Breger's list contains five of the stars in our area of
study, and two of these (HIP~13834 and 18692) are considered by us as
members. The former does not occur in the later Eggen lists, although
it has the appropriate metallicity for membership according to Breger,
and is well matched to the cluster kinematics despite its relatively
large distance from the cluster centre (20.5~pc).  The latter object
also occurs in Table~6, is far from the cluster centre, and is only
just contained within our kinematical selection limit.  With a low
metallicity according to Breger, further studies may well confirm it
as a non-member.

\begtabfull
\tabcap{6} {Stars from Eggen's (1982) list of Hyades group
stars, classified according to distance from our adopted cluster 
centre, $d$, $\Delta v$, and ${\bf z'}\pmb{$\Sigma$}^{-1}{\bf z}$.}
\centerline{
\vbox{
\def\p{\phantom{0}}
{\sevenrm \baselineskip 8pt
\vbox {\tabskip 1em plus 2em minus 0.5em
\halign {\hfil#\hfil& \hfil#\hfil& \hfil#\hfil& \hfil#\hfil \cr
\noalign{\hrule}
\noalign {\smallskip}
HIP&	$\scriptstyle d$& $\scriptstyle \Delta v$&
$\scriptstyle  {\bf z'}\pmb{$\Sigma$}^{-1}{\bf z}$ \cr
&	(pc)&	(km~s$^{-1}$)&     \cr
\noalign {\smallskip}
\noalign{\hrule}
\noalign {\smallskip}
12184&  21.2&    13.0&    22.09\cr
12189&  21.1&    15.8&    33.05\cr
12828&  25.7&   \p7.3&    18.25\cr
16813&  49.4&    19.4&    51.11\cr
18692&  45.9&   \p6.3&    13.67\cr
18975&  12.7&   \p5.5&    14.75\cr
22697&  63.4&   \p5.4&   \p3.32\cr
26382&  16.2&   \p3.0&   \p2.07\cr
\noalign {\smallskip}
\noalign{\hrule}
}}}
}}
\endtab

While our present membership analysis is based only on kinematical
arguments, metallicity determinations clearly contribute important
additional information to the membership, especially farther from the
cluster centre. Meanwhile, our findings underline a plausible
connection between the Hyades stars and escaping Hyades members.

We summarise the various kinematical populations evident from this
study as consisting of the core, the corona (extending out to the
tidal radius $r_t$, Kholopov 1969), the halo (with $r>r_t$ but still
dynamically bound to the cluster), and the moving group population
(with $r\ga 2r_t$, and with similar kinematics signifying remnants of
past membership).

\titlea{Dynamical properties of the cluster}

\titleb{Potential and density}

In the previous sections we have demonstrated confidence in the
spatial distributions derived for the cluster members, and our
objective now is to derive a description of the smoothed mass
distribution within the cluster, in order to assess its dynamical
behaviour, and its interaction with the Galactic potential.

The form of the cluster potential determines the distances at which it
behaves dynamically as a spherical system or point mass. The potential
at a position $\bf r$ is given by:
$$\Psi=\sum_{k=1}^N {m_k \over |{\bf r}-{\bf r}_k|} \eqno{(20)}$$
where the potential is expressed in units such that the gravitational
constant is equal to~1, masses are expressed in solar masses, and
distances are expressed in~pc. Potentials were computed on a regular
mesh of points within a 20~pc radius sphere, centred at the centre of
the cluster. To avoid local irregularities due to stars too close to
one of the points where $\Psi$ was computed, the effect of a star was
ignored if it was within 0.1~pc of that point (introducing a softening
parameter to model the potential at smaller separations has a
negligible effect on the conclusions). The values on the mesh were
interpolated to obtain the coordinates of points with specific rounded
values of the potential. Finally, equipotential surfaces were
determined by a least-squares fitting of an ellipsoid through points
with the same potential.

\begfig 6cm
\vskip-6cm
\centerline{\psfig{figure=fig18_kov.ps,width=6.6cm,angle=270}}
\figure{18}{The gravitational potential as a function of radial
distance, $\Psi(r)$, for the objects within 20~pc of the cluster centre,
constructed as described in the text.}
\endfig

Other than close to the centre of the cluster ($r<1.5$~pc), the
ellipsoids could be approximated by spheres to better than 2~per cent
indicating that, dynamically, the cluster has a high degree of
spherical symmetry.  Fig.~18 shows the value of the potential as a
function of distance $r$ from the centre. For $r>9$~pc, $\Psi(r)$ is
well represented numerically by:
$$\Psi(r)={240\over r} \eqno(21)$$
where $\Psi$ is expressed in M$_\odot$~pc$^{-1}$, assuming GM~=~1 for
the Sun.  Beyond 9~pc from its centre the cluster can therefore be
approximated dynamically by a point mass, with a mass (corresponding
to all of the Hipparcos stars considered here) of
$m_0=240$~M$_\odot$. This result is a consequence of the low space
density of the stars farther out from the cluster centre.

The results presented in Fig.~18 and Eq.~(21) are representative only
of the 180 stars selected within the 20~pc radius sphere. If we make
the assumption that the Hyades stars not contained in the Hipparcos
Catalogue have the same distribution, and that the total mass within
this sphere is $M_0$, the ordinates of Fig.~18, and the coefficient of
Eq.~(21), would have to multiplied by $M_0/m_0$ -- the total mass of
the cluster has been variously estimated at between 300~M$_\odot$
(Pels et al.\ 1975, Oort 1979), 400~M$_\odot$ (Gunn et al.\ 1988), and
460~M$_\odot$ (Reid 1992), according to the corrections made for
duplicity, faint stars, and white dwarfs.

\begfig 12cm
\vskip-12cm
\centerline{\psfig{figure=fig19a_kov.ps,width=6.6cm,angle=270}}
\centerline{\psfig{figure=fig19b_kov.ps,width=6.6cm,angle=270}}
\figure{19}{(a) The mean density distribution within the cluster, $\rho(r)$,
in M$_\odot$~pc$^{-3}$; and (b) the mass within the sphere of radius $r$,
in M$_\odot$, constructed as described in the text.}
\endfig

To represent the mean density of the cluster, each star of mass $m$
was replaced by a sphere with the same mass, but with a spatially
extended and continuous density distribution. The adopted model is of
a constant density within a sphere of radius $s/2$, linearly
decreasing up to $s$, with $s=6$~pc.  Several density distributions
were tested, all giving similar results provided that the radius of
the sphere $s$ is not significantly different from this value. The
same procedure as for the potential was then followed to determine
equidensity surfaces. In practice, these were very close to spheres,
confirming the spherical structure of the cluster out to a radial
distance of about 7--8~pc. Beyond this radius, the density is too
small for the method to yield significant results. The resulting
density distribution $\rho(r)$ (in M$_\odot$~pc$^{-3}$) and the
cumulative mass distribution (in M$_\odot$) are shown in Figs~19(a,b).
As for the potential, the results scale as $M_0/m_0$ in order to take
account of missing stars.  >From the values of the potential, the
escape velocity of the star can in principle be determined although,
as demonstrated in Sect.~7.2, in most cases the velocity errors are
too large for individual conclusions to be drawn.

The resulting density distribution has been compared to both a Plummer
model and a King model. For the former the best fit values of the core
radius and central density are $2.9$~pc and $1.8$~M$_\odot$~pc$^{-3}$,
corresponding to a central velocity dispersion of
$0.21$~km~s$^{-1}$. In the case of the King model, the best fit model
has a core radius of $2.6$~pc, a central density of
$1.8$~M$_\odot$~pc$^{-3}$ and a value of $\Psi(0)/\sigma^2=2.6$, where
$\Psi(0)$ is the central potential (Binney \& Tremaine 1987,
Sect.~4.4). This corresponds to a true central velocity dispersion of
$0.24$~km~s$^{-1}$.

For both models the central velocity dispersion is in good agreement
with the value derived in Sect.~7.2. However, results inferred from
these particular models should be viewed with some caution. The masses
in both models fall short of the total mass observed in the
Hyades. This is in part due to the fact that we are missing the faint
end of the luminosity function in the sample under study. Adding
fainter members of the Hyades may change the overall density
distribution, while significant mass may also be present in the outer
regions of the cluster. This is supported by the fact that the
observed half-mass radius ($\sim 5.7$~pc) is larger than the half mass
radius for the models above ($3.7$ and $3.0$~pc, respectively).

\begtabfull
\tabcap{7} {Moments of inertia (in M$_\odot$~pc$^2$) with respect to 
the three coordinate planes (Galactic coordinates). $r$ is the 
radius of the sphere (pc) and $N$ the number of stars included.}
\centerline{
\vbox{
{\sevenrm \baselineskip 8pt
\vbox {\tabskip 1em plus 2em minus 0.6em
\halign {\hfil #& \hfil #& \hfil #& \hfil #& \hfil #&\hfil #& \hfil 
#& \hfil #\cr
\noalign {\vskip 5pt}
\noalign{\hrule}
\noalign {\smallskip}
$\scriptstyle r$ & 
\quad $\scriptstyle N$& 
\qquad $\scriptstyle xy$ & $\scriptstyle \sigma$ & 
\qquad $\scriptstyle xz$ & $\scriptstyle \sigma$ & 
\qquad $\scriptstyle yz$ & $\scriptstyle \sigma$ \cr
\noalign {\smallskip}
\noalign{\hrule}
\noalign {\smallskip}
    6&     98&   490&    60&     590&    10&     810&    170\cr
    8&    123&   870&    70&    1110&    20&    1420&    250\cr
   10&    133&  1110&    80&    1420&    30&    2020&    300\cr
   12&    151&  1360&   100&    2080&    40&    3430&    410\cr
   14&    155&  1430&   210&    2450&   110&    4010&    450\cr
   16&    163&  1880&   220&    2660&   120&    4970&    500\cr
   18&    169&  2970&   270&    3310&   130&    5900&    560\cr
   20&    180&  3490&   330&    4720&   150&    7560&    710\cr
\noalign {\smallskip}
\noalign{\hrule}
}}}
}}
\endtab

\titleb{Dynamical shape of the outer region}

The previous results concern primarily the central part of the
cluster. The star density in the outer region is rather low, and these
stars do not significantly modify the local potential, as demonstrated
by the $1/r$ form of the potential in the outer regions. The dynamical
approach to the study of the large-scale structure is through the
computation of the moments and products of inertia of the
cluster. Table~7 gives the value of the moments of inertia with
respect to the three principal coordinate planes, in Galactic
coordinates, computed within spheres with increasing radii $r$.

The evolution of the moments of inertia within 6~pc is rather chaotic,
although demonstrating a tendency for a spherical dynamical
shape. Beyond $r=8$~pc, the moments of inertia with respect to the $y$
and $z$ axes become prominent, confirming the extension of the star
distribution essentially along the $x$ axis. The determination of the
principal moments of inertia confirms this: the principal axis is very
close to the $x$ axis (within 10$^\circ$), while stars lie both in the
direction of the Galactic centre and anticentre.

The moment of inertia with respect to the $xy$ plane is systematically
smaller than that with respect to the $xz$ plane, indicating that the
elongated part of the cluster is slightly flattened, and that the
spread is slightly smaller in the direction perpendicular to the
Galactic plane, as found by Oort (1979) and confirmed by Terlevich
(1987).

The angular momentum of the cluster, derived from the individual space
velocities, is small and insignificant in the central region, but
grows as more distant stars are taken into account, with a major axis
tending towards the direction $\ell=125^\circ$, $b=50^\circ$.  That
the observed angular momentum is negligible for the main dynamical
core of the cluster lends further support to the conclusions of
Sect.~7, that the velocity residuals are a consequence of the
observational errors, and further studies would be required to
demonstrate the existence or absence of significant angular momentum
or rotation.

\titleb{The effect of the Galactic potential}

The cluster is immersed in the potential of the Galaxy, so that the
equipotential surfaces, unlike the spherical surfaces for isolated
clusters, are distorted and eventually become open.

In the disk, the location of these Lagrangian points in an open cluster 
can be calculated from the Oort constants:
$$x_{\rm L}= \left({GM_c \over 4A(A-B)} \right)^{1/3}   \eqno{(22)}$$
where $M_c$ is the total mass of the cluster (King 1962, Eq.~24) and
$A$ and $B$ are Oort's constants. This distance is referred to as the
tidal radius of the cluster, $r_t$ (although the volume defined by the
equipotential surface is not spherical).  Adopting $A=14.8$~km
s$^{-1}$ kpc$^{-1}$ and $B=-12.4$~km s$^{-1}$ kpc$^{-1}$ (Feast \&
Whitelock 1997) and using $M_c=400$~M$_\odot$ gives
$r_t=10.3$~pc. Since this distance is beyond the limit where the
cluster potential has the form of Eq.~(21), we can conclude that the
tidal radius is of order $r_t=10$~pc, the precise value depending on
the value of $M_c$.  This is consistent with the consequences of the
asymmetry in the evolution of the moments of inertia with $r$: beyond
this distance the stars behave like companions of the cluster but
under the predominant forces of the Galactic gravitational field.

\titlea{The Hyades Hertzsprung-Russell Diagram} 

We have established well-defined spatial and velocity criteria for the
assignment of cluster membership, at least within the central
10--20~pc region, and we now use these members to refine the
Hertzsprung-Russell diagram with the objective of presenting a
consistent picture of the observational (i.e.\ colour versus absolute
magnitude) and theoretical (i.e.\ bolometric magnitude versus
effective temperature) relationships. Clusters provide an important
environment for testing associated stellar evolutionary theories,
representing a number of stars which are considered, as a first
approximation at least, to be at the same distance, co-eval, and of a
constant metallicity. The Hyades, as the nearest moderately rich
cluster, has been studied in detail for these reasons.

While stellar models have been highly successful in matching the
overall features of cluster colour-magnitude diagrams, unambiguous
detailed model fitting has proved more elusive. The major problems
compounding these studies in the specific case of the Hyades have been
the uncertain distance modulus of the cluster, and its associated
depth, both conspiring to make the transformation to absolute
magnitudes uncertain; the uncertainty in the mean metal content and in
particular the discrepancies between photometric and spectroscopic
determinations (e.g.\ Cayrel de Strobel 1982, 1990) leaving open the
initial conditions for the stellar evolutionary models; the
contribution to both coordinates of the observational HR diagram due
to the contribution of (undetected) binary systems; and additional
complications associated with all such models such as the
transformation from theoretical to observational quantities (requiring
accurate bolometric corrections and colour conversions) and remaining
theoretical uncertainties, primarily those associated with the theory
of convection, including the value of the mixing-length parameter and
the possibility of significant convective overshooting (e.g.\ Maeder
\& Mermilliod 1980, VandenBerg \& Bridges 1984).

Although the distance modulus of the cluster as a whole may have been
assigned a small standard error in any given study, the depth of the
cluster is such that the contribution to the intrinsic scatter of the
observational main sequence from significantly different distances of
the individual members may be substantial (see, e.g., Cayrel de
Strobel 1982). The direct result of this depth effect is a spread in
individual distance moduli of member stars leading to a main-sequence
population in the $(V, B-V)$ plane less sharply defined than those of
more distant clusters (cf.\ Figs~2 and 21). The consequence of this
observational scatter is that it has been difficult to make a reliable
estimate of the helium abundance of the cluster members on the basis
of model fitting which has, in turn, precluded unambiguous matching of
evolutionary models to features such as the cluster turn-off.

With the trigonometric parallaxes from Hipparcos, we are in a position
to construct the observational HR diagram with an accuracy of about
0.1~mag on individual values of $M_V$ or $M_{\rm bol}$ (these errors
are still dominated by the standard error of the parallaxes; in
comparison, the error on $V$ is typically 0.01~mag or smaller, and may
be neglected). Our goals in this section are to derive an optimally
constructed observational HR diagram for the cluster, and to use these
observations in isochrone fitting to determine the cluster age.

These goals are achieved in two successive steps: (i) using stellar
evolutionary model fits to a carefully constructed $M_{\rm bol}$
versus $T_{\rm eff}$ diagram for the lower part of the main sequence,
we will define the locus of the Hyades zero-age main sequence (ZAMS)
and hence, through modelling, an estimate of the cluster's helium
content; (ii) the observations over the complete part of the HR
diagram in combination with theoretical isochrones will be used to
determine the cluster age, using models with and without convective
overshooting in the core. In both cases, it is important to suppress
known binary systems from the observational diagram, and to take into
account the possible remaining biases in both coordinates caused by
unrecognised duplicity. The influence of interstellar reddening
effects has been considered to be negligible for the Hyades stars
(Crawford 1975, Taylor 1980) and is neglected in the following
discussions.

\titleb{The Hyades ZAMS and He abundance}

The physical parameters (effective temperature, spectroscopic gravity,
and metallicity) of a significant number of stars in the cluster have
been derived over the last 25~years, and with advances provided by
recent solid-state detectors, excellent spectroscopic data are now
available. We have selected 40 stars, with bolometric magnitudes in
the range 3--6~mag, for which high-resolution, high S/N spectra are
available (Cayrel de Strobel 1980, Branch et al.\ 1980, Cayrel et al.\
1984, Cayrel et al.\ 1985, Boesgaard 1989, Boesgaard \& Friel
1990). From these studies, the metal content and effective temperature
are known with high accuracy for each star, with $T_{\rm eff}$
determined to typically 50~K, or even better for some objects. Further
details of the determination of the [Fe/H] is given by Cayrel et al.\
(in preparation).

\begfig 14cm
\vskip-14cm
\psfig{figure=fig20_zams.ps,width=8.5cm,angle=270}
\figure{20}{Hertzsprung-Russell diagram ($M_{\rm bol}$, $\log T_{\rm eff}$) 
for the 40 dwarfs listed in Table~8. The symbols follow the
multiplicity flags given in Table~2: objects which are spectroscopic
binaries or radial velocity variable are indicated `*'; objects which
are resolved by Hipparcos or known to be double systems are shown as
circles; one object (HIP~18658) with detected photocentric
acceleration, and one object (HIP~19504) possibly resolved in
photometry, are shown by triangles. For the remaining objects, error
bars correspond to $\pm50$~K in $T_{\rm eff}$ for $\log T_{\rm
eff}\le3.78$ and to $\pm75$~K for $\log T_{\rm eff}>3.78$, and to
$\sigma_\pi$ in $M_{\rm bol}$. ZAMS loci as a function of mass were
constructed as described in the text, and are given for the Hyades
(dashed line) and solar (dotted line) metallicities given in
Table~10. The location of the Sun is also shown.  }
\endfig

The relevant data are listed in Table~8. Bolometric magnitudes were
calculated from the $V$ magnitude given in the Hipparcos Catalogue,
the Hipparcos parallax, and applying the appropriate bolometric
corrections of Bessel et al.\ (1997). The error in $M_v$, and
therefore $M_{\rm bol}$, is still dominated by the error on the
trigonometric parallax, rather than by the apparent magnitude (which
can be seen from an inspection of the standard errors on the mean
magnitudes for these objects given in the Hipparcos
Catalogue). Lutz-Kelker-type corrections have been ignored in view of
the small values of $\sigma_\pi/\pi$ (see Sect.~6.3). Fig.~20 shows
the resulting positions in the ($M_{\rm bol}$, $\log T_{\rm eff}$)
diagram.  For the subset of 20~stars for which no evidence of binarity
is indicated in Table~2, error bars are given corresponding to the
standard error on the trigonometric parallax contributing to the
standard error in $M_{\rm bol}$.  The errors on $T_{\rm eff}$ are
harder to quantify. Although the less massive stars were all observed,
reduced, and analysed with the same methods (observations and
modelling), the more massive stars were observed, reduced and analysed
by different authors, with different model atmospheres and different
effective temperature scales. We have assigned error bars
corresponding to $\pm50$~K in $T_{\rm eff}$ for $\log T_{\rm
eff}\le3.78$, and to $\pm75$~K for $\log T_{\rm eff}>3.78$, the latter
in part taking into account the effect of the higher rotational
velocity.

\begtabfullwid
\tabcap{8} {Data for the 40 stars for which high-resolution spectra 
provide accurate values for $T_{\rm eff}$ and $M_{\rm bol}$ for the 
metallicity determination of the main sequence. The columns are: (1) Hipparcos 
Catalogue number; (2) van Bueren number; (3) $V$~magnitude from the 
Hipparcos Catalogue; (4) Hipparcos parallax; 
(5) parallax standard error; (6) absolute visual magnitude, $M_v$;
(7) error on $M_v$ due only to the parallax error; (8)
log of effective temperature; (9) bolometric correction from Bessel et 
al.\ (1997); (10) resulting absolute bolometric magnitude; (11) [Fe/H]; (12)
reference for [Fe/H]: 
BLT~=~Branch, Lambert \& Tomkin (1980); 
B~=~Boesgaard (1989);
BB~=~Boesgaard \& Budge (1988); 
BF~=~Boesgaard \& Friel (1990); 
CCC~=~Cayrel, Cayrel de Strobel \& Campbell (1985); 
CCS~=~Chaffee, Carbon \& Strom (1971); 
F~=~Foy (1975).}
\centerline{
\vbox{
\def\p{\phantom{1}}
\def\pp{\phantom{11}}
{\sevenrm \baselineskip 8pt
\vbox {\tabskip 1em plus 2em minus 0.5em
\halign {\hfil#\hfil& \hfil#\hfil& \hfil#\hfil& \hfil#\hfil& \hfil#\hfil& 
\hfil#\hfil& \hfil#\hfil& \hfil#\hfil& \hfil#\hfil& \hfil#\hfil& 
\hfil#\hfil& #\hfil \cr
\noalign{\hrule}
\noalign {\smallskip}
  HIP&    vB&   V&   $\scriptstyle \pi$& 
$\scriptstyle \sigma_\pi$&  $\scriptstyle M_v$&  
$\scriptstyle \sigma_{M_v}$&  $\scriptstyle \log T_{\rm eff}$&  
BC&     $\scriptstyle M_{\rm bol}$&  [Fe/H]&  Ref.\cr
  &        &   (mag)&	(mas)&  (mas)&     (mag)&      (mag)&
   &      (mag)&     (mag)&    &   \cr
(1)& (2)& (3)& (4)& (5)& (6)& (7)& (8)& (9)& (10)& (11)& (12)\cr
\noalign {\smallskip}
\noalign{\hrule}
\noalign {\smallskip}
\noalign{\vskip 3pt}
15310& \pp2& 7.78& 21.64& 1.33& 4.46& 0.13& 3.772& --0.057& 4.40& 0.22& BLT\cr
18170& \pp6& 5.97& 24.14& 0.90& 2.88& 0.08& 3.852& --0.007& 2.88& 0.30& BB\cr 
18658& \pp8& 6.35& 25.42& 1.05& 3.38& 0.09& 3.827& --0.014& 3.36& 0.20& BB\cr 
19148& \p10& 7.85& 21.41& 1.47& 4.50& 0.15& 3.768& --0.062& 4.44& 0.12& CCS\cr 
19261& \p11& 6.02& 21.27& 1.03& 2.66& 0.11& 3.836& --0.011& 2.65& 0.10& BF\cr 
19504& \p13& 6.61& 23.22& 0.92& 3.44& 0.09& 3.828& --0.014& 3.43& 0.18& BF\cr 
19554& \p14& 5.71& 25.89& 0.95& 2.78& 0.08& 3.848& --0.008& 2.77& 0.08& BF\cr 
19781& \p17& 8.45& 21.91& 1.27& 5.15& 0.13& 3.746& --0.103& 5.05& 0.10& CCC\cr 
19786& \p18& 8.05& 22.19& 1.45& 4.78& 0.14& 3.761& --0.072& 4.71& 0.23& CCS\cr 
19793& \p15& 8.05& 21.69& 1.14& 4.73& 0.11& 3.756& --0.079& 4.65& 0.19& CCS\cr 
\noalign{\vskip 3pt}
19796& \p19& 7.11& 21.08& 0.97& 3.73& 0.10& 3.799& --0.029& 3.70& 0.18& BF\cr 
19877& \p20& 6.31& 22.51& 0.82& 3.07& 0.08& 3.836& --0.011& 3.06& 0.27& BB\cr 
19934& \p21& 9.14& 19.48& 1.17& 5.59& 0.13& 3.724& --0.139& 5.45& 0.09& CCC\cr 
20215& \p29& 6.85& 23.27& 1.14& 3.68& 0.11& 3.778& --0.049& 3.63& 0.23& CCS\cr 
20357& \p37& 6.60& 19.46& 1.02& 3.05& 0.11& 3.833& --0.012& 3.03& 0.16& BF\cr 
20480& \p42& 8.84& 20.63& 1.34& 5.41& 0.14& 3.733& --0.108& 5.30& 0.10& CCC\cr 
20491& \p44& 7.18& 20.04& 0.89& 3.69& 0.10& 3.817& --0.018& 3.67& 0.13& B\cr 
20492& \p46& 9.11& 21.23& 1.80& 5.74& 0.18& 3.713& --0.186& 5.56& 0.07& CCC\cr 
20557& \p48& 7.13& 24.47& 1.06& 4.07& 0.09& 3.796& --0.032& 4.04& 0.11& BF\cr 
20567& \p51& 6.96& 18.74& 1.17& 3.32& 0.14& 3.819& --0.017& 3.31& 0.16& B\cr 
\noalign{\vskip 3pt}
20577& \p52& 7.79& 20.73& 1.29& 4.37& 0.14& 3.766& --0.065& 4.31& 0.05& CCC\cr 
20661& \p57& 6.44& 21.47& 0.97& 3.10& 0.10& 3.804& --0.025& 3.07& 0.11& BF\cr 
20712& \p62& 7.36& 21.54& 0.97& 4.03& 0.10& 3.791& --0.035& 3.99& 0.14& BF\cr 
20719& \p63& 8.04& 21.76& 1.46& 4.73& 0.15& 3.766& --0.066& 4.66& 0.05& F\cr 
20741& \p64& 8.10& 21.42& 1.54& 4.75& 0.16& 3.761& --0.073& 4.68& 0.14& CCC\cr 
20815& \p65& 7.41& 21.83& 1.01& 4.11& 0.10& 3.792& --0.035& 4.07& 0.12& B\cr 
20899& \p73& 7.83& 21.09& 1.08& 4.45& 0.11& 3.771& --0.058& 4.39& 0.14& CCC\cr 
20935& \p77& 7.02& 23.25& 1.04& 3.85& 0.10& 3.801& --0.028& 3.82& 0.10& BB\cr 
20948& \p78& 6.90& 21.59& 1.09& 3.57& 0.11& 3.814& --0.020& 3.55& 0.12& BF\cr 
20951& \p79& 8.95& 24.19& 1.76& 5.87& 0.16& 3.719& --0.166& 5.70& 0.14& CCC\cr 
\noalign{\vskip 3pt}
21008& \p81& 7.09& 19.94& 0.93& 3.59& 0.10& 3.811& --0.022& 3.57& 0.13& BF\cr 
21066& \p86& 7.03& 22.96& 0.99& 3.83& 0.09& 3.812& --0.021& 3.81& 0.12& BF\cr 
21152& \p90& 6.37& 23.13& 0.92& 3.19& 0.09& 3.829& --0.013& 3.18& 0.13& BB\cr 
21317& \p97& 7.90& 23.19& 1.30& 4.73& 0.12& 3.768& --0.063& 4.66& 0.10& CCC\cr 
21474&  101& 6.64& 22.99& 0.95& 3.45& 0.09& 3.822& --0.016& 3.43& 0.19& BB\cr 
21543&  102& 7.53& 23.54& 1.29& 4.39& 0.12& 3.760& --0.078& 4.31& 0.05& CCS\cr 
22496&  119& 7.10& 22.96& 1.17& 3.90& 0.11& 3.776& --0.049& 3.86& 0.17& CCS\cr 
22524&  121& 7.29& 19.30& 0.95& 3.72& 0.11& 3.802& --0.027& 3.69& 0.15& BF\cr 
22550&  122& 6.79& 20.15& 1.14& 3.31& 0.12& 3.778& --0.049& 3.26& 0.16& CCS\cr 
23214&  128& 6.75& 23.09& 0.83& 3.57& 0.08& 3.817& --0.018& 3.55& 0.13& BF\cr 
\noalign {\smallskip}
\noalign{\hrule}
}}}
}}
\vskip 10pt
\endtab

Table~8 yields a metallicity for the Hyades of
[Fe/H]~=~$0.14\pm0.05$. The observational quantity, [Fe/H], the
logarithm of the number abundances of iron to hydrogen relative to the
solar value, is related to the metallicity Z, in mass fraction,
through [Fe/H]~=~$\log({\rm Z/X})-\log({\rm Z/X})_\odot$ for a solar
mixture of heavy elements (X~is the hydrogen abundance by mass).  With
the solar value ${\rm (Z/X)}_\odot~=~0.0245$ of Grevesse \& Noels
(1993a) and the adopted Hyades [Fe/H] we obtain ${\rm
Z/X}~=~0.034\pm0.007$ which is slightly, but significantly, higher
than the solar value. This is the observational quantity to be used in
the models. The error of 0.007 on (Z/X) includes the error on the
solar (Z/X) of about 11~per cent (Anders \& Grevesse 1989).

Earlier discrepancies in the determination of [Fe/H] for the Hyades,
in particular the differences determined from photometric and
spectroscopic observations, have been summarised by Cayrel de Strobel
(1982, 1990).  The present value is in reasonable agreement with
(although is not independent from) more recent determinations, e.g.\
Cayrel et al.\ (1985) ([Fe/H]~=~$+0.12\pm0.03$, or 0.14--0.15
accounting for the difference in activity between the Sun and the
Hyades dwarfs); VandenBerg \& Poll (1989) ([Fe/H]~=~+0.15). Note that
vB52 was rejected as representative of the cluster mean by Cayrel et
al.\ (1985) on the basis of its colour anomaly, possibly related to
the strong emission of H$\alpha$.

The lower part of the Hyades main sequence relevant to this study is
populated by low mass stars which are only slightly evolved, and
therefore rather close to their `zero age' position, the zero-age main
sequence (ZAMS) defining the locus on the HR diagram where the stars
become fully supported by core hydrogen burning. Being of intermediate
or low temperature, their spectra do not show He-lines, and their
photospheric He abundance cannot be determined
spectroscopically. Theoretical computations of internal structure have
shown that the ZAMS locus depends on the initial chemical composition,
both in terms of metallicity and He content. Comparison with
theoretical models is therefore used to estimate the He abundance.

In order to fit the data with improved theoretical calculations, new
zero-age main sequence models have been calculated by one of us (YL)
with the CESAM stellar evolutionary code (Morel 1993, 1997). Updated
input physics, appropriate to the mass interval covered by this sample
of Hyades stars (from about 0.8--1.6~M$_\odot$) has been used: updated
OPAL opacities (Iglesias \& Rogers 1996) complemented at low
temperatures by the Alexander \& Ferguson data (1994), nuclear
reaction rates from Caughlan \& Fowler (1988), a solar mixture of
heavy elements from Grevesse \& Noels (1993a) corresponding to the
mixture used in opacity calculations and the CEFF equation of state
(Christensen-Dalsgaard 1991).

The mixing-length parameter used for all models was
$\alpha=l/H_p=1.64$ pressure scale-heights ($l$ is the mixing-length
and $H_p$ the pressure scale-height), derived from the calibration in
radius of the solar model with the same input physics (see below). The
same value was applied to the Hyades following investigations of
visual binary systems with known masses and metallicity (Fernandes et
al., in preparation) resulting in similar values of $\alpha$ for a
wide range of ages and metallicities. A value of $\alpha=1.5$ was
suggested by VandenBerg \& Bridges (1984), although a somewhat higher
value was also not ruled out. Models using standard mixing-length
theory are very sensitive to the precise choice of $\alpha$, although
the region of sensitivity is restricted.  For a summary of the effects
of the major uncertainties associated with the use of such models,
including the physical terms such as the reaction rates, the
opacities, the universality of the mixing-length parameter, and the
relevance of overshooting, see Lebreton et al.\ (1995). In the sample
of low mass stars considered here, stars are believed to be
essentially homogeneous on the ZAMS so that modelling uncertainties
associated with convective core overshooting are avoided.

As the models strongly depend on chemical composition (Y, Z in mass
fraction), a first set of calculations was made with a wide range of
chemical compositions yielding a grid of theoretical ZAMS with
different helium and metal contents. Interpolation between these ZAMS
then yielded the value of the helium abundance giving the best fit to
the low-mass stars for the mean observational value of~(Z/X) given in
Table~9.

A final ZAMS (see Table~10) was computed for Y~=~0.260 and Z~=~0.024,
corresponding to the mean Hyades metallicity. As shown in Fig.~20 this
ZAMS fits the observational data rather satisfactorily, with the
larger scatter for $\log T_{\rm eff}>3.78$ possibly originating from
underestimated errors on the derived effective temperatures for the
more massive stars, as noted above. We provide for comparison, in
Table~10 and Fig.~20, the corresponding results which, with the same
physical assumptions, give a proper calibration of the observed
luminosity ($L_\odot=3.846\,(1\pm0.005)10^{33}$~erg~s$^{-1}$) and
radius ($R_\odot=6.9599\, 10^{10}$~cm) of the Sun at an age of 4.75
10$^9$ years\fonote{The age derived for the Sun by different authors
ranges from 4.5~Gyr (Guenther 1989) to 4.75~Gyr (Christensen-Dalsgaard
1982).  Guenther's comparisons of two solar models with ages of 4.5
and 4.7~Gyr did not yield significant differences in the derived solar
helium content nor in the mixing-length parameter.}  using
Z~=~$0.0175\pm0.0015$ (Grevesse \& Noels 1993b).

The resulting ZAMS for the Hyades lies significantly above that of the
Sun. Although the values of~Y found for the Sun (0.2659) and the
Hyades (0.260) are rather close, the higher metallicity of the Hyades
implies that the relative helium to metal enrichment ratio is smaller
than that obtained for the Sun (for the relevance of the assessment of
this enrichment ratio, see Pagel 1995). The value of Y~=~0.26 is close
to the value used by VandenBerg \& Bridges (1984) and would appear to
rule out the suspicions that the Hyades is helium deficient compared
with field stars (Str\"omgren et al.\ 1982). A refined value of the He
content of the Hyades will be presented in a forthcoming paper,
combined with the analysis of a number of Hyades binaries, whose
masses are very sensitive to the He content.

The uncertainty on our final estimate of Y has two sources. The
observational uncertainty on [Fe/H] gives an error on Y of 0.02, as
shown in Table~9. Additional uncertainties arise from the
determinations of $T_{\rm eff}$ and $M_{\rm bol}$.  >From three
different ZAMS models computed with Z~=~0.024, and with Y~=~0.25,
0.26, and 0.27, we infer that an error of 0.1~mag on $M_{\rm bol}$
leads to an error of about 0.025 on~Y. For the lower part of the main
sequence, and a given $M_{\rm bol}$, an error of 50~K on $T_{\rm eff}$
leads to an error of about 0.025 on~Y. Assuming that our individual
errors on $M_{\rm bol}$ and $T_{\rm eff}$ are uncorrelated, we
estimate that the observational scatter in the overall $T_{\rm
eff}$/$M_{\rm bol}$ diagram results in an error of only $\pm0.01$
in~Y. Combined with the independent error arising from the uncertainty
in~Z, we infer that the final error in~Y is dominated by the mean
value of [Fe/H] used for the Hyades, and is Y~=~$0.26\pm0.02$.

We note that the `single' objects apparently located $1-2\sigma$ above
the resulting main sequence have a metallicity close to the mean value
of the 40~stars. They may be undetected binaries, although their
velocity residuals with respect to the mean cluster space motion (see
Fig.~9) show no evidence for systematic departures which might be
expected as a result of their perturbed radial velocity. Since these
objects are not outliers in a colour-colour diagram, the $T_{\rm eff}$
values may also be slightly suspect.

\begtabfull
\tabcap{9} {Correspondence between [Fe/H] and Y, Z and Z/X used for 
the theoretical models described in the text.}
\centerline{\vbox {
\sevenrm \baselineskip 8pt
\tabskip 1em plus 2em minus 0.5em
\halign {\hfil#\hfil& \hfil#\hfil& \hfil#\hfil& \hfil#\hfil \cr
\noalign{\hrule}
\noalign {\smallskip}
$\scriptstyle [{\rm Fe/H}]$ & $\scriptstyle {\rm Y}$& $\scriptstyle {\rm Z}$&
$\scriptstyle {\rm Z/X}$ \cr
\noalign {\smallskip}
\noalign{\hrule}
\noalign {\smallskip}
0.14&    0.26&  0.024&  0.034\cr
0.09&    0.24&  0.020&  0.027\cr
0.19&    0.28&  0.028&  0.041\cr
\noalign {\smallskip}
\noalign{\hrule}
}}}
\endtab

\begtabfull
\tabcap{10} {$\log ({\rm L/L}_\odot)$ and $\log T_{\rm eff}$ as a function of 
mass for the Hyades (Y~=~0.260, Z~=~0.0240) and solar (Y~=~0.2659, Z~=~0.0175) 
zero-age main sequences derived from the CESAM code, as described in 
the text (see also Fig.~20).}
\centerline{
\vbox{
\def\p{\phantom{0}}
{\sevenrm \baselineskip 8pt
\vbox {\tabskip 1em plus 2em minus 0.5em
\halign {\hfil#\hfil& \hfil#\hfil& \hfil#\hfil& \hfil#\hfil& 
\hfil#\hfil& \hfil#\hfil& \hfil#\hfil \cr
\noalign{\hrule}
\noalign {\smallskip}
&\quad&	Hyades ZAMS\span\omit& \quad& Solar ZAMS\span\omit\cr
M/M$\scriptstyle _\odot$&&   
	log(L/L$\scriptstyle _\odot)$&   log $\scriptstyle T_{\rm eff}$&&
        log(L/L$\scriptstyle _\odot)$&   log $\scriptstyle T_{\rm eff}$\cr
\noalign {\smallskip}
\noalign{\hrule}
\noalign {\smallskip}
0.80&&  --0.695&   3.6632&&  --0.598&   3.6859\cr
0.90&&  --0.461&   3.6986&&  --0.362&   3.7216\cr
0.95&&  --0.354&   3.7144&&  --0.255&   3.7368\cr
1.00&&  --0.254&   3.7285&&  --0.161&   3.7498\cr
1.05&&  --0.164&   3.7408&&  --0.065&   3.7614\cr
1.10&&  --0.073&   3.7516&&   +0.032&   3.7717\cr
1.15&&   +0.018&   3.7615&&   +0.138&   3.7832\cr
1.20&&   +0.121&   3.7727&&   +0.231&   3.7934\cr
1.30&&   +0.301&   3.7932&&   +0.400&   3.8124\cr
1.40&&   +0.458&   3.8113&&   +0.551&   3.8309\cr
1.50&&   +0.598&   3.8286&&   +0.687&   3.8534\cr
1.60&&   +0.725&   3.8485&&   +0.810&   3.8800\cr
\noalign {\smallskip}
\noalign{\hrule}
}}}
}}
\endtab

\titleb{Isochrone fitting to the Hyades and the cluster age}

If the chemical composition of a cluster is known, the observational
HR diagram in combination with theoretical isochrones allow the
cluster age to be determined. Fig.~21 shows the HR diagram ($M_V$
versus $B-V$) for 131 stars where, in order to ensure minimal
contamination from non-cluster members, objects have been retained
only if they satisfy our kinematical membership criteria, and are
drawn from the objects lying within the $r<10$~pc radius of the
cluster centre (Table~2). From this sample, we have eliminated a few
with $\sigma_{B-V}>0.05$~mag. Error bars correspond to the standard
error in $M_V$, estimated from the standard error in the parallax, and
in $B-V$. The apparent magnitudes $V$, the $B-V$ colours, and the
standard errors in parallax and $B-V$ were taken from the Hipparcos
Catalogue.  Among these stars 72, indicated as filled circles in
Fig.~21, are not classified as (suspected) double or multiple
(Table~2) nor variable (as compiled in the Hipparcos Catalogue). The
open circles represent the double or multiple systems, spectroscopic
binaries, and variable stars, and these will not be used in further
discussions of the main sequence modelling. The main sequence has
significantly reduced scatter in comparison with the $V$, $B-V$
diagram of Fig.~2, and may be compared with recent determinations from
ground-based observations, e.g.\ Schwan (1991).\fonote{Dravins et al.\
(1997) have been able to further reduce the scatter in the Hyades HR
diagram by constraining the radial velocities and parallaxes of the
members according to the hypothesis of uniform space motion of the
cluster. Although a model-dependent approach, the reduced scatter in
their HR diagram, especially towards the faint end, supports our
conclusions about the absence of systematic velocity structure within
the cluster, whilst confirming our estimate of the internal velocity
dispersion, and providing further evidence that the majority of
binaries, at least with a not too large $\Delta m$, have been
identified.}

\begtabfull
\tabcap{11} {Stellar rotation, $v\sin i$, for stars in the turnoff region.}
\centerline{\vbox {
\sevenrm \baselineskip8pt
\tabskip 1em plus 2em minus 0.5em
\halign {\hfil#\hfil& \hfil#\hfil& \hfil#\hfil& \hfil#\hfil \cr
\noalign{\hrule}
\noalign {\smallskip}
  HIP&        $\scriptstyle M_V$&   $\scriptstyle B-V$&
$\scriptstyle v\sin i$  \cr
&                  &        &           (km~s$^{-1}$) \cr
\noalign {\smallskip}
\noalign{\hrule}
\noalign {\smallskip}
20542&         1.55&      0.15&       \phantom{0}40  \cr
20635&         0.85&      0.14&       \phantom{0}75  \cr
21029&         1.54&      0.17&       \phantom{0}70  \cr
21683&         1.23&      0.15&                 115  \cr
23497&         1.13&      0.16&                 115  \cr
\noalign {\smallskip}
\noalign{\hrule}
}}}
\endtab

As expected, most of the stars falling significantly above the main
sequence are double or multiple stars. There is no sub-dwarf sequence,
a fact already noted in previous work (Hanson \& Vasilevskis 1983,
Griffin et al.\ 1988). One star, HIP 17962, is located below the main
sequence. This is V471~Tau (WD~0347+171), an eclipsing binary
consisting of a K0V star and a white dwarf (Nelson \& Young 1970). In
the turnoff region ($B-V<0.2$~mag), there are five stars with no
indications of duplicity or variability: HIP~20542 (vB47), 20635
(vB54), 21029 (vB82), 21683 (vB108) and 23497 (vB129). The star with
the bluest colour index is HIP 20648 (vB56), a known blue straggler
(Abt 1985, Eggen 1995).

After elimination of the identified binaries, the lower part of the
main sequence has an increased scatter compared to that seen for the
bluer stars, a fact partly attributable to the lower accuracy of the
corresponding parallaxes (as illustrated by the error bars), but
possibly partially related to chromospheric activity. Campbell (1984)
has shown that many red Hyades dwarfs show colour anomalies which are
found to correlate with various indicators of chromospheric
activity. The four third-magnitude red giants $\gamma$ (vB28, HIP
20205), $\delta^1$ (vB41, HIP 20455), $\epsilon$ (vB70, HIP 20889) and
$\theta^1$ (vB71, HIP 20885) Tau represent the cluster's giant
branch. The stars vB41 and vB71 are known spectroscopic binaries.  The
star vB28 was reported to be double from the speckle result of Morgan
et al.\ (1982), although this result was not confirmed in the speckle
duplicity survey of the Hyades cluster of Mason et al.\ (1993),
possibly because the star was observed when the seeing was
approximately 2~arcsec.

\begfigwid 12cm
\vskip-12cm
\centerline{\psfig{figure=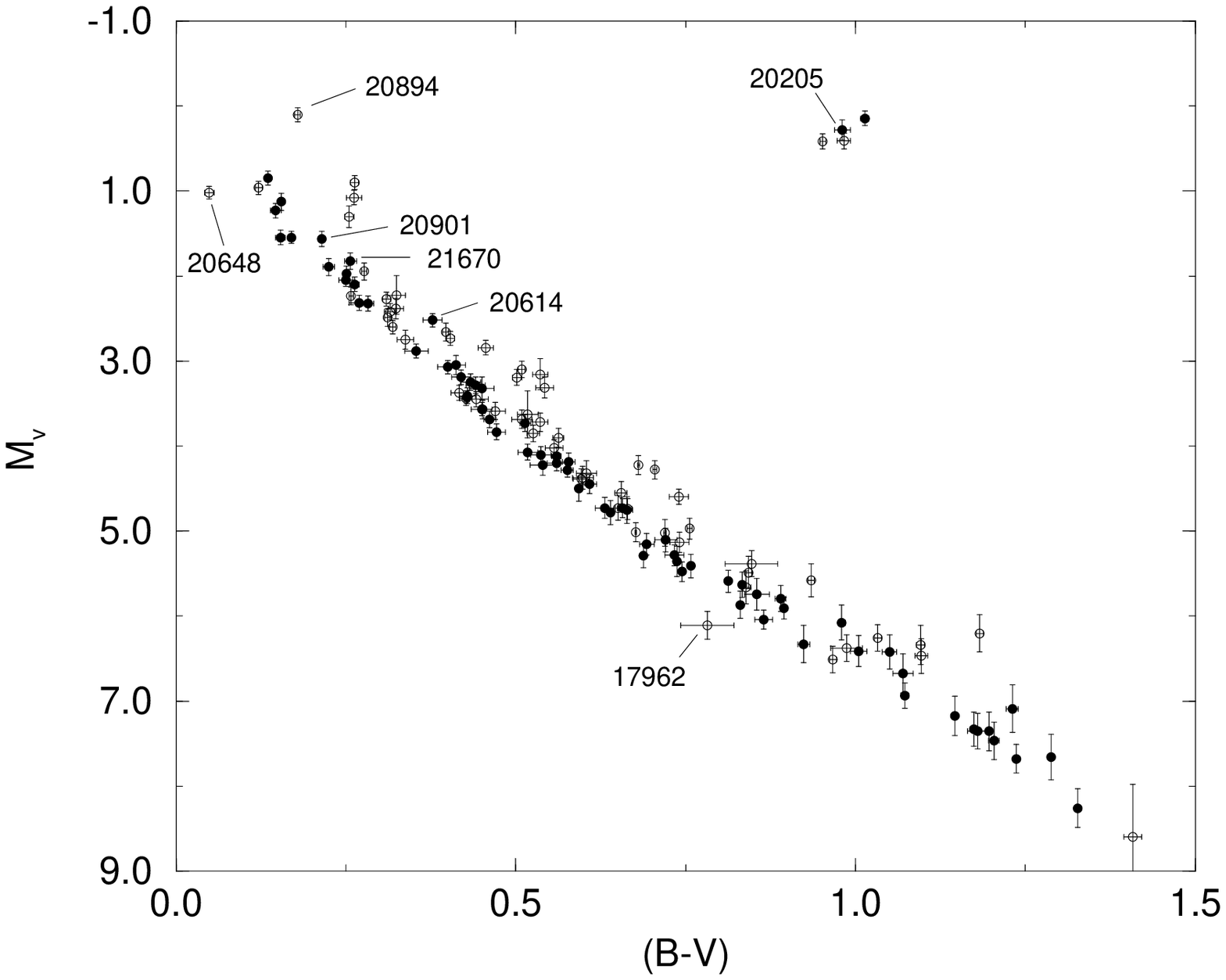,height=12cm}}
\vskip -20pt
\figure{21}{Absolute magnitude, $M_V$, versus $B-V$, for the stars 
considered as reliable cluster members within $r<10$~pc. Filled circles
indicate objects which are not classified as (suspected) double or 
variable. Error bars correspond to the standard error in the Hipparcos
parallaxes and $B-V$ colour indices. Specific objects indicated are 
discussed in the text.
}
\endfig

After accounting for known binaries and variable stars, the diagram
still appears to be possibly contaminated by unrecognised binaries. In
particular, the stars HIP 20901 (vB74), 21670 (vB107) and 20614 (vB53)
lie significantly above the cluster main sequence.  HIP 20901 and
21670 are apparently Am type stars (Abt \& Morrell 1995), amongst
which the high frequency of spectroscopic binaries is well established
(Abt 1961, Jaschek \& Jaschek 1987); the possibility that these stars
are binaries cannot therefore be ruled out. HIP 20614 is a fast
rotator ($v\sin i=145$~km~s$^{-1}$) (Abt \& Morrell 1995), and
photometry indicates a possible binary (Eggen 1992). These three stars
have been omitted in the subsequent fitting of the main sequence.

Rotation affects the colours of the stars, the effect depending on the
equatorial velocity and on the inclination of the rotational axis with
respect to the line of sight, leading to shifts of a few hundredths in
$B-V$ and a few tenths in $M_V$, generally towards the red and to
higher luminosities. Values of $v\sin i$ taken from Abt \& Morrell
(1995) for the stars in the turnoff region are given in Table~11.

In order to compare the observational HR diagram with theoretical
isochrones, evolutionary models were calculated using the same input
physics as for the ZAMS models from which the initial helium abundance
was estimated (Y~=~0.26, Z~=~0.024, $\alpha=1.64$). Sequences were
determined for masses of 0.8, 1.0, 1.2, 1.4, 1.7, 2.0, 2.5, 3.0 and
4.0~M$_\odot$, from the ZAMS to the beginning of the red-giant
branch. For each mass, two evolutionary sequences were calculated: a
standard sequence, and one taking into account an overshooting of the
mixed convective core which extends the size of the convective core
over a distance of 0.20 pressure scale-heights (the latter
significantly modifying the shape of the resulting isochrones and
hence the estimated ages -- the reference value of 0.2 comes from
Schaller et al.\ 1992). Representative evolutionary states for each
mass were extracted, and the Geneva isochrone program used to obtain
isochrones with ages in the range 500--750~Myr in steps of 50~Myr,
with and without overshooting.

One of the main difficulties in the transformation of isochrones from
the theoretical plane ($M_{\rm bol}$, $\log T_{\rm eff}$) to the
observational plane ($M_V$, $B-V$) lies in the colour/temperature
transformation. As a temperature indicator, the $B-V$ index has the
disadvantage of also being sensitive to metallicity. Moreover, since
the stellar surface gravity varies along the position on the
isochrone, the influence of gravity on the $B-V$ index has to be taken
into account. We adopted the calibration from Alonso et al.\ (1996)
which allows derivation of the $B-V$ colour as a function of $T_{\rm
eff}$ and [Fe/H]. This calibration, valid in the range 4000~K $<T_{\rm
eff}<$ 8000~K, was extrapolated to higher $T_{\rm eff}$ according to
the results of Haywood (1997, private communication). The adopted
calibration yields, for the Sun ($T_{\rm eff}=5780$~K), a $B-V$ index
of 0.62, in good agreement with the recent estimation of
$0.628\pm0.009$ by Taylor (1997). The transformed $B-V$ colours were
then corrected for the influence of gravity according to the
relationships given by Arribas \& Mart\'{\i}nez Roger (1988). Finally,
in order to estimate $M_V$ from $M_{\rm bol}$ the $V$ bolometric
corrections from Bessel et al.\ (1997) were adopted.

\begfig 8cm 
\vskip -8cm
\centerline{
\hskip -0.5cm \psfig{figure=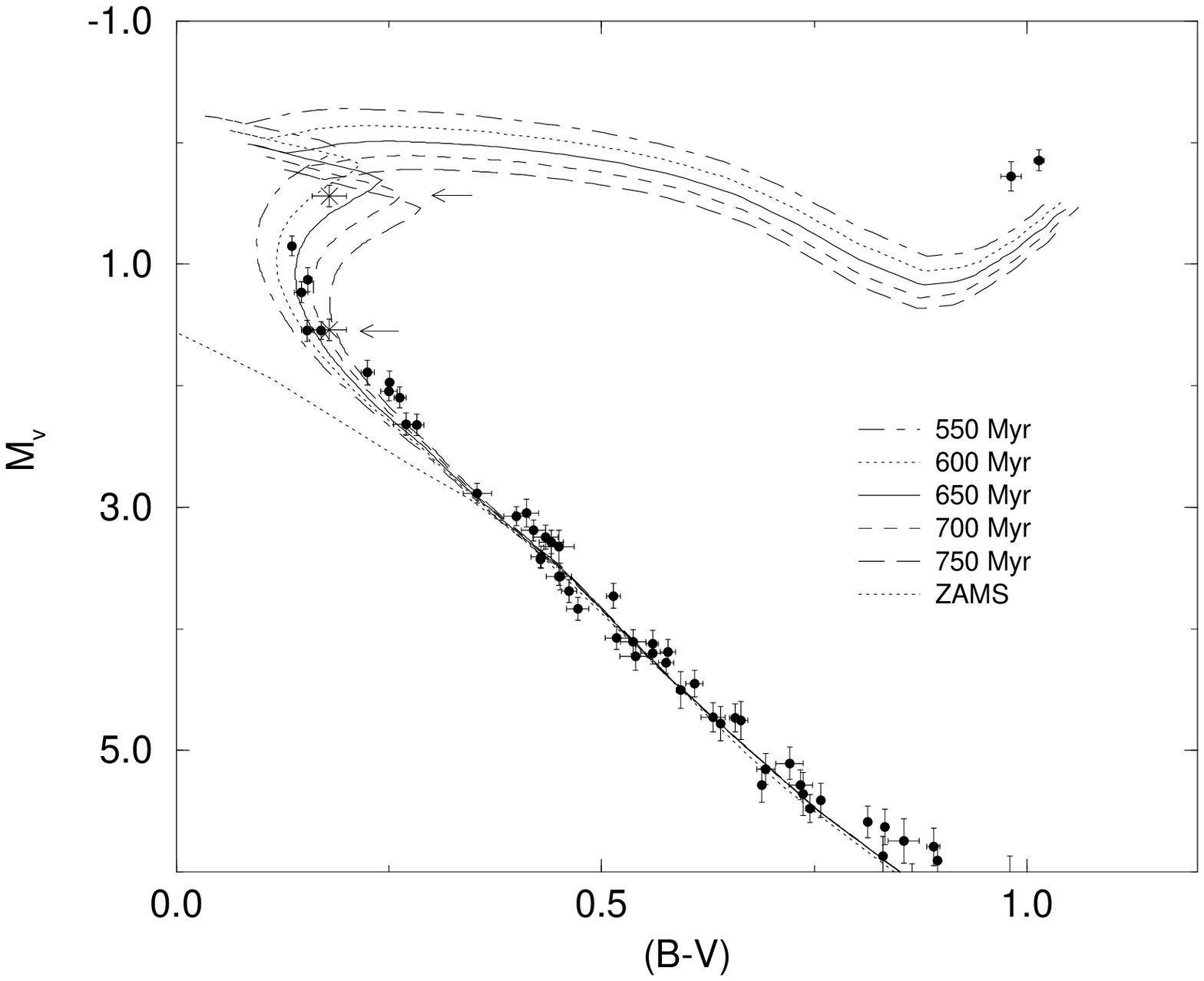,width=9.5cm} \hskip -0.5cm}
\vskip -12pt
\figure{22}{The 69 single stars with the location of the ZAMS, and 
with isochrones corresponding to the range 550--750~Myr 
calculated with overshooting. The two objects indicated by different
symbols (arrowed) are discussed in the text. 
}
\endfig

Fig.~22 shows the theoretical isochrones corresponding to 550, 600,
650, 700 and 750~Myr calculated with overshooting, superimposed on the
observational HR diagram. Fig.~23 shows isochrones corresponding to
500, 550, 600, 650, 700~Myr calculated without overshooting.  On both
figures the zero-age main sequence is also indicated.  Our present
results show that the five stars located in the turn-off region, for
which we have no evidence of duplicity, can be reasonably modelled
either with an isochrone of 550~Myr without overshooting, or with an
isochrone of about 650~Myr calculated with overshooting. As the
turnoff region is sparsely populated, we have included in the HR
diagram the star HIP~20894 ($\theta^2$ Tau, vB72). This is the
brightest star in Fig.~21, not counting those in the giant branch, and
is the brighter component of a wide visual pair (with $\theta^1$ Tau)
and a well-known spectroscopic binary. A difference in $V$~magnitude
of $1.10\pm0.01$ and in $B-V$ colour of $0.006\pm0.005$ between the
components of the SB star were estimated by Peterson et al.\ (1993).
We adopted these values in combination with the Hipparcos parallax to
place these two stars in Figs~22--23, where they are indicated as
different symbols (arrowed).

Using this system as an additional constraint, it appears that the
primary component of the system ($\theta^2$ Tau~A) does not lie on the
isochrones between 500 and 600~Myr without overshooting, while an
agreement with the isochrone grid including overshooting remains very
satisfactory. Although not included in Fig.~22 to avoid crowding, we
estimate that the theoretical isochrone corresponding to 625~Myr,
calculated with overshooting, provides an optimum fit to the present
observational data.

\begfig 8cm 
\vskip -8cm
\centerline{
\hskip -0.5cm \psfig{figure=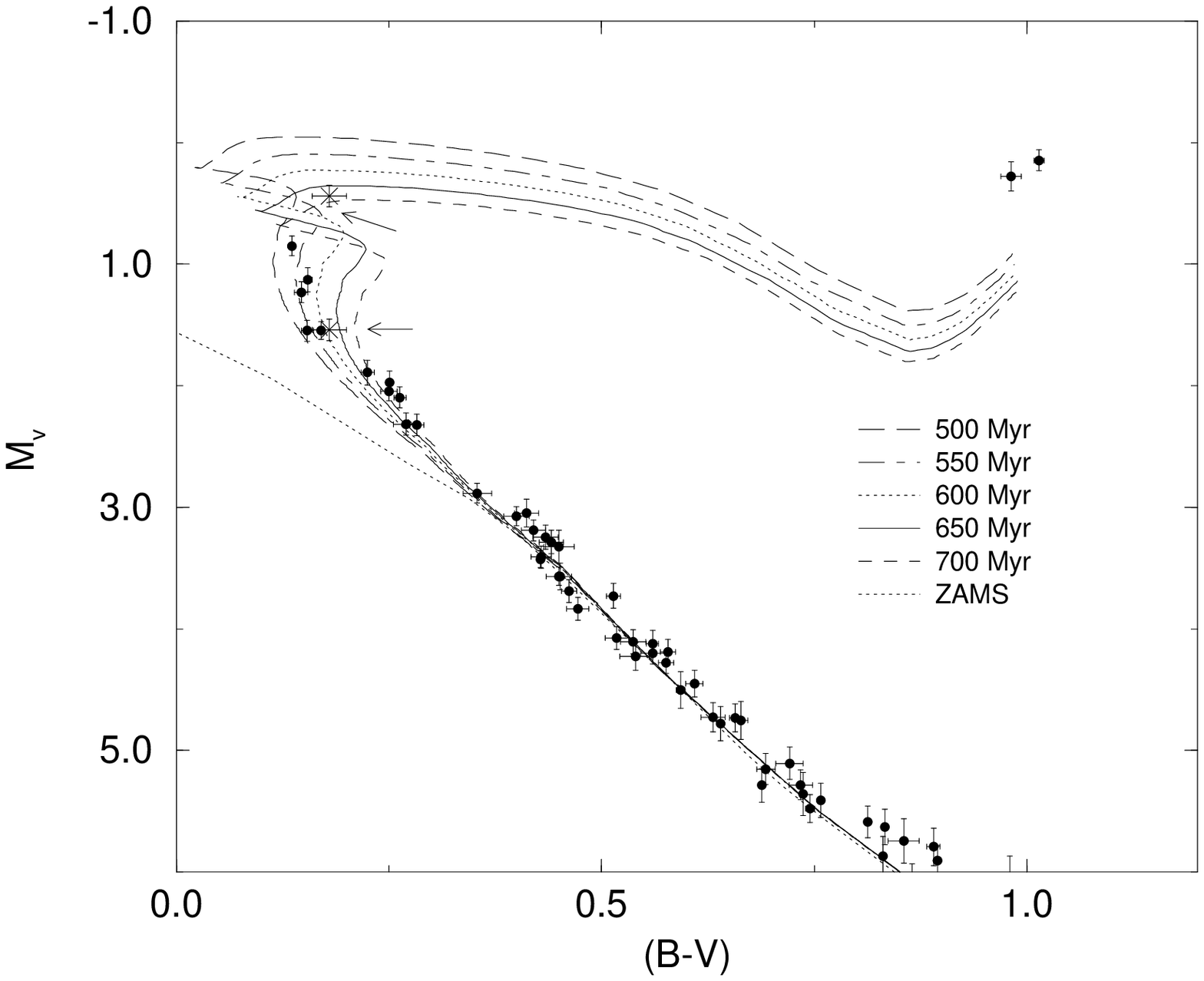,width=9.5cm} \hskip -0.5cm}
\vskip -15pt
\figure{23}{The 69 single stars with the location of the ZAMS, and 
with isochrones corresponding to the range 500--700~Myr 
calculated without overshooting. The two objects indicated by different
symbols (arrowed) are discussed in the text. 
}
\endfig

For a given set of models (with or without overshooting) we estimate
an age uncertainty of about 30~Myr coming from the visual fitting of
an isochrone to the observations. However, the cluster age
determination is model dependent, and an uncertainty of about 15~per
cent arises from the uncertainty on the amount of overshooting adopted
in the calculations. Another significant source of uncertainty comes
from the relationship adopted in the transformation between $T_{\rm
eff}$ and $B-V$, where the corresponding uncertainty in the age
determination could reach some 20~per cent. The effect of rotation,
relatively unimportant in the Hyades (as shown in Table~11), is
unlikely to lead to an overestimation of the age by 50~Myr (Maeder
1971). From all of these considerations, and taking into account
uncertainties coming from the adopted models, the transformation
between $T_{\rm eff}$ and $B-V$, and the effect of undetected
binaries, it is difficult to assign a very meaningful estimate of the
uncertainty of our age determination, which may reach 100~Myr,
although an uncertainty of about 50~Myr may be a more realistic
estimate. In summary, our results suggest a cluster age of
$625\pm50$~Myr, with observational evidence for the presence of
convective overshooting.  Support for overshooting from cluster main
sequence fitting was already presented by Maeder \& Mermilliod (1980),
who compared observational HR diagrams for 34 clusters in the age
range spanning the Pleiades to the Hyades, and found that agreement
could be obtained if the stellar convective cores are extended by a
certain amount due to convective overshooting or other physical
mechanisms.

Previous estimates of the age of the Hyades from isochrone fitting
ranged between 500 and 900~Myr (Barry et al.\ 1981).  Cayrel de
Strobel (1990) gave 655~Myr from the mean value of different age
determinations. Recently, Torres et al.\ (1997a) found 600~Myr.
Kroupa (1995) estimated a dynamical age of about 500~Myr.  Direct
comparison of our result with the isochrone-based ages quoted in the
literature is complicated by differences in the models used. All of
the previous estimates rely on models with solar composition, or
interpolated from models having different metallicities. In this work,
we have calculated models specifically for the Hyades abundance.

We stress that distances to star clusters based on main sequence
fitting to the Hyades must be corrected for chemical composition
differences: thus the fitted $m-M$ value for a system with solar
abundance should be reduced by roughly 0.13~mag to allow for the fact
that [Fe/H]$_{\rm Hyades}=0.14$.

\titlea{Conclusions}

The Hipparcos parallaxes and proper motions together provide a
consistent picture of the Hyades distance, structure and dynamics.
They yield a cluster convergent point motion consistent with the
individual trigonometric parallaxes, and together explain the larger
distance modulus derived from the most recent ground-based proper
motion investigations as originating from differences in the magnitude
of the adopted cluster space motion, and small systematic effects in
the ground-based proper motions. Conversely, the smaller distance
modulus traditionally derived from a variety of ground-based
trigonometric parallax programmes are attributed to errors in these
ground-based parallaxes, a conclusion supported by the most recent
distance modulus derived from consideration of the GCTSP parallaxes by
van Altena (1997b), in good agreement with our present results. There
is good agreement with determinations using high-precision radial
velocities (Stefanik \& Latham 1985, Gunn et al.\ 1988).  Recent
distance determinations to individual objects in the cluster, most
notably the results of Torres et al.\ (1997a,b,c), are in excellent
agreement with the Hipparcos trigonometric parallaxes, although their
extrapolation to a corresponding mean cluster distance is again
affected by systematic effects in the ground-based proper motions
used.

The combination of the Hipparcos astrometry with radial velocity
measurements from ground-based programmes provides three-dimensional
velocities allowing candidate membership selection to be based on
positional and kinematical criteria. A number of new cluster members
have been found within 20~pc of the cluster centre, and candidates can
be classified as escaping members on the basis of their velocity
residuals. No evidence for systematic internal velocity structure is
found; rather, the results are fully consistent with a uniform cluster
space motion with an internal velocity dispersion of about
0.3~km~s$^{-1}$. Spatial distribution, mass segregation, and binary
distributions are consistent with $N$-body simulations.

The cluster has a tidal radius of $r_t\simeq10$~pc.  Outside this
region, the stellar distribution is elongated along the direction of
the Galactic centre and anti-centre, and is slightly flattened in the
direction perpendicular to the Galactic plane.  Inside this region,
the cluster has spherical symmetry with a core radius of
$r_c\simeq2.7$~pc, and a half-mass radius of 5.7~pc.  The presence of
objects closely linked kinematically with the cluster core, but well
beyond the tidal radius, probably originates from stellar encounters
and diffusion beyond the Lagrangian points.

The well-defined observational main sequence has been transformed into
a theoretical $M_{\rm bol}$ versus $T_{\rm eff}$ diagram, from which
fitting of the cluster zero-age main sequence yields a helium
abundance of Y~=~$0.26\pm0.02$. Theoretical isochrones matching the
helium and metal content provide observational evidence for convective
overshooting, and yield a cluster age of $625\pm50$~Myr.

With the caveat that the primary importance of the Hipparcos results
is to provide individual distances to cluster members, rather than an
estimated distance to the cluster centre of mass (a concept meaningful
only in the restricted context of the cluster members contained in the
Hipparcos Catalogue), our estimated distance to the observed centre of
mass for the objects within 10~pc of the cluster centre is
$46.34\pm0.27$~pc, corresponding to a distance modulus
$m-M=3.33\pm0.01$~mag. This mean distance is, in practice, only
marginally modified (formally by about 0.4~pc) for the derived centre
of mass for Hipparcos objects within $r<20$~pc of the cluster centre.

\acknow{We thank the members of the Hipparcos scientific consortia
who participated in the creation of the Hipparcos Catalogue on which
this study has been based. We thank E.~H{\o}g, L.~Lindegren, R.S. Le
Poole, and H.~Schrijver for valuable comments and contributions.
M.~Mayor (Geneva) provided radial velocities from the Coravel
systematic survey of Hipparcos stars in advance of publication, and
S.~van Eck (Brussels) kindly obtained measurements for about half of
the newly-measured stars. We are grateful to W.~van Altena for
guidance in interpreting discrepancies between the Hipparcos and
ground-based parallaxes. Our particular thanks are due to the referee,
I.R.~King, for his careful and critical reading of the manuscript, and
for his numerous valuable and substantial suggestions for improvements
of the paper.}

\vskip -10pt
\begref{References}

\ref Aarseth, S.J., 1973, Vistas in Astr., 15, 13 

\ref Abt, H.A., 1961, ApJSS, 6, 37

\ref Abt, H.A., 1985, ApJ, 294, L103

\ref Abt, H.A., Morell, N.I., 1995, ApJSS, 99, 135

\ref Alexander, D.R., Ferguson, J.W., 1994, ApJ 437, 879

\ref Alonso, A., Arribas, S., Mart\'{\i}nez Roger, C., 1996, A\&A, 313, 873

\ref van Altena, W.F., 1974, PASP, 86, 217

\ref van Altena, W.F., Lee, J.T., Hoffleit, E.D., L\'opez, C.E., 1991,
General Catalogue of Trigonometric Parallaxes, Yale Univ. Press

\ref van Altena, W.F., Lee, J.T., Hoffleit, E.D., 1993,
in `Workshop on Databases for Galactic Structure', Davis Philip, A.G., 
Hauck, B., Upgren, A.R. (eds), L.~Davis Press, 65

\ref van Altena, W.F., Lee, J.T., Hoffleit, E.D., 1994,
in `Galactic and  Solar System Astronomy: Observation
and Application', Morrison, L.V., Gilmore, G.F., D. Reidel, 
Dordrecht, 50

\ref van Altena, W.F., Lee, J.T., Hoffleit, E.D., 1995,
General Catalogue of Trigonometric Stellar Parallaxes, Fourth Edition, Yale 
Univ.\ Obs.

\ref van Altena, W.F., Lu, C.L., Lee, J.T., et al., 1997a, ApJ Letters, in press

\ref van Altena, W.F., Lee, J.T., Hoffleit, E.D., 1997b, Baltic Astron., 6(1), 27

\ref Anders, E., Grevesse, N., 1989, Geochim.\ Cosmochim.\ Acta, 53, 197

\ref Arenou, F., Lindegren, L., Fr{\oe}schl\'e, M., G\'omez, A.E.,
Turon, C., Perryman, M.A.C., Wielen, R., 1995, A\&A, 304, 52

\ref Arribas, S., Mart\'{i}nez Roger, C., 1988, A\&A, 206, 63

\ref Battinelli, P., Capuzzo-Dolcetta, R., 1991, In `The Formation and 
Evolution of Star Clusters', Janes, K. (ed.), ASP Conf.\ Ser., 13, 139
 
\ref Barry, D.C., Cromwell, R.H., Hege, K., Schoolman, S.A., 
1981, ApJ, 247, 210

\ref Bessel, M., Castelli, F., Plez, B., 1997, A\&A, in press

\ref Binney, J., Tremaine, S., 1987, Galactic Dynamics, Princeton
University Press

\ref Blaauw, A., Gum, C.S., Pawsey, J.L., Westerhout, G., 1960, MNRAS, 121, 123

\ref Boesgaard, A.M., 1989, ApJ, 336, 798

\ref Boesgaard, A.M., Budge, K.G., 1988, ApJ, 332, 410

\ref Boesgaard, A.M., Friel, E.D., 1990, ApJ, 351, 467

\ref Boss, L., 1908, AJ, 26, 31

\ref Branch, D., Lambert, D.L., Tomkin, J., 1980, ApJ, 241, L83

\ref Breger, M., 1968, PASP, 80, 578

\ref Brosche, P., Denis-Karafistan, A.I., Sinachopoulos, D., 1992, A\&A 253, 113

\ref Brown, A., 1950, ApJ, 112, 225

\ref Buchholz, M., 1977, A\&A, 58, 377

\ref van Bueren, H.G., 1952, Bull.\ Astron.\ Inst.\ Neth., XI, 385

\ref Cameron, L.M., 1985, A\&A, 152, 250

\ref Campbell, B., 1984, ApJ, 283, 209

\ref Casertano, S., Iben, I., Shiels, A., 1993, ApJ, 410, 90

\ref Caughlan, G.R., Fowler, W.A., 1988, Atomic Data Nuc.\ Data Tables, 40, 284

\ref Cayrel de Strobel, G., 1980, In `Star Clusters', Hesser, J.E.\ (ed.), 
IAU Symp.\ 85, Reidel, 91

\ref Cayrel de Strobel, G., 1982, In `The Scientific Aspects of the 
Hipparcos Space Astrometry Mission', ESA SP--177, 173

\ref Cayrel de Strobel, G., 1990, Mem.\ Soc.\ Astron.\ Ital., 61, 613

\ref Cayrel de Strobel, G., 1996, A\&A Rev., 7, 243

\ref Cayrel, R., Cayrel de Strobel, G., Campbell, B., D\"appen, W., 1984, ApJ, 283, 205

\ref Cayrel, R., Cayrel de Strobel, G., Campbell, B., 1985, A\&A, 146, 249

\ref Chaffee, F.H., Carbon, D.F., Strom, S.E., 1971, ApJ, 166, 593

\ref Christensen-Dalsgaard, J., 1982, MNRAS, 199, 735

\ref Christensen-Dalsgaard J., 1991, in `Challenges
to Theories of the Structure of Moderate-Mass Stars', 
Gough D.O., Toomre J. (eds), Springer-Verlag, 11

\ref Corbin, T.E., Smith, D.L., Carpenter, M.S., 1975, Bull.\ Am.\ Astron.\ Soc., 7, 337

\ref Crawford, D.L., 1975, AJ, 80, 955

\ref Dame, T.M., Ungerechts, H., Cohen, R.S., de Geus, E.J., Grenier, I.A.,
May, J., Murphy, D.C., Nyman, L.-\AA ., Thaddeus, P., 1987, ApJ 322, 706

\ref Detweiler, H.L., Yoss, K.M., Radick, R.R., Becker, S.A., 1984, AJ, 89, 1038

\ref Dommanget, J., Nys, O., 1994, Comm.\ Obs.\ R.\ de Belg., Serie~A, No.~115

\ref Dravins, D., Lindegren, L., Madsen, S., Holmberg, J. 1997, 
in HIPPARCOS Venice '97, ESA SP--402, 733

\ref Eddington, A.S., 1913, MNRAS, 73, 359

\ref Eggen, O.J., 1960, MNRAS, 120, 540

\ref Eggen, O.J., 1967, Ann.\ Rev.\ Astron.\ Ast., 5, 105

\ref Eggen, O.J., 1969, ApJ, 158, 1109

\ref Eggen, O.J., 1982, ApJSS, 50, 221

\ref Eggen, O.J., 1992, AJ, 104, 1482

\ref Eggen, O.J., 1993, AJ, 106, 1885

\ref Eggen, O.J., 1995, AJ, 110, 823

\ref ESA, 1997, The Hipparcos and Tycho Catalogues, ESA SP-1200

\ref Feast, M.W., Whitelock, P.A., 1997, MNRAS, in press

\ref Foy, R., 1975, In `Abundance Effects in Classification', 
Hauck, B. Keenan, P.C.\ (eds), IAU Symp.\ 72, Reidel, 209

\ref Fuente Marcos, R. de la, 1995, A\&A, 301, 407

\ref Gatewood, G., Castelaz, M., de Jonge, J.K., Persinger, T.,
Stein, J., Stephenson, B., 1992, ApJ, 392, 710

\ref Golay, M., 1972, In `Problems of Calibration of Absolute Magnitudes
and Temperatures of Stars', Hauck, B., Westerlund, B.E.\ (eds), 
IAU Symp.\ 54, Reidel, 27

\ref Grevesse N., Noels A., 1993a, in `Origin and Evolution of the Elements',
Prantzos,  N., Vangioni-Flam E., Cass\'e, M. (eds), Cambridge University Press

\ref Grevesse N., Noels A., 1993b, `Association Vaudoise des
Chercheurs en Physique', `La Formation des Elements Chimiques', Hauck, B., 
Plantani, S., Raboud, D., (eds)

\ref Griffin, R.F., Gunn, J.E., Zimmerman, B.A., Griffin, R.E.M., 1988, AJ, 96, 172

\ref Guenther, D.B., 1989, ApJ 339, 1156

\ref Gunn, J.E., Griffin, R.F., Griffin, R.E.M., Zimmerman, B.A., 1988, AJ, 96, 198

\ref Hanson, R.B., 1975, AJ, 80, 379

\ref Hanson, R.B., 1977, BAAS, 9, 585

\ref Hanson, R.B., 1980, In `Star Clusters', Hesser, J.E.\ (ed.), 
IAU Symp.\ 85, Reidel, 71

\ref Hanson, R.B., Vasilevskis, S., 1983, AJ, 88, 844

\ref Hardorp, J., 1981, A\&A, 105, 120

\ref Hauck, B., 1981, A\&A, 99, 207

\ref Heckmann, O., L\"ubeck, K., 1956, Z.\ Astrophys., 40, 1

\ref Heintz, W.D., 1988, PASP, 100, 839

\ref Helfer, H.L., 1969, AJ, 74, 1155

\ref H\'enon, M., 1970, A\&A, 9, 24

\ref Henry, T.J., McCarthy, D.W., Jr., 1993, AJ, 106, 773

\ref Hertzsprung, E., 1909, ApJ, 30, 135

\ref Hodge, P.W., Wallerstein, G.W., 1966, PASP, 78, 411

\ref Ianna, P.A., McNamara, B.R., Greason, M.R., 1990, AJ, 99, 415 

\ref Iben, I., 1967, Ann.\ Rev.\ Astron.\ Ast., 5, 571

\ref Iben, I., Tuggle, R.S., 1972, ApJ, 173, 135

\ref Iglesias C. A., Rogers F. J., 1996, ApJ 464, 943

\ref Innanen, K.A., Harris, W.E., Webbink, R.F., 1983, AJ, 88, 338

\ref Jaschek, C., Jaschek, M., 1987, The Classification of Stars, Cambridge 
University Press

\ref Johnson, H.L., Mitchell, R.I., Iriarte, B., 1962, ApJ, 136, 75

\ref Jones, D.H.P., 1971, MNRAS, 152, 231

\ref Kholopov, P.N., 1969, Soviet AJ, 12, 4

\ref King, I.R., 1962, AJ, 67, 471

\ref King, I.R., 1966, AJ, 71, 64

\ref Klemola, A.R., Harlan, E.A., McNamara, B., Wirtanen, C.A., 1975, AJ, 80, 642

\ref Koester, D., Weidemann, V., 1973, A\&A, 25, 437

\ref Kovalevsky, J., Lindegren, L., Froeschl\'e, M., et al., 1995, A\&A, 304, 34

\ref Kroupa, P., 1995, MNRAS, 277, 1522

\ref Lebreton, Y., Michel, E., Goupil, M.J., Baglin, A., Fernandes, J., 1995,
In `Astronomical and Astrophysical Objectives of Sub-Millarcsecond
Optical Astrometry', H\o g, E., Seidelman, P.K., (eds), IAU Symp.\ 166,
Kluwer, 135

\ref Lindegren, L., 1989, In `The Hipparcos Mission', 
ESA SP--1111, Vol.~III, Chapter~18.

\ref Lindegren, L., Kovalevsky, J., 1995, A\&A, 304, 189

\ref Lindegren, L., 1995, A\&A, 304, 61

\ref Lindegren, L., R\"oser, S., Schrijver, H., et al., 1995, A\&A, 304, 44

\ref Loktin, A.V., Matkin, N.V., Fedorov, V.V., 1987, Sov.\ Astron., 31, 582

\ref Loktin, A.V., Matkin, N.V., 1989, Astron.\ Nachr., 310(3), 231

\ref Luri, X, Arenou, F., 1997, in HIPPARCOS Venice '97, ESA SP--402, 449

\ref Lutz, T.E., 1970, AJ, 75, 1007

\ref Lutz, T.E., Kelker, D.H., 1973, PASP, 85, 573

\ref Ma, C., et al., 1997, IERS Technical Note 24, Obs.\ de Paris

\ref Maeder, A., 1971, A\&A, 10, 354

\ref Maeder, A., Mermilliod, J.-C., 1980, A\&A, 93, 136

\ref Mason, B.D., McAlister, H.A., Hartkopf, W.I., Bagnuolo, W.G., 1993, 
AJ, 105, 220

\ref Mathieu, R.D., 1985, In `Dynamics of Star Clusters', 
Goodman, J., Hut, P. (eds), IAU Symp.\ 113, Reidel, 427

\ref McAllister, H.A., 1977, AJ, 82, 487

\ref McClure, R.D., 1982, ApJ, 254, 606

\ref Mermilliod, J.C., 1995, In `Information and On-line Data in Astronomy',
Egret, D., Albrecht, M.A. (eds), Kluwer, 127

\ref Morel P., 1993, IAU Coll.\ 137, In `Inside the Stars', 
Weiss, W.W., Baglin, A. (eds), ASP Conf.\ Ser., 40, 445

\ref Morel, P., 1997, A\&ASS, in press

\ref Morgan, B.L., Beckmann, G.K., Scaddan, R.J., Vine, H.A., 1982, 
MNRAS, 198, 817

\ref Morris, S.C., 1992, JRASC, 86, 292

\ref Morris, S.C., Luyten, W.J., 1983, BAAS, 15, 683

\ref Murray, C.A., 1989, A\&A, 218, 325

\ref Murray, C.A., Harvey, G.M., 1976, R.\ Greenwich Obs.\ Bull., 182, 15

\ref Nelson, B., Young, A., 1970, PASP, 82, 699
 
\ref Oort, J.H., 1979, A\&A, 78, 312

\ref Pagel, B.E.J., 1995, In `Astronomical and Astrophysical Objectives of 
Sub-Millarcsecond Optical Astrometry', H\o g, E., Seidelman, P.K., (eds), 
IAU Symp.\ 166, Kluwer, 181

\ref Patterson, R.J., Ianna, P.A., 1991, AJ, 102, 1091

\ref Pearce, J.A., 1955, PASP, 67, 23

\ref Pels, G., Oort, J.H., Pels-Kluyver, H.A., 1975, A\&A, 43, 423

\ref Perryman M.A.C., Lindegren, L., Kovalevsky, J., et al., 1995, A\&A, 304, 69

\ref Peterson, D.M., Solensky, R., 1987, ApJ, 315, 286

\ref Peterson, D.M., Stefanik, R.P., Latham, D.W., 1993, AJ, 105, 2260

\ref Reid, N., 1992, MNRAS, 257, 257

\ref Reid, N., 1993, MNRAS, 265, 785

\ref Scarfe C.D., Batten A.H., Fletcher J.M., 1990, Publ.\ DAO XVIII, 21
  
\ref Schaller, G., Schaerer, D., Meynet, G., Maeder, A., 1992, A\&SS, 96, 269 

\ref Schwan, H., 1990, A\&A, 228, 69

\ref Schwan, H., 1991, A\&A, 243, 386

\ref Seares, F.H., 1944, ApJ, 100, 255 

\ref Seares, F.H., 1945, ApJ, 102, 366

\ref Sears, R.L., Whitford, A.E., 1969, ApJ, 155, 899

\ref Smart, W.M., 1938, Stellar Dynamics, Cambridge University Press

\ref Smart, W.M., 1939, MNRAS, 99, 168

\ref Smith, H., Eichhorn, H., 1996, MNRAS, 281, 211

\ref Spitzer, L. 1975, In `Dynamics of Stellar Systems', 
Hayli, A. (ed.), IAU Symp.\ 69, Reidel, 1

\ref Stefanik, R.P, Latham, D.W., 1985, In `Stellar Radial Velocities',
Davis Philip, A.G., Latham, D.W.\ (eds), IAU Coll.\ 88, L.~Davis Press, 213

\ref Str\"omberg, G., 1922, ApJ, 56, 265

\ref Str\"omberg, G., 1923, ApJ, 57, 77

\ref Str\"omgren, B., Olsen, E.H., Gustafsson, B., 1982, PASP, 94, 5

\ref Taylor, B.J., 1980, AJ, 85, 242

\ref Taylor, B., 1997, In `Fundamental Stellar Properties: 
the Interaction between Observation and Theory', IAU Symp.\ 189, in press

\ref Terlevich, E., 1987, MNRAS, 224, 193

\ref Thackeray, A.D., 1967, In `Determination of Radial Velocities and 
their Application', Batten, A.H., Heard, J.F.\ (eds), IAU Symp.\ 30, 
Academic Press, 163

\ref Theuns, T., 1992a, A\&A, 259, 493

\ref Theuns, T., 1992b, A\&A, 259, 503

\ref Torres, G., Stefanik, R.P., Latham, D.W., 1997a, ApJ, 474, 256

\ref Torres, G., Stefanik, R.P., Latham, D.W., 1997b, ApJ, 479, 268

\ref Torres, G., Stefanik, R.P., Latham, D.W., 1997c, ApJ, in press

\ref Turner, D.G., Garrison, R.F., Morris, S.C., 1994, JRASC, 88, 303

\ref Upgren, A.R., 1974, AJ, 79, 651

\ref Upgren, A.R., Weis, E.W., Fu, H.-H., Lee, J.T., 1990, AJ, 100, 1642

\ref Upton, E.K.L., 1970, AJ, 75, 1097

\ref Upton, E.K.L., 1971, AJ, 76, 117

\ref VandenBerg, D.A., Bridges, T.A., 1984, ApJ, 278, 679

\ref VandenBerg, D.A., Poll, H.E., 1989, AJ, 98, 1451

\ref Wallerstein, G., Hodge, P.W., 1967, ApJ, 150, 951

\ref Wayman, P.A., 1967, PASP, 79, 156

\ref Wayman, P.A., Symms, L.S.T., Blackwell, K., 1965, Royal Obs.\ Bull., No.~98

\ref Weidemann, V., Jordan, S., Iben, I., Casertano, S., 1992, AJ, 104, 1876

\ref Wielen, R., 1967, Veroff.\ Astr.\ Rechen-Inst., Heidelberg, 19

\ref Wielen, R., 1974, Proc. First European Astronomical Meeting, Mavridis,
L.N. (ed.), 2, 326

\ref Wielen, R., 1997, A\&A, in press
\endref

\bye